\documentclass[aps,prd,twocolumn,superscriptaddress,nofootinbib]{revtex4}
\usepackage{graphicx}
\usepackage{epsfig}
\usepackage{bm}
\usepackage{latexsym,amssymb,amsmath,amsfonts,amssymb,txfonts,pxfonts,wasysym,float}
\usepackage{enumitem}
\usepackage{color}
\usepackage{mathrsfs}
\usepackage{wasysym}

\usepackage{pifont}

\newcommand{\postscript}[2]{\setlength{\epsfxsize}{#2\hsize}
   \centerline{\epsfbox{#1}}}

\newcommand{\comment}[1]{}

\usepackage[usenames,dvipsnames]{xcolor}
\definecolor{orange}{cmyk}{0,0.5,1,0}
\definecolor{rossoCP3}{cmyk}{0,.88,.77,.40}
\definecolor{graa}{rgb}{0.8,0.8,0.8}
\definecolor{blaa}{rgb}{0.2,0.2,0.6}

\begin{document}

\title{\color{rossoCP3}{Understanding  the wicked world that surrounds you}}

\author{Luis A. Anchordoqui}
\affiliation{Department of Physics \& Astronomy,  Lehman College, City University of
  New York, NY 10468, USA}
\affiliation{Department of Physics,
 Graduate Center, City University
  of New York,  NY 10016, USA}
\affiliation{Department of Astrophysics,
 American Museum of Natural History, NY
 10024, USA}
\author{Gabriela V. Anchordoqui}
\affiliation{Sociedad Odontol\'ogica La Plata, Calle 13 No. 680, La
  Plata 1900, Argentina}
\date{Fall 2017}

\begin{abstract}
  \noindent There is a new urgency for wider segments of the
  non-expert population to join in shaping the future direction of our
  technological society. Certain decisions being made now in the
  development of energy resources will profoundly affect the lives of
  the next few generations. This course is oriented toward providing
  an educational base for participating in these decisions. At this
  level, there is no need for any mathematics beyond simple arithmetic,
  nor for any previous knowledge of physics or chemistry. 
\end{abstract}

\maketitle

\section{Conservation of Energy}
\label{sec:1}

{\bf \S~What is energy?}
\begin{equation}
{\rm Energy} \equiv \left \{\begin{array}{l} {\rm makes \ things \ happen} \\
{\rm does \, work}\\
{\rm does \ not \ change \ as \ things  \ happen} \end{array} \right.  \, .
\label{energy}
\end{equation}
The last statement in (\ref{energy}) is a profound and subtle truth, a
great discovery of the 19th century; namely, that energy is not only
an intuitive vagary (``I have lots of energy'') but can be measured,
and once you have a certain amount in a close system (e.g., Earth) you
keep it, but it does change form. This is called the {\it first law of
  thermodynamics:}  energy changes in form but not in
  amount~\cite{Clausius}.  Our world is full of transformation of
energy, e.g.,
\begin{itemize}
\item Stored chemical energy in muscles $\Rightarrow$ mechanical
  energy of movement and heat energy in muscles.
\item Stored gravitational energy of water in a dam~$\Rightarrow$~energy of falling water~$\Rightarrow$~mechanical energy of
  generators $\Rightarrow$ electrical energy.
\item Stored nuclear energy in uranium $\Rightarrow$ kinetic energy of
  fission fragments $\Rightarrow$ heat energy in
  water~$\Rightarrow$~electrical energy.
\end{itemize}
In all these process, there is a quantity whose amount is
the same before and after the process. That quantity is energy.\\ 

The {\it second law} is more subtle. It says that not only can't you
increase the amount of energy in the system, {\it but in any process
you always degrade part of it to a form which is less useful}~\cite{Carnot}. For
example, the chemical energy stored in the muscles is the source of
some useful work, but part of it is also converted to heat which can
never be entirely recovered for useful work. (Usually, it is entirely
wasted.) Many times these losses are due to friction.\\

\mbox{{\bf \S~If energy is a quantity, how do you measure it?}} This is a long story, and as a result of the length of the story, we have many
different units. In the U.S., a common unit for {\it electrical
energy} is a kilowatt-hour (kWh), and for  {\it heat energy} either the British
thermal unit (Btu), or kilocalorie (kcal or Cal). The latter is the one
used in diet planning. Since heat and electrical energy can be changed
into one another without loss, there is a fixed relation between the
units; namely, 1~kWh = 860~Cal, 1~kWh = $3,412$~Btu, 1~Cal = 3.967~Btu. \\

{\bf \S~What do these units mean?}~A Btu raises the temperature of a
pint (1~lb) of water $1^\circ$F. So to
bring 1 quart (2~lb) of H$_2$O from $42^\circ$F to boiling
($212^\circ$F) takes 
\begin{equation}
2 \times (212 - 42)^\circ{\rm F}= 340~{\rm Btu} \, .
\end{equation}
This by itself is not very enlightening; but when you are
told that every time a cubic foot of natural gas is burnt, $1,000$~Btu of
heat (photon energy)  are released, and therefore 3 quarts of water
can be boiling; it is only then that you can plan how many cubic feet
of gas you need to power a city and for better or worse, planning is
absolutely essential in our modern density populated society. It is no
longer possible for everyone to go out and chop a tree on a moment's
notice. A Cal raises the temperature of 1 liter of water $1^\circ$~C.

The kWh is most often used as a measure of electrical energy, and is
the energy supplied by a 10 ampere-current from a 100 volt line
running for 1 hour. This is the energy released in an hour's use of a
toaster, or 10 hours' use of a 100 watt bulb.

A $1,000$ watt heater, running for one hour uses 1~kWh, and
shows up as a 12\cent{} cost on your month bill. Burning a gallon of
gasoline releases 36 kWh of heat, or $125,000$~Btu or about
$31,000$~Cal. Your daily food consumption (about $2,200$~Cal) is about
2.5~kWh. If you didn't get rid of the heat, you would boil in 3
days. Compare the price of 1~kWh of electricity to that of food, which
is about \$3.00. The kWh is incredibly cheap for what it does. The trouble is
we don't really appreciate what it does. Imagine having no electric
lighting, and then someone offers you the use of a 60 watt bulb for
two hours a night for a week, all for a nickel -- you take it!
\begin{figure*}[t]
    \postscript{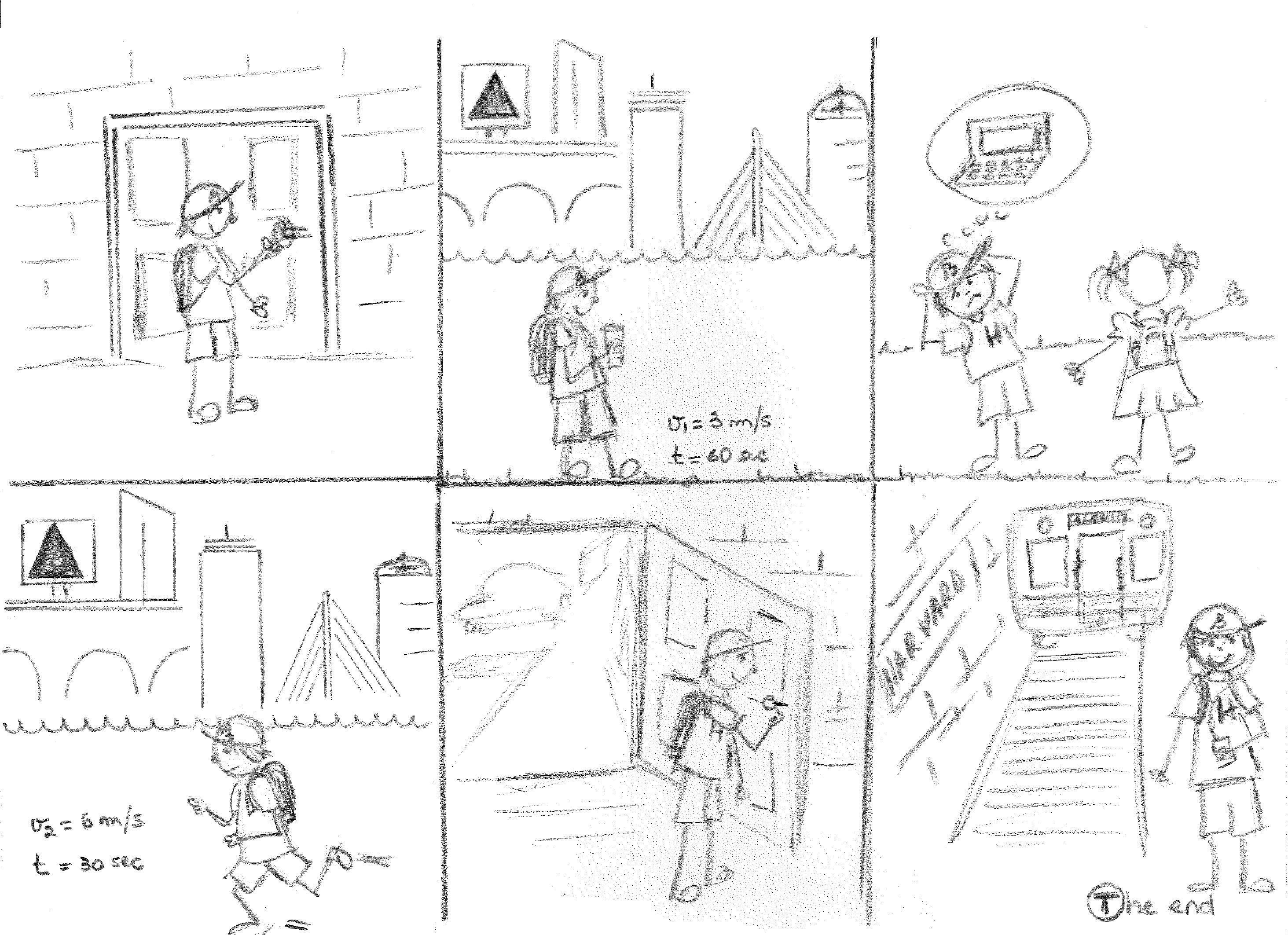}{0.8}
\caption{When Harry met Sally.
\label{fig:0}}
\end{figure*}

A kWh is about $3,400$~Btu. This means that the 100 watt bulb going for
10 hours submerged under 100~lb of water will raise the temperature of
that water by $34^\circ$F. 

 If energy is used at 1~kWh/hour, we simply say the {\it
power usage} is 1~kW (1~kilowatt). So an electric bulb burns at 0.1~kW
(100~watts).\\

{\bf \S~Relative motion interlude.} Everything is in motion. Even the stuff that
appears to be motionless moves, but (of course) the motion is
relative. For example, while you are reading these notes, you are
moving at about $107,000$~kilometers per hour relative to the Sun, and
you are moving even faster relative to the center of the Galaxy. When
we discuss the motion of something we describe the motion with respect
to something else. In other words, to describe the motion of something
we need a reference point, sometimes called the origin, and a
reference of time, e.g.: A long time ago in a galaxy far, far away...

The {\it position} defines an object's location in space with respect
to the origin. To determine the object's location you need a measuring
stick. The {\it displacement} defines the change in position that
occurs over a given period of time. Note that it is not the same to
move 3 meters north from the origin than 3 meters south. To
distinguish a displacement to the north of the origin from that to the
south, or the east, or the west, or up, or down, we say the
displacement has a magnitude (3 meters in the example above) and a
direction (north, south, east, west, up, or down). Objects with
magnitude and direction are generically called {\it vectors}.

Consider the situation of the cartoon in Fig.~\ref{fig:0}. One
morning, on his way to work, Harry left home on time, and he walk at
3~m/s east towards downtown. After exactly one minute, {\it when Harry
  met Sally}, he realized that he had left his computer at home, so he
turns around and runs, at 6~m/s, back to get it. He was running twice
as fast as he walked, so it took half as long (30 seconds) to get home
again.

There are several ways to analyze those 90 seconds between the time
Harry left home and the time he arrived back again. One number to
calculate is his {\it average speed}, which is defined as the total
distance covered divided by the time. If he walked for 60~seconds at
3~m/s, he covered 180~m. Then he covered the same distance on the way
back, so he went 360~m in 90~seconds,
\begin{equation}
{\rm average \ speed} = \frac{\rm distance}{\rm elapsed \ time} =
4~{\rm m/s} \,
.
\end{equation}
The {\it average velocity}, on the other hand, is given by:
\begin{equation}
{\rm average \ velocity} = \frac{\rm displacement}{\rm elapsed \ time}  
\, .
\end{equation}
Note that for the situation described above Harry's average velocity for
the round trip is zero, because he was back where he started, so the
{\it displacement} was zero.

We usually think about speed and velocity in terms of their
instantaneous values, which tell us how fast, and, for velocity, in
what direction an object is traveling at a particular instant. The
{\it instantaneous velocity} is defined as the rate of change of position
with time, for a very small time interval.  The {\it instantaneous speed} is
simply the magnitude of the instantaneous velocity.

An object accelerates whenever its velocity changes. The {\it
  acceleration vector} measures the rate at which an object speeds up,
slows down, or changes direction. Any of these variations constitutes a
change in velocity. Going back to the example we used above, let's say
instead of instantly breaking into a run the moment Harry turned around,
he steadily increased his velocity from 3~m/s west to 6~m/s west in
a 10 second period. If his velocity increased at a constant rate, he
experienced a constant acceleration of 0.3~m/s per second (or, 0.3
m/s$^2$). We can figure out the average velocity during this time. If
the acceleration is constant, which it is in this case, then the
average velocity is simply the average of the initial and final
velocities. The average of 3~m/s west and 6~m/s west is 4.5~m/s
west. This average velocity can then be used to calculate the distance
he traveled during the acceleration period, which was 10~seconds
long. The distance is simply the average velocity multiplied by the
time interval, so 45~m.

Similar to the way the average velocity is related to the
displacement, the {\it average acceleration} is related to the change in
velocity: the average acceleration is the change in velocity over the
time interval (in this case a change in velocity of 3~m/s in a time
interval of 10 seconds). The {\it instantaneous acceleration} is is defined
as the rate of change of velocity with time, for a very small time
interval. As with the instantaneous velocity, the time interval is
very small (unless the acceleration is constant, and then the time
interval can be as big as we feel like making it).

On the way out, Harry traveled at a constant velocity, so his
acceleration was zero. On the trip back his instantaneous
acceleration was 0.3~m/s$^2$ for the first 10 seconds, and then zero
after that as he maintained the top speed. Just as Harry arrived back
at his front door, his instantaneous acceleration would be negative,
because his velocity dropped from 6~m/s west to zero in a small time
interval. If he took 2 seconds to come to a stop, his acceleration
was $a = -3~{\rm m/s^2}$. To return, Harry decided to take the subway
to ensure he was going to arrive on time.\\

{\bf \S~Types of energy.} There are two types of energy: {\it potential} and
{\it kinetic}. Potential energy is energy that is stored in an object due to
its position. This type of energy is not in use, but is available to do
work. For example, a book possesses potential energy when it is
stationary on the top of the bookshelf. Kinetic energy is energy that
is possessed by an object due to its motion. For example, if the book
were to fall off the shelf, it would possess kinetic energy as it
fell. The kinetic energy of an
object depends on the mass $m$ of the object as well as its speed $v$,
\begin{equation}
{\rm kinetic \, energy} \equiv K = \frac{1}{2} \, m \, v^2 \, .
\end{equation}
All energy is either potential or kinetic. The fantastic thing about
this commodity, energy, is that when the conversion from potential to
kinetic takes place it occurs in a definite, predictable way. For
example, a ton of water dropping over Niagara Falls {\it always}
yields the same amount of electrical energy. Or the combustion of a
gallon of oil {\it always} gives the same amount of heat (thermal
energy). Thus, all forms of energy have the same measure.

\section{Forms of Energy}
\label{sec:2}

{\bf \S~Mechanical energy.} It represents the energy that is possessed by
a mechanical system or device due to its motion or position. Stated
differently, mechanical energy is the ability of an object to do
work. Mechanical energy can be either kinetic (energy in motion) or
potential (energy that is stored). The sum of an object's kinetic and
potential energy equals the object's total mechanical energy. 

The difference between kinetic and mechanical energy is that kinetic
is a type of energy, while mechanical is a form that energy takes. For
instance, a bow that has been drawn and a bow that is launching an
arrow are both examples of mechanical energy. However, they do not
both have the same type of energy. The drawn bow is an example of
potential energy, because the energy necessary to launch the arrow is
only being stored in the bow; whereas the bow in motion is an example of
kinetic energy, because it is doing work. If the arrow strikes a bell,
some of its energy will be converted to sound energy. It will no
longer be mechanical energy, but it will still be kinetic energy.\\

{\bf \S~Thermal energy.} This is the kinetic energy of molecules
moving in a random way. The {\it faster} they move, the higher the
{\it temperature}. The thermal energy may be transfered from one body
(say the ocean) to another (say the air). This is called {\it
heating} something. The transfer occurs through the collision of the
speedy molecules in the warm body (the ocean) with the sluggish
molecules in the cold body (the air), resulting in a rise of
temperature of the air.\\

{\bf \S~Radiant energy (photons).} This is a form of kinetic energy
carried by gamma rays, X-rays, ultraviolet (UV) rays, light, infrared
(IR) rays, microwaves, and radio waves. The energy comes in tiny
packets called photons, and the energy in each packet depends on the
type of radiant energy: X-ray photons carry higher energy than UV
photons, which are in turn more energetic than light photons, 
etc~\cite{Einstein:1905cc}. Light photons themselves differ in the amount of energy they
carry: blue photons more than yellow, yellow more than red, etc. The
spread is enormous: an X-ray photon is about $10,000$ times as
energetic as a blue photon, which in turn is about 10 billion times as
energetic as a photon carrying an AM signal. Single photons are
detectable by the eye only under very special conditions. Typically, a
60 watt light bulb is emitting $10^{19}$ (10 billion billion) photons
per second of visible light. Even at a distance of 100 yards, your eye
would be intercepting about 5 billion of these per second. Only at a
distance of about 600 miles from the light bulb does your eye
intercept an average of only one of these photons per second.\\

{\bf \S~Newtonian dynamics interlude.}~The newtonian idea of {\it
  force} is based on experimental
observation~\cite{Newton:1687}. Experiment tells us that everything in
the universe seems to have a preferred configuration: e.g., {\it (i)}
masses attract each other; {\it (ii)} magnets can repel or attract one
another. The concept of a force is introduced to quantify the tendency
of objects to move towards their preferred configuration. If objects
accelerate very quickly towards their preferred configuration, then we
say that there is a big force acting on them. If they don't move (or
move at constant velocity), then we say there is no force.

We cannot see a force; we can only deduce its
  existence by observing its effect.  More specifically, forces are
  defined through Newton's laws of motion:
\begin{enumerate}[start=0]
 \item A {\it
    particle} is a small mass at some position in space. 
\item When the sum of the forces acting on a particle is zero,
  its velocity is constant.
\item The sum of forces acting on a
  particle of constant mass is equal to the product of the mass of the
  particle and its acceleration:
\begin{equation}
{\rm force} = {\rm mass} \times {\rm acceleration} \, .
\end{equation}
\item The forces exerted by two
  particles on each other are equal in magnitude and opposite in
  direction.
\end{enumerate}
The standard unit of force is the newton, given the symbol N.  The
newton is a derived unit, defined through Newton's second law of
motion: one newton is the force needed to accelerate one kilogram (kg
= 2.2~lb) of mass at the rate of one meter per second squared in
direction of the applied force. 

Now, a point worth noting at this
juncture is that forces are vectors, which evidently have both
magnitude and direction.  For example, the {\it gravitational force} is a force that attracts any two
objects with mass~\cite{Newton:1687}.  The magnitude of this force is
directly dependent upon the masses of both objects $m$ and $M$ and
inversely proportional to the square of the distance  $r$ that separates
their centers,
\begin{equation}
{\rm gravitational \ force} = F_g = \frac{G \, M \, m}{r^2} \, ,
\label{GF}
\end{equation}
where $G = 6.673 \times 10^{-11}~{\rm N \, m^2/kg}$ is the
proportionality constant. The direction of the force is along the line
joining the two objects. Near the Earth's surface, the acceleration
due to gravity is approximately constant,
\begin{equation}
{\rm gravitational \ acceleration} = g = \frac{G \, M_\oplus}{R_\oplus^2} \approx 9.8~{\rm
  m/s^2} \,  ,
\label{Earth-g}
\end{equation}
where $M_\oplus = 1.3 \times 10^{25}~{\rm lb}$ is the mass of the
Earth and \mbox{$R_\oplus = 3,959~{\rm miles}$} its radius. 

So, the Earth pulls on the Moon because of gravity? Why doesn't the
moon get pulled into the Earth and crash?  To answer this provocative 
question we first note that an object can move around in a circle with a
constant speed yet still be accelerating because its direction is
constantly changing. This acceleration, which is always directed in
toward the center of the circle, is called centripetal
acceleration. The magnitude of this acceleration is written as
\begin{equation}
{\rm centripetal \ acceleration} = \frac{v^2}{r}
\, ,
\end{equation}
where $v$ is the speed of the object and $r$ the radius of the circle.
If a mass is being accelerated toward the center of a circle, it must
be acted upon by an unbalanced force that gives it this
acceleration. The {\it centripetal force} is the net force required to
keep an object moving on a circular path. It is always directed inward
toward the center of the circle. So, we can say that the Moon 
continuously {\it falls} towards Earth due to gravity, but does not get
any closer to Earth because its motion is an orbit. In other words,
the Moon {\it is} constantly trying to fall upon the Earth, due to the
force of gravity; but it is {\it
  constantly missing}, due to its {\it tangential velocity}.\\

  {\bf \S~Gravitational energy.}~This is energy stored whenever two
  masses are separated, and is recoverable as kinetic energy when they
  fall together. Examples are: {\it (i)}~water to go over a waterfall
  (water is separated from earth); {\it (ii)}~high jumper at top of
  jump; {\it (iii)}~interstellar dust before it comes together to form
  a star. The general expression for gravitational potential energy
  arises from the law of gravity and is equal to the work done against
  gravity to bring a mass to a given point in space.

  Indeed, the general form of the gravitational potential energy
  follows from (\ref{GF}) and,  for
  a particle of mass $m$, is given by
\begin{equation}
 {\rm gravitational \ potential \ energy} =  U =  - \frac{G
  \, M \, m}{r} \,,
\label{U1}
\end{equation}
where $M$ is the mass of the attracting body, and $r$ is the distance
between their centers.  Note that the gravitational force (\ref{GF}) approaches
zero for large distances, and consequently it makes sense to choose
the zero of gravitational potential energy at an infinite distance
away. This means that the gravitational potential energy near a planet is 
negative. This negative potential energy is indicative of a {\it
  bound state}; once a mass is near a large body, it is trapped until
something can provide enough energy to allow it to escape. 
 
Since the zero of
gravitational potential energy can be chosen at any point  (like the
choice of the zero of a coordinate system), the potential energy at a
height $h$ above that point is equal to the work which would be required
to lift the object to that height with no net change in kinetic
energy. We define the work done by a constant
force as the product of the force and the distance moved in the
direction of the force. With this in mind, the gravitational potential
energy of an object of mass $m$ near the Earth's surface is found to be
\begin{equation}
U=  {\rm gravitational \ force} \times {\rm distance} = m \, g \, h \, ,
\label{U2}
\end{equation}
where $h$ is the height above the zero-point energy.

With regard to the sign difference in (\ref{U1}) and (\ref{U2}) we
might note in passing that it relates to the choice of zero-point
energy, which is taken at infinity in (\ref{U1}) and at the surface of
the Earth in (\ref{U2}). This choice is completely arbitrary because
the important quantity is the difference in potential energy, and this
difference will be the same regardless of the choice of zero
level. However, once this position is chosen, it must remain fixed to
describe a given physical phenomenon.

In summary, the gravitational energy is {\it stored}
energy. It is not at all obvious that it is present. However, it can
be called on and used when needed, mostly to be converted to some form
of kinetic energy.\\

{\bf \S~Nuclear energy.}~This is energy which is stored in certain
(almost) unstable nuclei (such as uranium), and which is released when
the unstable nucleus is disturbed (in much the same way as a stretched
rubber band which is snipped). This is called fission and takes place in
nuclear reactors. It is also energy which is stored when
two nuclei which {\it want} to come together are allowed to do
so. (Such as a stretched rubber band which is allowed to
contract). This is called fusion and takes place in
stars.\\

{\bf \S~Chemical energy.} This is a repeat of the nuclear story, but with
much lower energy content per gram of material. Some chemicals release
energy when they are disturbed (TNT, nitroglycerin) some when
they combine (carbon and oxygen). The potential energy is said to be
stored in the carbon (oil, coal, etc.). More correctly, it is stored
in the electric field between the carbon and the oxygen. When the C
and O come together, the electric field gives up its energy in the
form of a photon (kinetic energy).  \\

{\bf \S~Forms of energy in a steady flow.} Water is everywhere around
us, covering 71\% of the Earth's surface. The water content of a human
being can vary between 45\% and 70\% of body mass. Water can exist
in three states (a.k.a. phases) of matter: solid (ice), liquid, or
gas. Matter in the solid state has a definite shape and size. Matter
in the liquid phase has a fixed volume, but can be of any
shape. Matter in the gas phase can be of any shape and also can be
easily compressed. Liquids and gases both flow, and this is why they
are called {\it fluids}.

The density of a  homogeneous amount of matter is defined to be the
amount of mass $M$ divided by the volume $V$ of material,
\begin{equation}
{\rm density} = \rho = \frac{M}{\rm V} \, .
\end{equation}
The unit for density is the kilogram per cubic meter, that is
kg/m$^3$.  Before proceeding we note that the mass of an object is a
measure of how much {\it matter} (or material) the object has. It is
measured in kg.  The weight of an object, on the other hand, is a
measure of how large the force of gravity is on that object.  It is
measured in N.  The mass does not depend on the location of the
object, e.g., a 1 liter bottle of water has a mass of 1~kg. If this
bottle were taken to the surface of Mars, its mass would still be 1~kg
(as long as no water is taken out of the bottle). However, since there
is less gravity on Mars, the weight of the bottle would be less on Mars than
here on Earth.

When you swim underwater  you can feel the water pressure against your
eardrums. The deeper you swim, the greater the pressure. The pressure
you feel is due to the weight of the water above you. As you swim
deeper, there is more water above you and therefore greater
pressure.  The pressure a liquid exerts then depends on its depth.
This is because the pressure is the force applied perpendicular to the
surface of an object per unit area over which that force is
distributed. Then, at a depth $h$,
\begin{equation}
{\rm liquid \ pressure} = P = \rho g h \,,
\end{equation}
because $F_g = mg$ and $m = \rho A h$, i.e., $P$ depends on the area
$A$ over which the force $F_g$ is distributed. The unit for pressure
is the Pascal, $1~{\rm Pa} = 1~{\rm N/m^2}$.

\begin{figure}[tbp]
\postscript{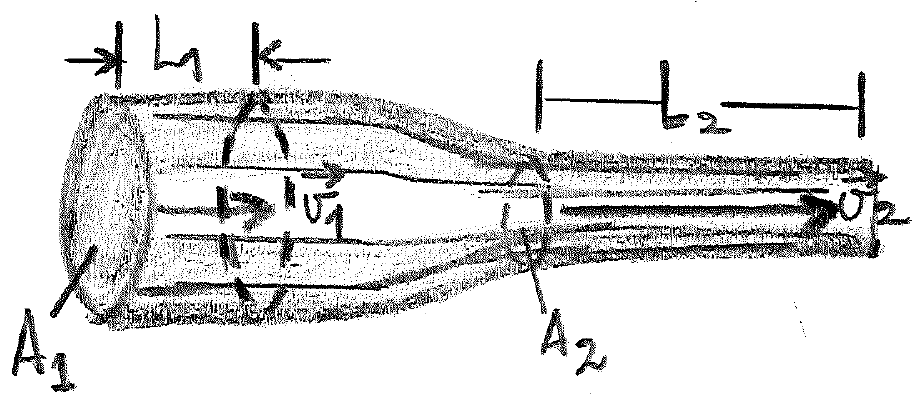}{0.8}
\caption{A fluid moving with steady flow through a pipe of varying
  cross-sectional area. The volume $V_1 = A_1 \times L_1$ of fluid flowing through area $A_1$
  in a time interval 􏰏$\Delta t$ must equal the volume  $V_2 = A_2
  \times L_2$  flowing through area $A_2$
  in the same time interval. Therefore,
  $A_1 v_1 = A_2 v_2$. \label{fig:0a}}
\end{figure}

\begin{figure}[tbp]
\postscript{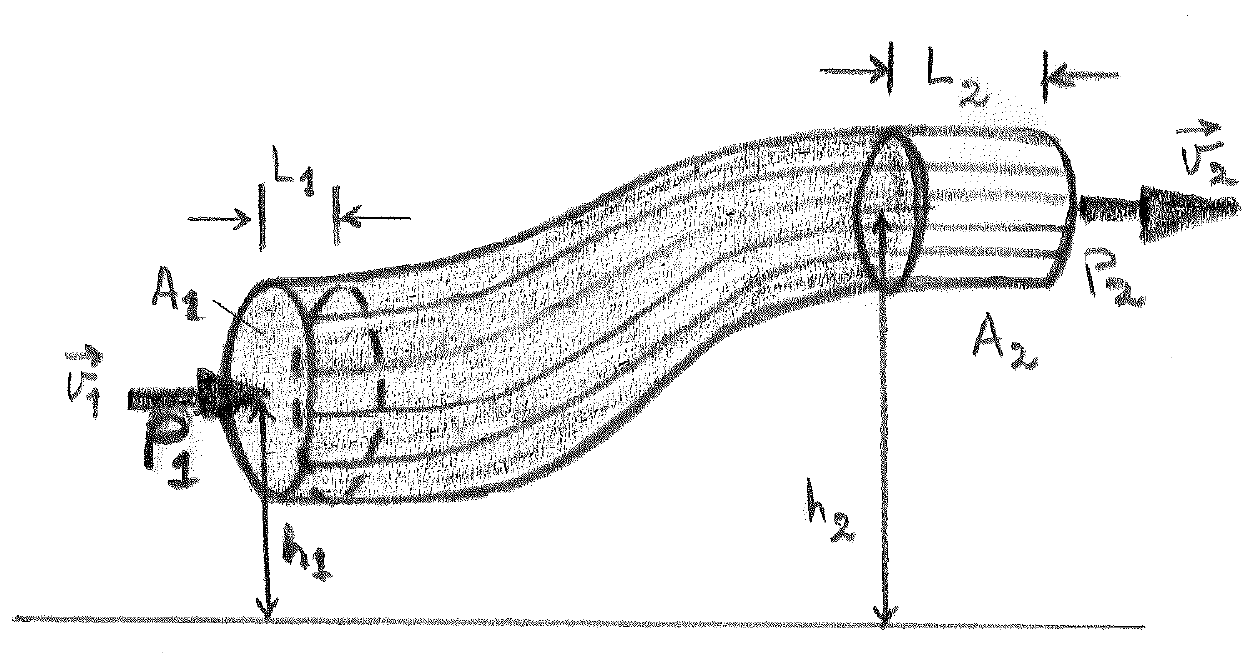}{0.8}
\caption{A fluid in steady flow through a constricted pipe. The volume
  $A_1\times L_1$ on the left is equal to the volume $A_2 \times L_2$
  on the right. For an interval of time $\Delta t$, the 
  velocity on the left $v_1 = L_1/\Delta t$ is smaller than $v_2 =
  L_2/\Delta t$. \label{fig:0b}}
\end{figure}

Anyone who has ever lifted a heavy submerged object out of the water
is familiar with the concept of {\it buoyancy}, which is the apparent
loss of weight experinced by objects submerged in a liquid. This is
because, when an object is submerged, the water exerts an upward force
on it that it is exactly opposite to the direction of gravity's
pull. This upward force is called the buoyant force, and it is a
consequence of pressure increasing with depth. The macroscopic
description of the buoyant force, which results from a very large
number of collisions of the fluid molecules, is called Archimedes'
principle~\cite{Archimedes}.  It is stated as follows: {\it An
  immersed object is buoyed up by a force equal to the weight of the
  fluid it displaces.}

So far we have only discussed fluids at rest. In studying fluid
dynamics we focus our attention on what is happening to various fluid
particles at a particular point in space at a particular time. The
flow of the fluid is said to be {\it steady} if at any given point,
the velocity of each passing fluid particle remains constant in
time. This does not mean that the velocity at different points in
space is the same. The velocity of a particular particle may change as
it moves from one point to another. That is, at some other point the
particle may have a different velocity, but every other particle which
passes the second point behaves exactly as the previous particle that
has just passed that point. Each particle follows a smooth path, and
the path of the particles do not cross each other. The path taken by a
fluid particle under a steady flow is called a {\it streamline}.

Consider a {\it steady flow} of a fluid through an enclosed pipe. Because no
fluid flows in or out of the sides, the mass flowing past any point
during a short period of time must be the same as the mass flowing
past any other point, so
\begin{equation}
\rho_1 A_1 v_1 = \rho_2 A_2 v_2 \,,
\label{continuity1}
\end{equation}
where the subscript 1 and 2 refer to two different points along the
pipe shown in Fig.~\ref{fig:0a}. This equation is called the {\it continuity
  equation}. If the fluid is incompressible, then the density is the
same at all points along the pipe and (\ref{continuity1}) becomes
\begin{equation}
A_1 v_1  = A_2 v_2 \, .
\label{continuity2}
\end{equation} 
We see that if the cross sectional area is decreased, then the flow
rate increases. This is demonstrated when you hold your finger over
the part of the outlet of a garden hose. Because you decrease the
cross sectional area, the water velocity increases. 

The general expression that relates the pressure difference between
two points in a pipe to both velocity changes (kinetic energy change)
and elevation (height) changes (potential energy change) was first
derived by Bernoulli and is given by
\begin{equation}
P_1 + \frac{1}{2} \rho v_1^2 + \rho g h_1 = P_2 + \frac{1}{2} \rho
v_2^2 + \rho g h_2 \, ;
\end{equation}
see Fig.~\ref{fig:0b}~\cite{Bernoulli}. Since 1 and 2 refer to any two locations along the pipeline, we may
write the expression in general as
\begin{equation}
P + \frac{1}{2} \rho v^2 + \rho g h = {\rm constant} \, .
\end{equation}
Bernoulli's equation can be considered to be a statement of the
conservation of energy principle appropriate for a steady flow: {\it The
  work done by the pressure forces on the fluid particle is equal to
  the increase in the kinetic and gravitational potential energy of
  the particle.}

\section{Thermodynamics}

\begin{figure*}[t]
    \postscript{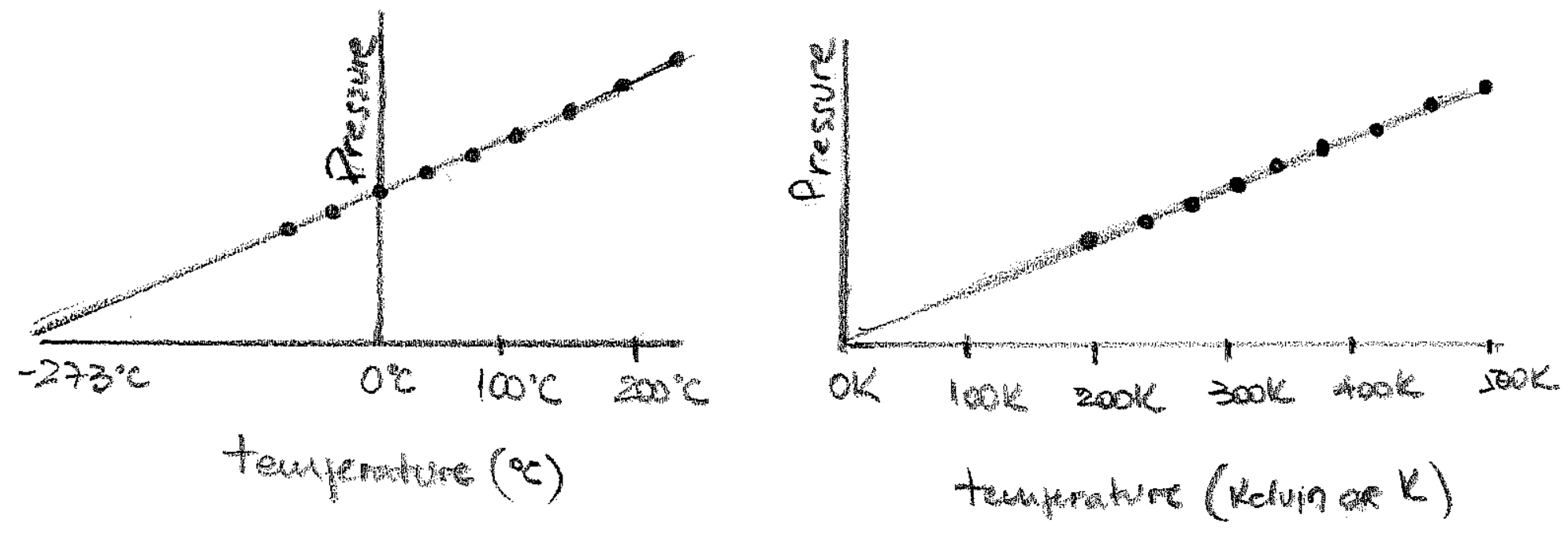}{0.8}
\caption{Temperature scales: Celsius (left) and Kelvin (right).
\label{fig:1}}
\end{figure*}

{\bf \S~What is heat?} The subjective sensation, familiar to everyone,
is that when you touch something hot, {\it heat} goes into your hand and
burns it. We talk of the transfer of heat from one object (the
radiator) to another (your hand). The fact that the radiator is hot is
sometimes expressed as ``the radiator is at high {\it temperature}.''

In the olden days (before 1800), people thought of heat as a fluid
(called caloric), which runs from a hot body to a cooler body. Each
body has a fixed amount of heat, or caloric, and the temperature was a
measure of the concentration of heat (a small body could be at a
higher temperature than a larger body, but could have less heat,
because the small body has a higher concentration of heat content, say
more per gram of material).

This theory, although in some ways appealing, fell apart with the
observation of Count Rumford: that by grinding away at the cannon
barrel he was able to produce unlimited amount of heat (that is, as
long as he kept grinding away!)~\cite{Thompson}. So it wasn't true that there was a
fixed amount of caloric, or heat, in the cannon: there was a way of
creating it by doing work~\cite{Joule}.

Heat and work are mutually convertible: you can do work and get heat,
or use heat and do work (a car engine or steam engine). The catch is,
that when you go from heat to useful work, you never achieve 100\%
conversion: there is always some heat wasted. The reason will be given
in later discussion. \\

{\bf \S~Absolute zero.} As mentioned in Sec.~\ref{sec:2}, the random
motion of molecules in a sample of matter constitutes a form of
kinetic energy, called thermal energy. The temperature is a measure of
the average kinetic energy/molecule. Temperatures are usually measured
in degrees Fahrenheit ($^\circ$F)~\cite{Fahrenheit}, or degrees Celsius
($^\circ$C)~\cite{Celsius}.  The temperature at which water freezes is
defined as $0^\circ$C. The temperature at which water boils is defined
as $100^\circ$C.  The relation between the Fahrenheit (F) and Celsius
(C) temperature scales is given by ${\rm F} = 9 {\rm C}/5 + 32.$

It will be convenient to talk about temperature in absolute
scale. That is, we would like the absolute zero of temperature to
correspond to that point where all molecular motion ceases. It turns
out that if we plot the pressure of a given volume of gas as a
function of the temperature, we find it looks like the one shown in
Fig.~\ref{fig:1}. The line, extrapolated to zero pressure, gives a value of
$-273^\circ$C (or $-460^\circ$F) for absolute zero.\ So let us make absolute zero
read $0^\circ$K (or 0~Kelvin), and therefore the freezing point of
water is $273^\circ$K~\cite{Kelvin}. Room temperature ($68^\circ$F) is $20^\circ{\rm
  C} = (273 + 20)^\circ{\rm K} = 293^\circ{\rm K}$. We can never reach
absolute zero ($0^\circ$K), but temperatures as low as $5 \times
10^{-10\circ}$K have been reached~\cite{Leanhardt}. Liquid nitrogen at
atmospheric pressure is at $75^\circ$K (about $-200^\circ$C, or
$-330^\circ$F). 

The coldest natural temperature ever directly measured on Earth is
$-89.2^\circ$C $= -128.6^\circ$F $= 184.0^\circ$K, which was recorded
at the Russian Vostok station in Antarctica ($78.5^\circ$S,
$106.9^\circ$E) on 21 July
1983~\cite{Turner}.  On 10 August 2010,  satellite
observations measured a surface temperature of $-93.2^\circ$C =
$-135.8^\circ$F = $180.0^\circ$K along a ridge between Dome Argus and Dome
Fuji ($81.8^\circ$S, $59.3^\circ$E)~\cite{Scambos}.\\

{\bf \S~So, what is heat?} Heat is {\it also} kinetic energy (moving
energy), but it is the kinetic energy of zillions of atoms in a
material vibrating (in a solid) or zapping around (in a gas) in a
random way, going nowhere in particular. {\it Heat is the random
  energy of atoms and molecules}. This is very different from
organized energy. It is like throwing a bag of hot popcorn (organized
energy) {\it vs.} letting them pop around inside the popper (random
energy). Popcorn popping very quickly has a lot of heat energy.

Now, the total heat energy of a material depends on two things: {\it
  (i)}~how many atomos there are, and {\it (ii)} how energetic each
atom is (on average). Then the  heat energy $Q$ can be expressed as
the product of the number of atoms $N$ and the average  random  energy
of  an  atom $\langle\varepsilon\rangle$,
\begin{equation}
 Q  = N \, \langle\varepsilon\rangle \, .
\end{equation}
The {\it temperature} is a direct measurement of the second factor,
the average energy of an atom. High temperature means peppy atoms; low
temperature means sluggish atoms. When the atoms are so sluggish that
they do not move, then you have hit bottom in the temperature scale:
{\it absolute zero} (which is $T = -459^\circ$F, very cold.) \\

{\bf \S~How do you get heat from mechanical work?} Well, when you push
your hand across the table (mechanical energy), you set the molecules
at the surface of the table into random vibration and create heat
energy (friction, in this case).\\

{\bf \S~How do you transfer heat?} You touch the hot poker; the madly
vibrating atoms on the surface of the poker bombard  the molecules of
your skin; these then start vibrating, causing all kind of
neurochemical reactions which tell your brain, ``It's hot, let go!''\\

{\bf \S~How much work makes how much heat?} The important thing to
know here is that the same work makes the same amount of heat {\it
  always}. The equivalence will be given by example: for each 50 pounds
of force which are pushed or dragged through one foot (say, across the
table), you make enough heat to raise the temperature of 1 ounce of
water by about $1^\circ$F. (You can also push with a force of 5~lb
through 10~ft, or with 10~lb through 5~ft, as long as force $\times$
distance = 50~lb ft). If you push twice as hard (100 lb) through one foot,
you make twice the heat. If this last part confuse you, then forget it
for now and just remember that
heat and mechanical work are interconvertible in definite
proportions. The proportions are $1,000~{\rm ft \, lb} = 1.28~{\rm Btu}
= 1/3~{\rm Cal}$, or $1~{\rm Btu} = 779~{\rm ft \, lb}$.\\ 

{\bf \S~First law of thermodynamics.}~We have seen in Sec.~\ref{sec:1}
that this is an expression of the conservation of energy. Energy can
cross the boundaries of a closed system in the form of heat $Q$ or
work $W$. By system, we mean a well-defined group of particles or
objects.\footnote{More specifically, a {\it thermodynamic system} is
  any body of matter or radiation large enough to be described by
  macroscopic parameters without reference to individual (atomic or
  subatomic) constituents. A complete specification of the system
  requires a description not only of its contents but also of its
  boundary and the interactions with its surroundings (a.k.a. the
  environment) permitted by the properties of the boundary. Boundaries
  need not be impenetrable and may permit passage of matter or energy
  in either direction or to any degree. An isolated system exchanges
  neither energy nor mass with its environment. A closed system can
  exchange energy with its environment but not matter, while open
  systems also exchange matter.}  Energy transfer across a system
boundary due solely to the temperature difference between a system and
its surroundings is called heat.  For a closed system, the first law
of thermodynamics is expressed as
\begin{equation}
E_{\rm in} - E_{\rm out} = \Delta E_{\rm system}
\end{equation}
where $E_{\rm in} = Q_{\rm in} + W_{\rm in}$ is the total energy entering the system, $E_{\rm
  out} = Q_{\rm out} + W_{\rm out}$ is the total energy leaving the system, and $\Delta E_{\rm
  system}= E_{\rm final} - E_{\rm initial} = E_f - E_i$ is the change in the total energy of the system. \\

\begin{figure}[tbp]
\postscript{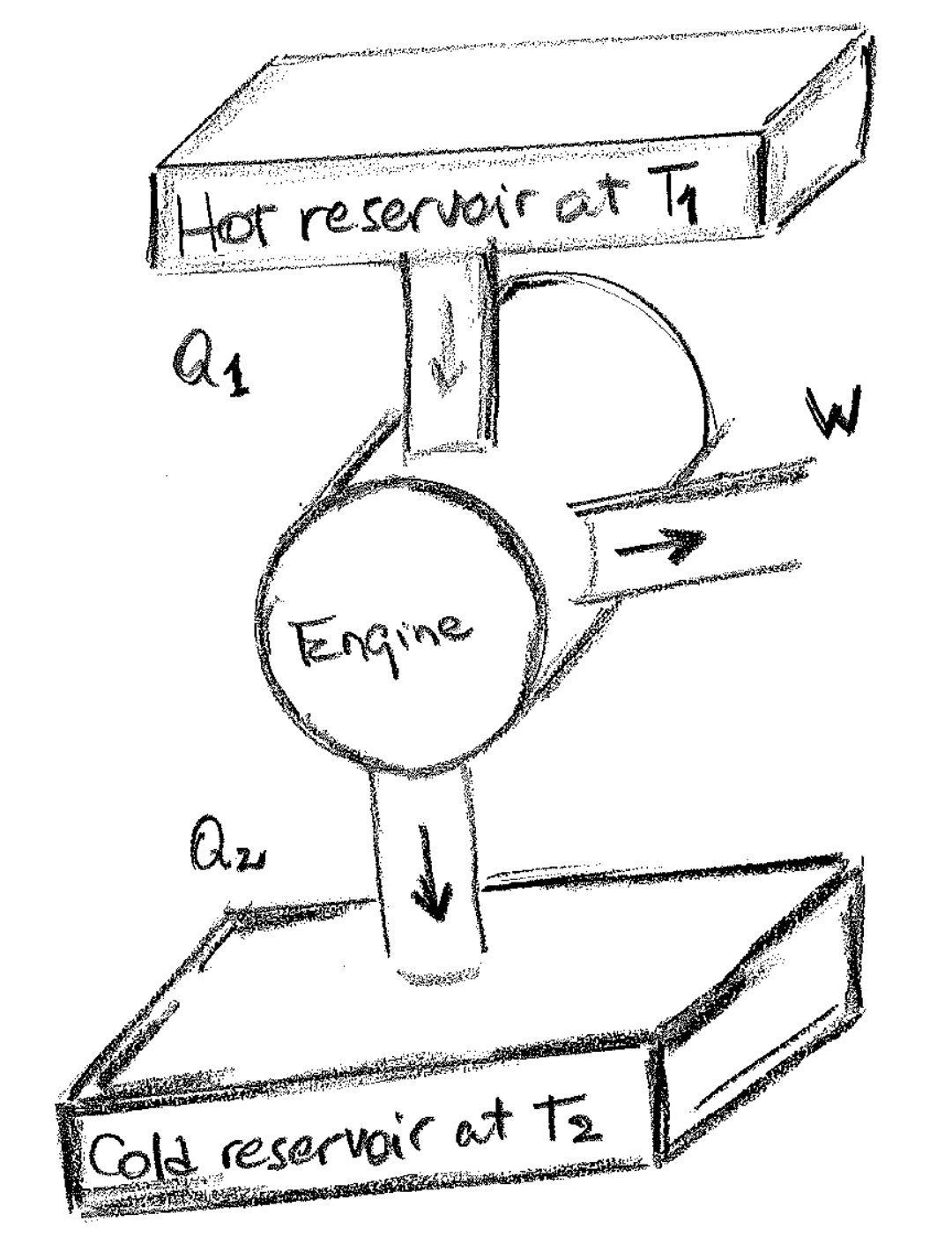}{0.8}
\caption{Schematic representation of a heat engine. The engine absorbs
energy $Q_1$ from the hot reservoir, expels energy $Q_2$ to the cold
reservoir, and does work $W$.
\label{fig:2}}
\end{figure}

\begin{figure}[tbp]
\postscript{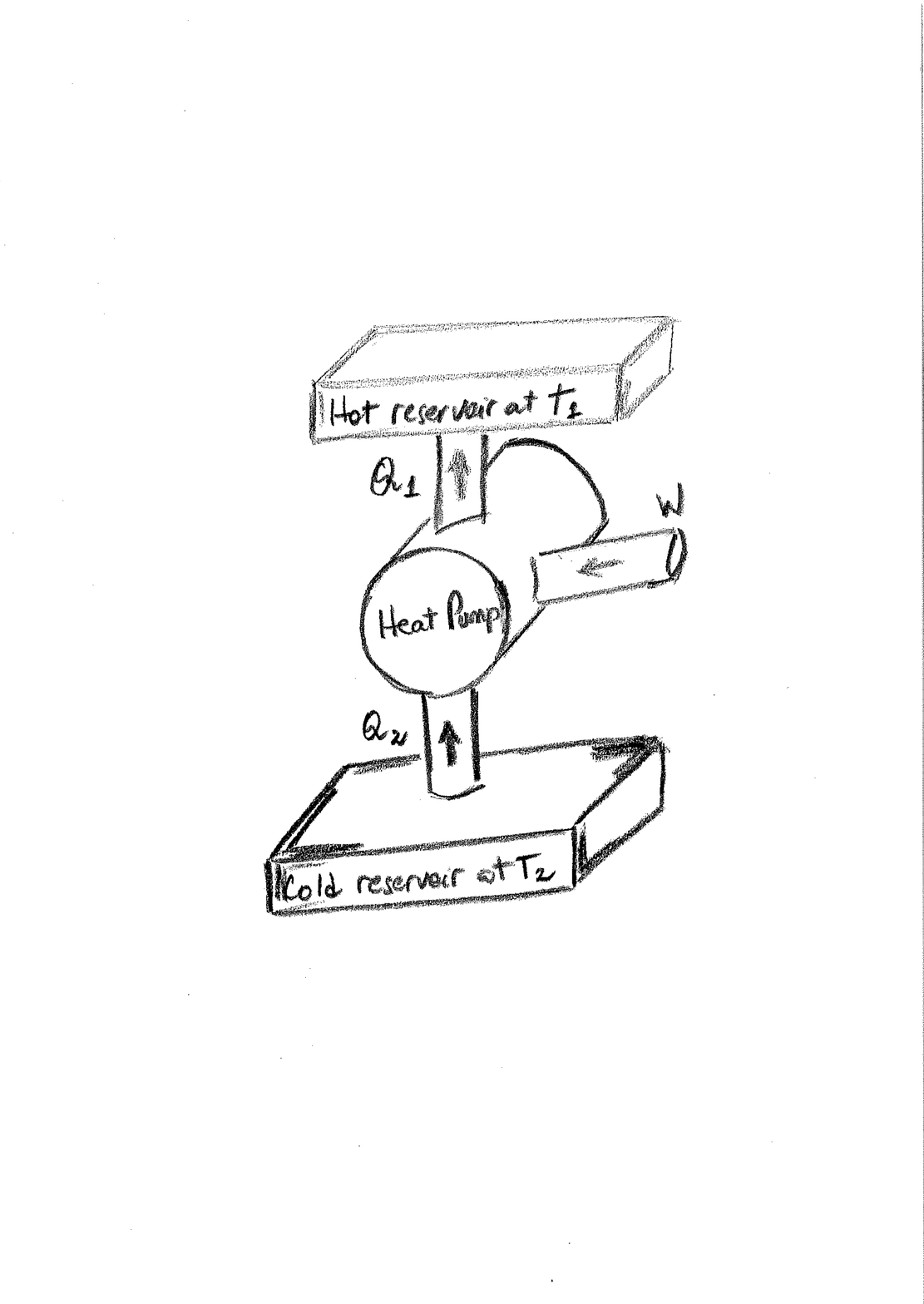}{0.8}
\caption{Schematic representation of a refrigerator. This 
device absorbes heat $Q_2$ from a cooler reservoir and a greater
amount of heat $Q_1$ is released to a warmer reservoir by
means of work $W$ supplied from some external source. \label{fig:3}}
\end{figure}

\begin{figure*}[t]
    \postscript{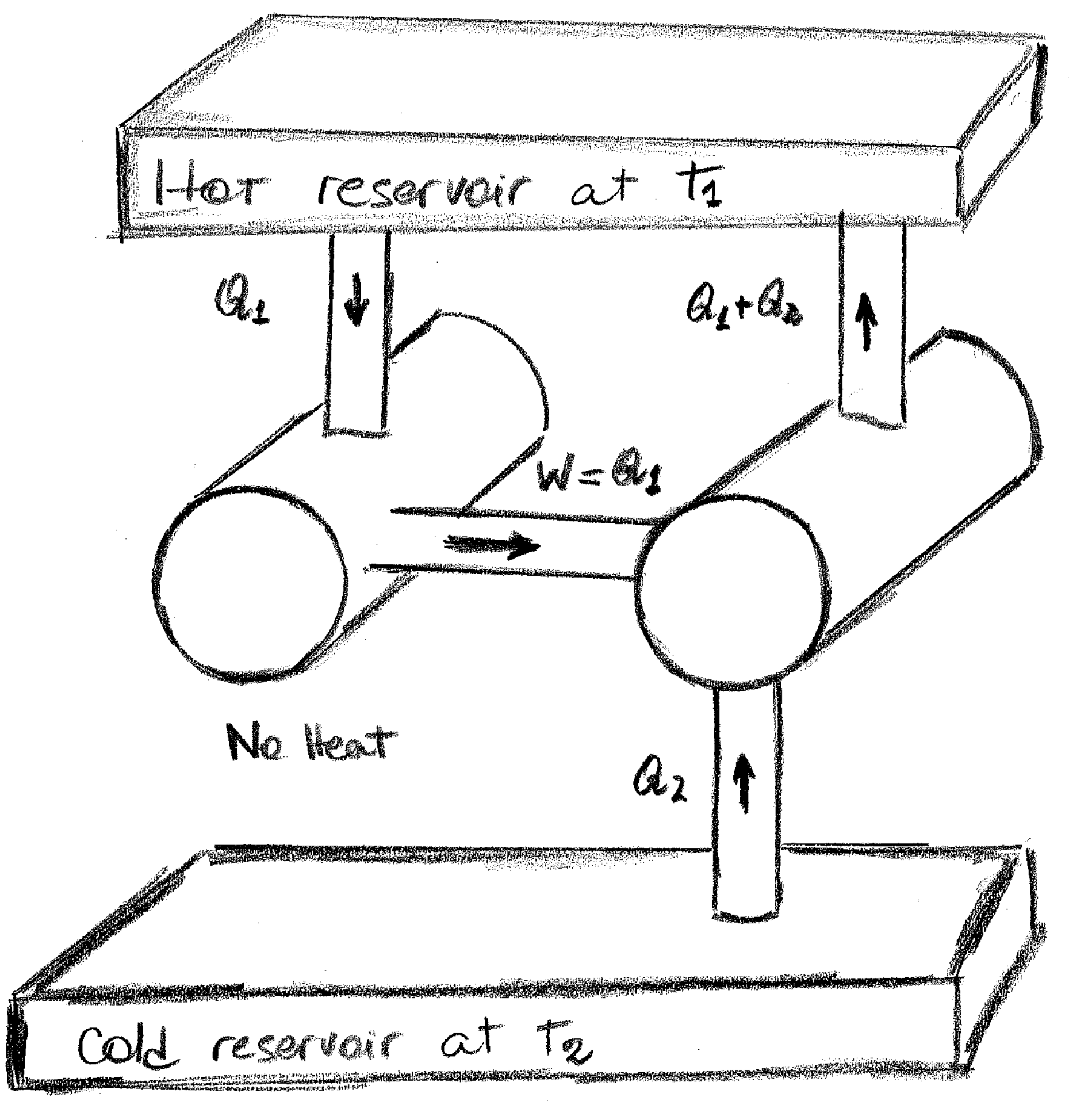}{0.8}
\caption{Schematic diagram of a heat engine that absorbs energy from a
  hot reservoir $Q_1$ and does an equivalent amount of work $W$. It is
  impossible to construct such a perfect engine, because we could use
  the work liberated by the engine to run a refrigerator. 
\label{fig:4}}
\end{figure*}

{\bf \S~Second law of thermodynamics.} Each cubic meter of air at room
temperature has about $3 \times 10^{25}$ molecules zipping around in a
random fashion.  The
average kinetic energy of each atom in the air is 
\begin{equation}
\langle\varepsilon\rangle = \frac{3}{2} \ k_B  \ T \,,
\label{k_b}
\end{equation}
where $k_B = 1.380 \times 10^{-23}~{\rm J} \, ^\circ{\rm K}^{-1}$ is
the Boltzmann constant, and one Joule (J) is the work done by a force
of one newton when its point of application moves one meter in the
direction of action of the force (equivalent to one $3,600$th of a
watt-hour). The total kinetic energy of all these molecules is
about 347~Btu. The room may then be thought of as a {\it reservoir} of
thermal energy at temperature $70^\circ$F (or $295^\circ$K). The
question we may well ask is the following: Is it possible to extract
{\it some} of this thermal energy and change it {\it entirely} into
useful work (such as turning a generator)? There is no contradiction
here with the conservation of energy, but it is a fact that no
such process is known. This negative statement, which is the result of
everyday experience constitutes the {\it second law of
  thermodynamics}: It is impossible to construct an engine that,
operating in a cycle, will produce no effect other than the extraction
of heat from a reservoir and the performance of an equivalent amount
of work~\cite{Carnot}.

Reduced to its simplest terms, the important characteristics of
heat-engine cycles may be summed up as follows: {\it (i)}~There is some
process during which there is an absorption of heat from a reservoir
at high temperature (called simply the {\it hot reservoir}). {\it
  (ii)}~After some work is done, there is some process during which
heat is rejected to a reservoir at lower temperature (called simply
the {\it cold reservoir}).  For a heat engine, $Q_1$ is the
heat taken from the boiler, $W$ is the work done, and $Q_2$ is the heat
transferred to the condenser, with $Q_1 = W + Q_2$; see Fig.~\ref{fig:2}. 
For a refrigerator (or air conditioner), $Q_2$ is the heat taken
from the refrigerator by coolant, $W$ is the energy pumped in by the
motor, and $Q_1$ is the heat given up to the surrounding air, with
$Q_1 = W + Q_2$;  see Fig.~\ref{fig:3}..
\begin{figure}[t]
    \postscript{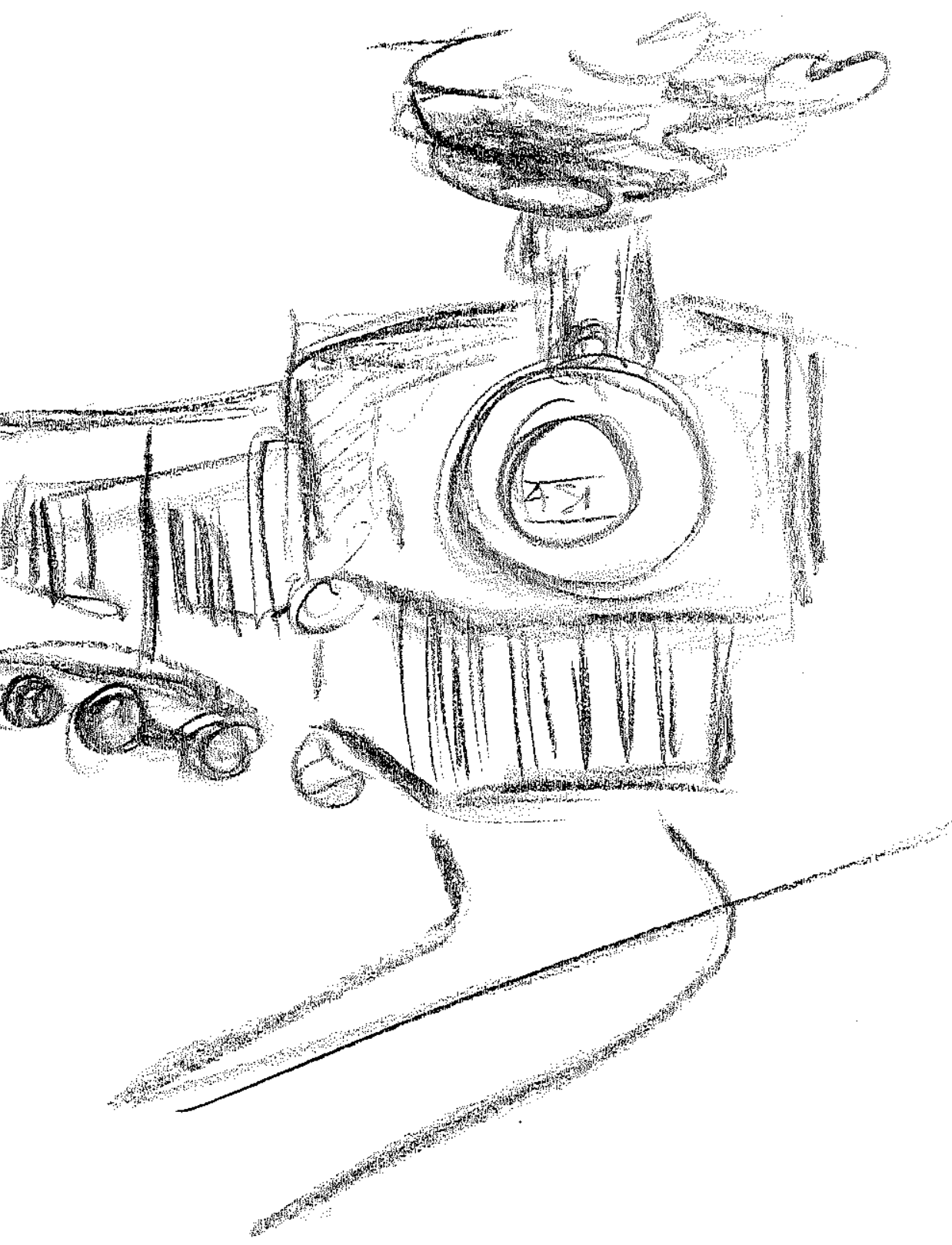}{0.8}
\caption{A steam engine is a heat engine that performs mechanical work
  using steam as its working fluid such as railway steam locomotives. 
\label{fig:5}}
\end{figure}

Now, using the idea of a refrigerator, we can show that a {\it
  violation} of the {\it second law} would make it impossible to construct a
device which, operating in a cycle, will produce {\it no effect}
other than transfer of heat from a cooler body to a warmer body. Since
this never happens (for statistical reasons), we can then see why the
{\it second law} works. The {\it proof} is as follows: Suppose we had an
engine (on the left in Fig.~\ref{fig:4}) which violates the second law
(rejects no heat to the cold reservoir). Then we could use the work
liberated by the engine to run a refrigerator which operates between
the same two reservoirs and takes heat $Q_2$ from the cold, and puts
$W + Q_2 = Q_1 + Q_2$ into the hot. The net result is a transfer of
heat from the cold to the hot reservoir, with nothing else
occurring. This is impossible so the engine of the left of Fig.~\ref{fig:4} is
impossible.\\

{\bf \S~Thermal efficiency.} So now we come to a practical
question: Given a hot reservoir at temperature $T_1$, and a cold
reservoir at temperature $T_2$, what is the highest efficiency with
which a heat engine may operate between the two? By efficiency one
simply means
\begin{equation}
\frac{{\rm work \ out}}{{\rm heat \ extracted \ from \ hot \ reservoir}}
= \frac{W}{Q_1} \, .
\end{equation}
It has been shown that the {\it maximum} efficiency is
\begin{equation}
{\rm maximum \ efficiency} = \eta_{\rm max} = \frac{T_1 - T_2}{T_1} \,,
\end{equation}
where $T_1$ and $T_2$ are in $^\circ$K.

\begin{figure*}[t]
    \postscript{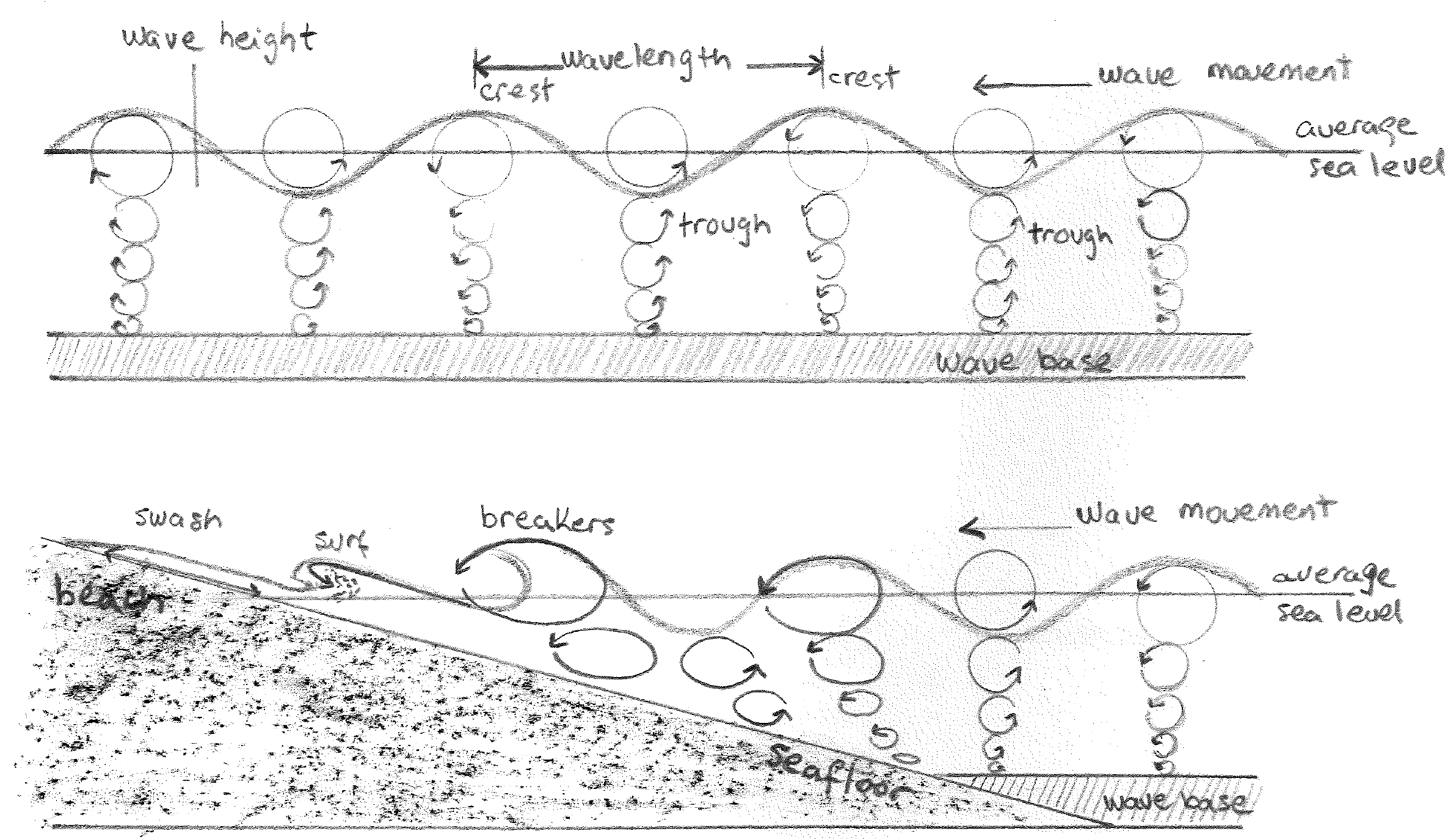}{0.9}
\caption{Evolution of waves in the surface of the ocean.  
In the upper panel we show the circular path
  of particles in the open ocean due to oscillations from passing
  waves. In the lower panel we show the waves of transition. The
  frictional interaction of the wave with the seabed causes orbital
  particle circles to stretch as the wave is dissipated. After a
  wave breaks, the remains of the wave moves as a chaotic surf until
  it spreads onto the beach as swash.
  \label{fig:5-6}}
\end{figure*}

\begin{figure*}[t]
\begin{minipage}[t]{0.49\textwidth}
    \postscript{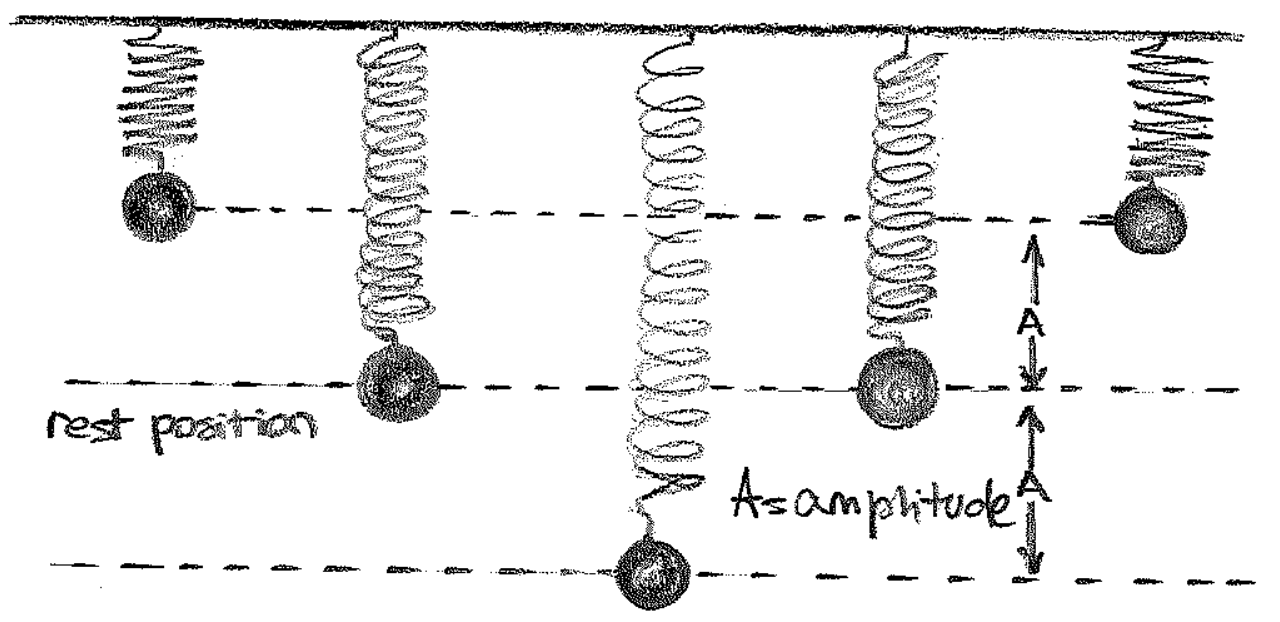}{0.999}
\end{minipage}
\begin{minipage}[t]{0.49\textwidth}
    \postscript{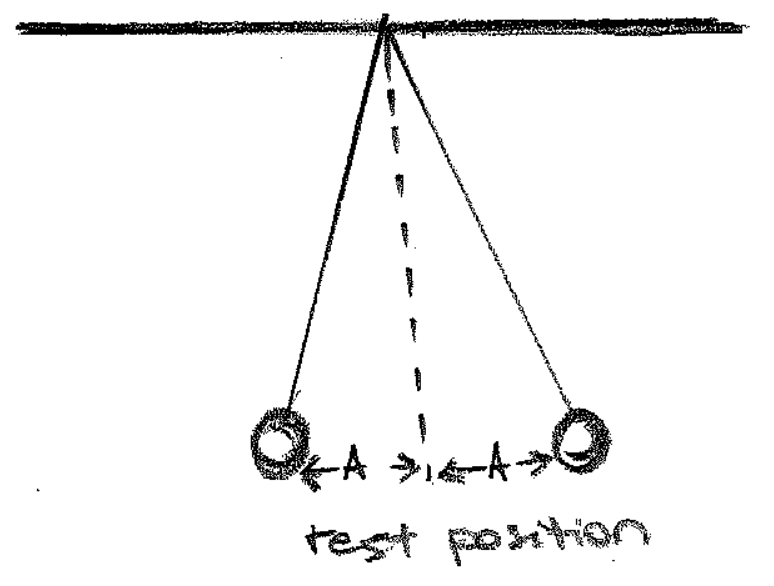}{0.64}
\end{minipage}
\caption{Periodic motion of a vibrating spring (left) and pendulum (right). The rest position is where the
  object will remain at rest. The object can move through its rest
  position. The distance in either direction from the rest position to
  a maximum displacement is called the amplitude.
 \label{fig:6}}
\end{figure*}

For example, the steam engine shown in Fig.~\ref{fig:5}, with a boiler heating
the steam to $400^\circ$C ($= 670^\circ$K), and a condenser operating
at $100^\circ$C ($= 370^\circ$K) will have a maximum theoretical
efficiency of
\begin{equation}
\eta_{\rm max} = \frac{670 - 370}{670} \simeq 45\% \, . 
\end{equation}
 
Similarly the coefficient of performance of a refrigerator, defined as
\begin{eqnarray}
{\rm CoP} & = & \frac{\rm heat \ removed \ from \ cold \
  reservoir}{{\rm work \ done \ (electrical \
  energy \ used})} \nonumber \\ & = & \frac{Q_2}{W}  =  \frac{Q_2}{Q_1 - Q_2} \end{eqnarray}
has a maximum value of
\begin{equation}
{\rm CoP}_{\rm max} = \frac{T_2}{T_1 - T_2} \, .
\end{equation}
For an air conditioner operating between a room at $70^\circ$F ($=
295^\circ$K) and the outdoors at $85^\circ$F ($=395^\circ$K), the
maximum theoretical coefficient of performance is
\begin{equation}
{\rm CoP}_{\rm max} = \frac{295}{309 - 295} = \frac{295}{14} = 20 \, .
\end{equation}
In practice, this turns out to be about 7 to 10. Commonly, one rates
air conditioners in terms of Btu/kWh. For a coefficient performance of
10, this is 10~kWh (heat)/1~kWh (electric) = 34,000~Btu/kWh.

\section{Waves as Energy Transfer}

\begin{figure}[t]
    \postscript{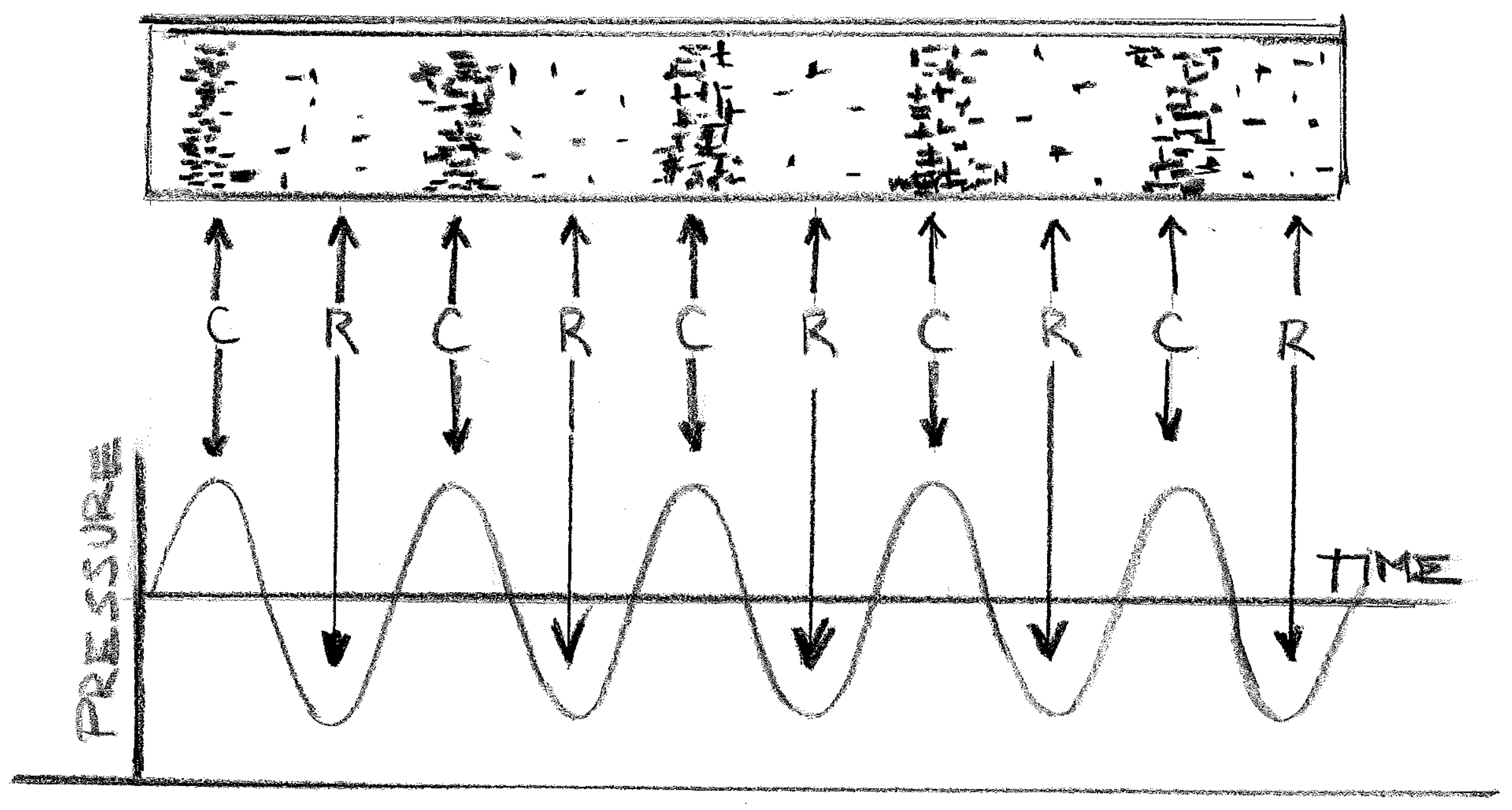}{0.9}
\caption{The sound is a longitudinal wave. Each molecule moves parallel to
  the energy of the wave. Molecules are pushed/pulled by air
  pressure. Air is compressed in some spots and stretched in
  others. Compressed areas are called compressions. Stretched areas are called rarefactions.
 \label{fig:7}}
\end{figure}

{\bf \S~Wave basics.} A wave is a type of energy transmission that
results from a periodic disturbance (vibration).  Waves transfer
energy from one place to another without transferring matter. They are
composed of a series of repeating patterns. There are two classes of
waves: {\it transverse} and {\it longitudinal}.  In the former the
vibration is perpendicular to the direction of motion of the wave,
whereas in the latter the vibration is in the same direction as the
direction of the wave.

Everyone has seen waves on the surface of water. You made them in the
bathtub when you were a child, and you have seen them in the ocean.
In the open ocean the water waves are transverse. A water wave can
travel hundreds of kilometers over the ocean, but the water just moves
up and down as the waves passes. Energy is transferred from one water
molecule to the next by the forces that hold the molecules
together. Near the shore the waves become also longitudinal: the small
distance from the surface to the bottom of ocean makes the difference,
see Fig.~\ref{fig:5-6}.

We live surrounded by waves. Some are visible, others are not. By
observing the visible waves (e.g., in water) we can describe some
characteristics that all waves, including invisibles ones, have in
common.\\

{\bf \S~Periodic motion.} Most waves originate from objects that are
vibrating so rapidly that they are difficult to observed with our
unaided senses. For the purposes of describing the properties of
vibrating objects,  we need a slowly moving device such as a mass
bouncing on a spring or a pendulum; see Fig.~\ref{fig:6}.

When an object repeats a pattern of motion, as a bouncing springs
does, we say the object exhibits periodic motion. The vibration, or
oscillation, of the object is repeated over and over with the same
time interval each time. When we describe the motion of a vibrating
object, we call one complete oscillation a {\it cycle}. The number of
cycles per second is called the frequency $\nu$.  The unit used to
measure frequency is the hertz (Hz). Another term used in describing
vibrations is the period ${\cal T}$, which is the time required for
one cycle. The period is usually measured in seconds.  Frequency and
period are reciprocals, i.e.,
\begin{equation}
\nu = \frac{1}{{\cal T}} \quad {\rm and} \quad {\cal T} = \frac{1}{\nu} \,
\end{equation}
If the frequency is 60~Hz, then the period is 1/60~s (or
0.017~s). If the period is 0.01~s, the frequency is 100~Hz.

As a pendulum of length $\ell$ swings, it repeats the same motion in equal time
intervals, 
\begin{equation}
{\cal T} = 2 \pi \sqrt{\ell/g} \, .
\label{pendulum-period}
\end{equation}
We say it exhibits periodic motion. Observing successive
swings, we find that the distances reached by the pendulum on either
side of the rest position are almost equal. In the same way a
vertically bouncing mass on a spring exhibits periodic motion, and it
too moves almost the same distance on either side of the rest
position. This is a property of all objects oscillating with periodic
motion. The distance in either direction from the rest position to
maximum displacement is called the amplitude $A$.\\

{\bf \S~What is sound?} We have seen that a wave is a transfer of
energy, in the form of a disturbance usually through a material, or
medium. Sound is a pressure wave, which is created by a vibrating
object. This vibrations set particles in the surrounding medium
(typical air) in vibrational motion, thus transporting energy through
the medium. Since the particles are moving in parallel direction to
the wave movement, the sound wave is referred to as a longitudinal
wave. The result of longitudinal waves is the creation of compressions and
rarefactions within the air, as shown in Fig.~\ref{fig:7}.

A sound wave, as any other wave can be characterized by its: {\it
  (i)}~amplitude $A$, which is the distance from the midpoint of the wave
to a crest or trough (maximum displacement from equilibrium); {\it
  (ii)}~frequency $\nu$, which is the number of repeating patterns (cycles)
per unit of time; {\it (iii)}~period ${\cal T}$,  which is the time for one
cycle; {\it (iv)}~wavelength $\lambda$, which is the distance (shown
in Fig.~\ref{fig:5-6}) from one crest (or
trough) to another crest (or trough), and {\it (v)}~speed, 
\begin{equation}
v = \lambda \, \nu \,  = \lambda/{\cal T} \, . 
\end{equation}
The wavelength has distance units, such as meters, and so
the speed is measured in m/s. The sound requires an elastic medium
(solid, liquid, or gas) for transmission.  The speed of sound in dry
air is about 330~m/s at $0^\circ$C, and increases 0.60~m/s for every
$^\circ$C increase, {\it i.e.},
\begin{equation}
v_{\rm sound} = \left[331.5+ 0.6 \, \left(\frac{T}{^\circ{\rm C}} \right)
\right]~{\rm m/s} \, .
\label{v-sound}
\end{equation}
The amount of work done to generate the energy that sets the particles
in motion is reflected in the degree of displacement which is measured
as the amplitude of a sound.  The frequency is measured as the number
of complete back-and-forth vibrations of a particle of the medium per
unit of time. The human ear can hear from 20 to $20,000~{\rm
  Hz}$. Infrasonic is below this frequency and ultrasonic above. The
softest audible sound modulates the air pressure by around
$10^{-6}~{\rm Pa}$, whereas the loudest (pain inflicting) audible
sound does it by $10^2~{\rm Pa}$. 

The sound can bounce off of objects, and the angle of incidence is
equal to the angle of {\it reflection}. The sound {\it reflection}
gives rise to echoes. Multiple echoes are called reverberations. The
study of the sound properties (especially reflections) is called
acoustics. The change of the speed of sound in different mediums can
bend a wave if it hits the different medium at a non $90^\circ$
angle. This is called {\it refraction}. The wave bends toward a slower
medium or away from a faster medium. Ultrasound imaging, bats, and
dolphins all use reflection and refraction of sound waves. 

Waves can superimpose, and
constructively and destructively interfere, increasing each other or
destroying each other.  Standing waves are formed when a wave is
reflected and constructively interferes such
that the wave appears to stand still.\\

{\bf \S~Doppler effect.}  When we observe a sound  wave from a
source at rest, the time between the arrival wave crests at our
instruments is the same as the time between crests as they leave the
source. However, if the source is moving away
from us, the time between arrivals of successive wave crests is
increased over the time between their departures from the source,
because each crest has a little farther to go on its journey to us
than the crest before. The time between crests
is just the wavelength divided by the speed of the wave, so a wave
sent out by a source moving away from us will appear to have a longer
wavelength than if the source were at rest. Likewise, if the source is
moving toward us, the time between arrivals of the wave crests is
decreased because each successive crest has a shorter distance to go,
and the waves appear to have a shorter wavelength.

\begin{figure}[t]
    \postscript{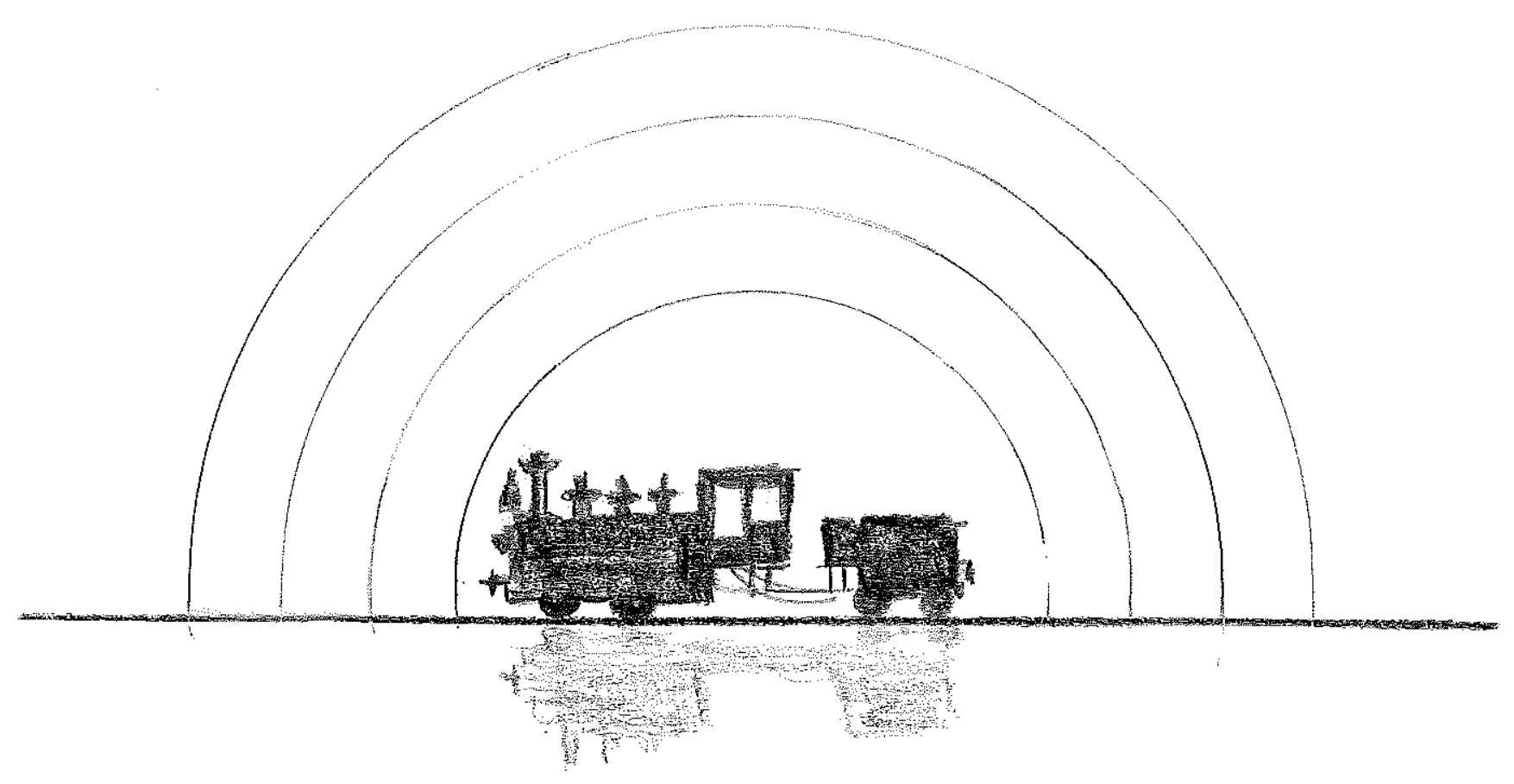}{0.9}
\caption{Sound waves emitted by a source at rest. The source emits
  $\nu_{\rm emitted}$
  ``crests'' per second. If the observer were at rest, he would detect
  $\nu_{\rm emitted}$ crests per second. If the observer (a.k.a. receiver) moves
  with velocity $V_{\rm receiver}$,  how many crests does he detect per second?
  \label{fig:8}}
\end{figure}
\begin{figure}[t]
    \postscript{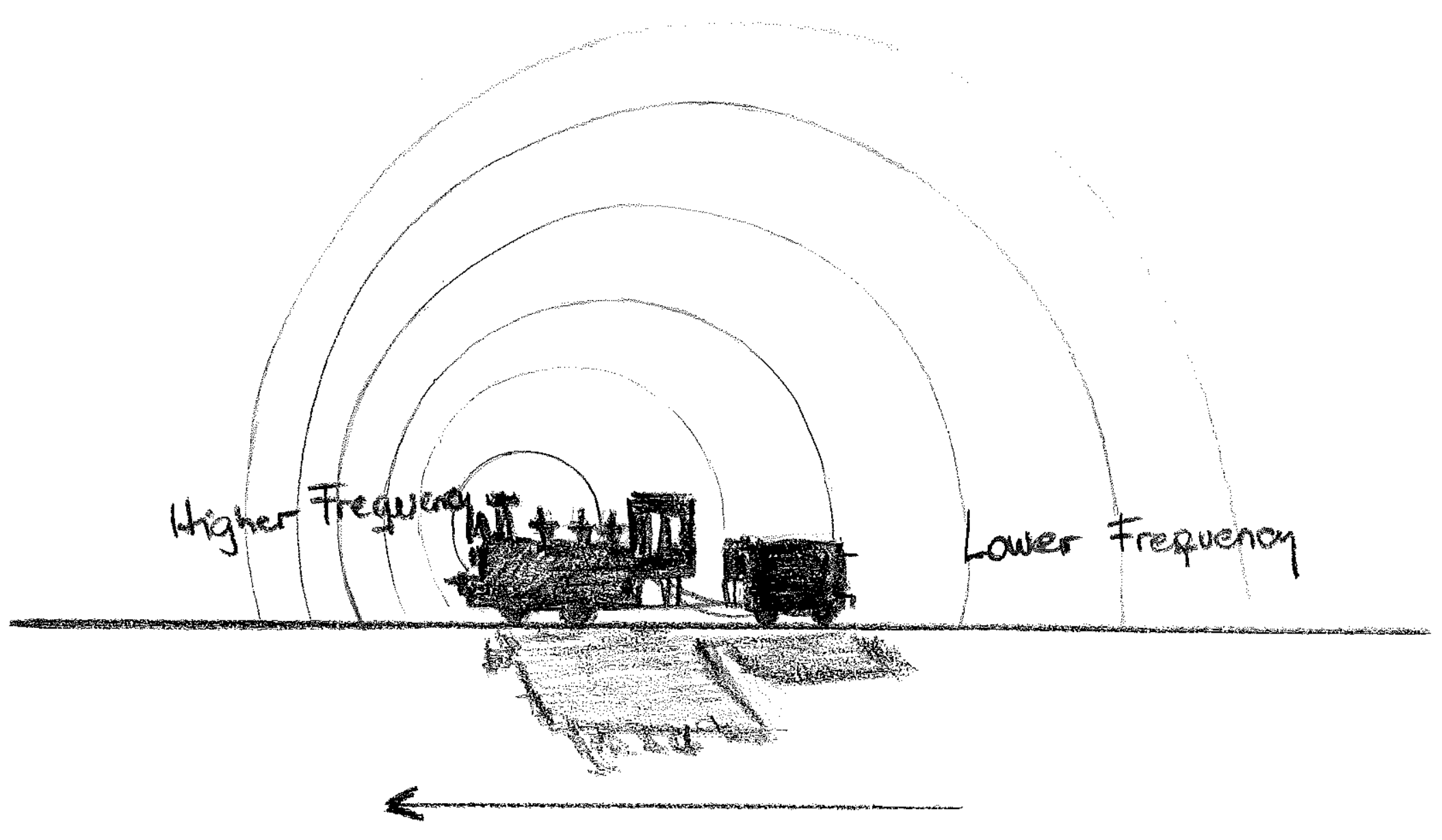}{0.9}
\caption{Sound waves emitted by a source moving to the left. The peaks in
pressure ahead of the source are more closely spaced than the peaks
behind the source.
 \label{fig:9}}
\end{figure}

A nice analogy was put forward by Weinberg~\cite{Weinberg:1977ji}. He
compared the situation with a travelling man that has to send a letter
home regularly once a week during his travels: while he is travelling
away from home, each successive letter will have a little farther to
go than the one before, so his letters will arrive a little more than
a week apart; on the homeward leg of his journey, each successive
letter will have a shorter distance to travel, so they will arrive
more frequently than once a week. 

The {\it Doppler effect} is the change in the observed frequency of a
source due to the relative motion between the source and the
receiver~\cite{Doppler}. The relative motion that affects the observed
frequency is only the motion in the line-of-sight between the source
and the receiver.

We first consider the relative motion of the receiver. If a source is
stationary, as the one exhibited in Fig.~\ref{fig:8}, it will emit
sound waves that propagate out from the source as shown in the
figure. As the receiver moves towards the source with velocity $V_{\rm
  reciever}$,
it will detect the sound coming from the source, but each successive
sound wave will be detected earlier than it would have if the receiver
were stationary, due to the motion of the receiver in the
line-of-sight. Hence, the frequency with which each successive wave
front would be detected will be changed by this relative motion
according to $\Delta \nu = V_{\rm receiver}/\lambda_{\rm emitted}$, where $\Delta \nu =
\nu_{\rm received} -\nu_{\rm emitted}$ is the change in the observed
frequency and $\lambda_{\rm emitted}$ is
the original wavelength of the source. Since the original frequency of
the source can be expressed in terms of the wavelength as $\nu_{\rm
  emitted} = v_{\rm sound}/
\lambda_{\rm emitted}$, the observed frequency becomes:
\begin{eqnarray}
\nu_{\rm received} & = & \nu_{\rm emitted} + \Delta \nu \nonumber \\
& = & \frac{v_{\rm sound} + V_{\rm receiver}}{\lambda_{\rm emitted}} \nonumber \\
& = & \nu_{\rm emitted} \left(\frac{v_{\rm sound} + V_{\rm
      receiver}}{v_{\rm sound}} \right) \, . 
\label{rolfi30}
\end{eqnarray}
Note that (\ref{rolfi30}) only works if the relative velocity of the
receiver, $V_{\rm receiver}$ is towards the source. If the motion is away from the
source, the relative velocity would be in the opposite direction and
(\ref{rolfi30}) would become:
\begin{equation}
\nu_{\rm received} = \nu_{\rm emitted} \left(\frac{v_{\rm sound} -
    V_{\rm receiver}}{v_{\rm sound}} \right) \, .
\label{rolfi31}
\end{equation}
(\ref{rolfi30}) and (\ref{rolfi31}) are usually combined and expressed as:
\begin{equation}
\nu_{\rm received} = \nu_{\rm emitted} \left(\frac{v_{\rm sound} \pm
    V_{\rm receiver}}{v_{\rm sound}} \right) \, .
\label{rolfi32}
\end{equation}

If the source is moving towards the receiver, the effect is slightly
different. The spacing between the successive wave fronts would be
less, as seen in Fig.~\ref{fig:9}. This would be expressed as: $\Delta
\lambda = V_{\rm source}/\nu_{\rm emitted}$, where $V_{\rm source}$ is the relative velocity of the
source. To calculate the observed frequency
\begin{eqnarray}
\nu_{\rm received} & = & \frac{v_{\rm sound}}{\lambda_{\rm emitted} + \Delta \lambda} \nonumber \\
& = & \nu_{\rm emitted} \left(\frac{v_{\rm sound}}{v_{\rm sound} -
    V_{\rm source}} \right) \, .
\label{rolfi33}
\end{eqnarray}
Note that this is only when the source is moving towards the receiver. If the source is moving away, (\ref{rolfi33}) would be changed to:
\begin{equation}
\nu_{\rm received} =  \nu_{\rm emitted} \left(\frac{v_{\rm
      sound}}{v_{\rm sound} + V_{\rm source}} \right) \, .
\label{rolfi34}
\end{equation}
When (\ref{rolfi33}) is combined with (\ref{rolfi34}), we have:
\begin{equation}
\nu_{\rm received} = \nu_{\rm emitted} \left(\frac{v_{\rm sound}}{v_{\rm
      sound} \mp V_{\rm source}} \right) \, .
\label{rolfi35}
\end{equation}
Notice that this time, the plus/minus symbol is inverted because the
sign on top is to be used for relative motion of the source towards
the receiver.

By combining (\ref{rolfi32}) and (\ref{rolfi35}), we obtain
\begin{equation}
\nu_{\rm received} = \nu_{\rm emitted} \left( \frac{v_{\rm sound} \pm
    V_{\rm receiver}}{v_{\rm sound} \mp V_{\rm source}} \right) \, .
\end{equation}
We stress again two important points. Firstly, that the quantities for
the velocity of the receiver $V_{\rm receiver}$ and the velocity of the source
$V_{\rm sound}$ are only the magnitudes of the relative velocities in (or along)
the line of sight. In other words, the component of the velocity of
the source and the receiver, that are perpendicular to the line of
sight do not change the received frequency. Secondly, that the top
sign in the numerator and the denominator are the sign convention to
be used when the relative velocities are towards the other. If the
source were moving towards the receiver, the sign to use in the
denominator would be the minus sign. If the source were moving away
from the receiver, the sign to use would be the plus sign.

One interesting application of the Doppler effect is the active
sonar. This is a system in which pulses of acoustic energy are
launched into the water for the purpose of producing echoes. By
examining the echoes of transmitted pulses, it affords the capability
of both detecting the presence of and estimating the range of an
underwater target. We must carefully define the {\it source} and {\it
  receiver} for both the outgoing active pulse and the returning
signal.  For the outgoing active pulse, the Doppler shifted frequency
of the active pulse when it hits the target (that is the receiver)
would be:
\begin{equation}
\nu_{\rm received}^{\rm target} = \nu_{\rm emitted} \left(\frac{v_{\rm sound} \pm
    V_{\rm target}}{v_{\rm sound} \mp V_{\rm source}} \right) \, .
\label{ptown1}
\end{equation}
For the return pulse, there would be a similar shift but now the {\it
  source} would be the {\it target}, and the {\it receiver} would be
the ship sending out the original active pulse, and the base frequency
would be the Doppler shifted frequency from above. The frequency of
the echo as measured by the ship is
\begin{equation}
\nu_{\rm echo} = \nu_{\rm received}^{\rm target} \left(\frac{v_{\rm sound} \pm
  V_{\rm source}}{v_{\rm sound} \mp V_{\rm target}} \right) \, .
\label{ptown2}
\end{equation} 
Substituting in (\ref{ptown2}) for $\nu_{\rm received}^{\rm target}$ as given by (\ref{ptown1}), we have
\begin{equation}
\nu_{\rm echo} = \nu_{\rm emitted} \left( \frac{v_{\rm sound} \pm
  V_{\rm target}}{v_{\rm sound} \mp V_{\rm source}} \right)
\left(\frac{v_{\rm sound} \pm V_{\rm source}}{v_{\rm sound} \mp V_{\rm
      target}} \right) \, .
\end{equation}
Again, the velocities are only the magnitudes of the velocity in the
line of sight, and we must take care to pick the correct sign to use
in front of each velocity.

\section{Electricity and Magnetism}
\label{sec5}

{\bf \S~Electric charge.} There are two types of observed electric
charge, which we designate as positive and negative. The convention
was derived from Franklin's experiments~\cite{Franklin}. He rubbed a glass
rod with silk and called the charges on the glass rod positive. He
rubbed sealing wax with fur and called the charge on the sealing wax
negative. Like charges repel and opposite charges attract each
other. The unit of charge is called the Coulomb (C).  The smallest
unit of  {\it free} charge known in nature is the charge of an electron or
proton, which has a magnitude of $e = 1.602 \times 10^{-19}~{\rm C}$.

Charge of any ordinary matter is quantized in integral multiples of
$e$. An electron carries one unit of negative charge $-e$, whereas a
proton carries one unit of positive charge $+e$. In a closed system,
the total amount of charge is conserved since charge can neither be
created nor destroyed. A charge can, however, be transferred from one
body to another.

Consider a system of two point charges, $q_1$ and $q_2$, separated by
a distance $r$ in vacuum. The magnitude of the force exerted by $q_1$
on $q_2$ is given by 
\begin{equation}
F_e = k_e \frac{q_1 q_2}{r^2} \,,
\end{equation}
where $k_e = 8.9875 \times 10^9~{\rm N \, m^2/C^2}$ is the Coulomb's constant~\cite{Coulomb}.\\

{\bf \S Electric field.}~An electric charge $q$ produces an electric
field everywhere. To quantify the strength of the field created by
that charge, we can measure the force a positive {\it test charge} $q_0$
experiences at some point. 
We must take $q_0$ to be infinitesimally small so that the field $q_0$
generates does not disturb the ``source charges.'' With this in mind,
the magnitude of the electric field $E$ is
defined as $E = F/q_0$.

\begin{figure}[t]
    \postscript{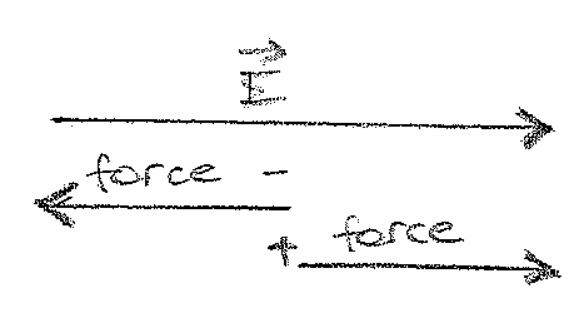}{0.9}
\caption{Force acting on a charged particle inside an electric field.
 \label{fig:10}}
\end{figure}

An electric field $\vec E$ would then exert a force on a charge, as
shown in Fig.~\ref{fig:10}. The
direction of the force is along the field if the charge is positive
(e.g. a proton) and opposite to the field if the charge is negative
(e.g. an electron). \\

\begin{figure}[t]
    \postscript{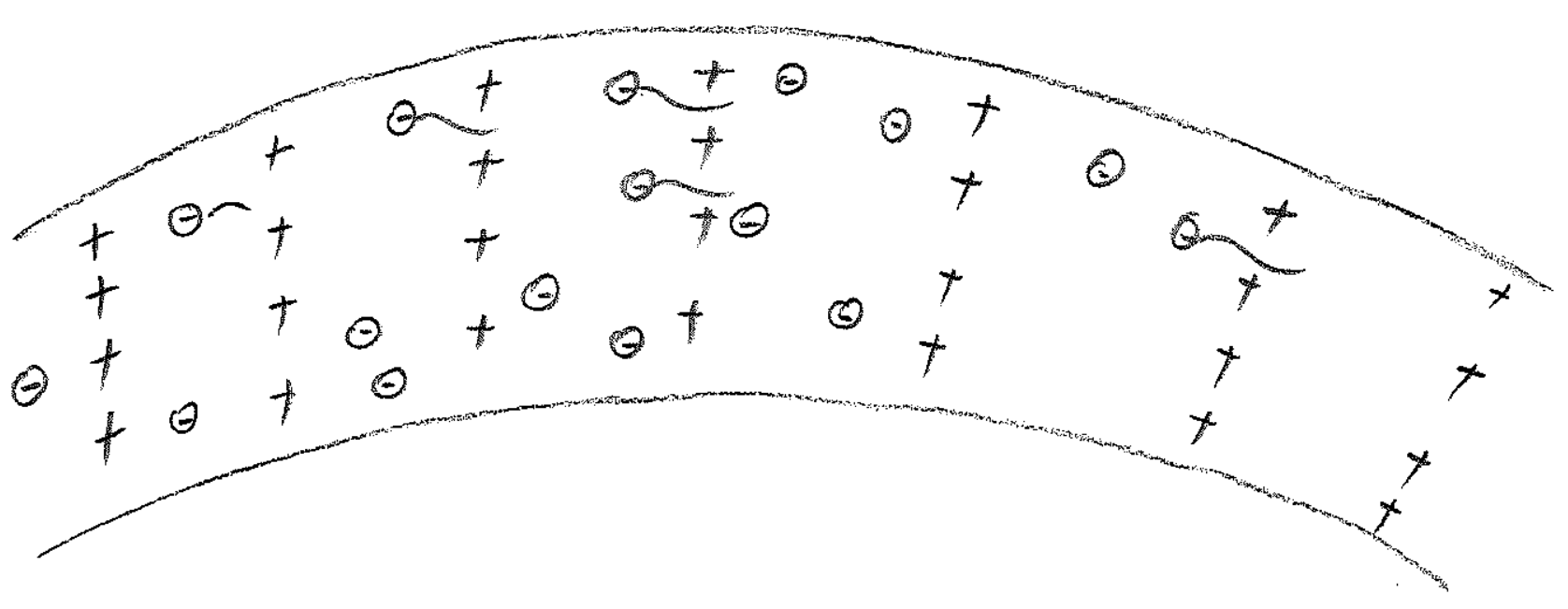}{0.9}
    \caption{Electric current in a copper wire. The charge carriers
      are the mobile electrons and the positively charged copper ions
      are essentially stationary in the metal array. \label{fig:13}}
\end{figure}

{\bf \S~Electric current.} The current $i$ is the rate of flow of the
electric charge in a wire. Seen through some super-microscope, a
copper wire carrying an electrical current looks like the illustration
in Fig.~\ref{fig:13}. The $+$ charges are fixed atoms arranged on a regular
array. They vibrate in place, but do not flow along the wire. The
$\ominus$ charges are electrons which flow along the wire, bumping
into the fixed atoms, losing their energy in this way  (i.e.,
``heating the wire''). The electrons are pushed by the battery or
generator, and the current can be thought as being produced by the
electrical pressure, or {\it voltage} of the power source.

We have seen that the electric charge is measured in coulombs. A
coulomb contains about 6 billion billion electrons. More conveniently,
we measure the rate of flow of electric charge, or {\it current}. If
one coulomb of charge past some point in a circuit in one second, then
the current is one ampere (A). Typically, the current through a light
bulb is 0.7 to 1 ampere.\\
  
{\bf \S~Electromotive force.} The energy given to each coulomb by the
power source is called the electromotive force (emf), or the voltage
$V$. Therefore, if there are $q$ coulombs, the total energy handed out
is proportional to $V$ times $q$. The 
\begin{equation}
{\rm power} = \frac{{\rm energy}}{{\rm time}} 
\end{equation}
is proportional to 
\begin{equation}
V \ \left(\frac{q}{t} \right) = {\rm volts} \times {\rm amperes} \, .
\end{equation} 
The units are such that the power (in watts) equals the voltage (in volts)
times the current (in amperes). A watt is a small amount of power: it
is $1/1000$ of a kilowatt, and the {\it kilowatt} is the power
generated at the rate of 1~kWh/hour.

Here is one example. A light bulb rated at 100~W (or watts) is connected to a line
voltage of 110~V (or volts). {\it (i)}~What is the current through the light
bulb?  
\begin{equation}
{\rm power \  (watts) = voltage \ (volts)}  \times {\rm current \
  (amperes)} . \nonumber
\end{equation}
This implies that 
\begin{equation}
100~{\rm watts} = 110~{\rm volts} \times {\rm current}  \,,
\end{equation}
and so
\begin{equation}
{\rm current} = \frac{100~{\rm watts}}{110~{\rm volts}} = 0.9~{\rm amperes} \, .
\end{equation}
{\it (ii)}~How much energy
(in kWh) is used by the light bulb in 24 hours?
\begin{equation}
{\rm energy \ (kWh)} = {\rm power \ (kilowatts)} \times {\rm time \
  (hours)} \, .
\end{equation}
Now, since
\begin{equation}
100~{\rm watts} = \frac{1}{10} (1,000~{\rm watts}) = \frac{1}{10}~{\rm
  kilowatt} \,,
\end{equation}
we have
\begin{equation}
{\rm energy} = \frac{1}{10}~{\rm kW} \times 24~{\rm hours} = 2.4~{\rm kWh} \, .
\end{equation}
At 12\cent/kWh this is 29\cent. {\it (iii)}~If the light bulb were
immersed in a bathtub full of water (500~lb, or 500~pints) initially
at $60^\circ$F, what would be the temperature of the water after 24
hours? The bulb releases 2.4~kWh in 24~hours, which is $2.4~{\rm kWh} \times
3,412~{\rm Btu/kWh} = 8,189$~Btu. Each Btu will heat 1 lb H$_2$O
$1^\circ$F. Therefore, to heat 500~lb $1^\circ$F takes
500~Btu. $8,189$~Btu will raise the temperature of the water in the tub
by $8,189/500 = 16.4^\circ$F.

Another example is as follows. A 100 MW electric power station serves
a metropolitan area of $2 \times 10^6$ people. How much electrical
energy (in kWh) is supplied per capita per month?
Since
\begin{eqnarray}
1,000~MW & = & 1,000 \times 10^6~{\rm watts} \nonumber \\
& = & 10^6 \times (1,000~{\rm watts}) \nonumber \\
& = & 10^6~{\rm kilowatts} \, . \nonumber
\end{eqnarray}
the energy  generated  in  a  month is 
\begin{eqnarray}
 {\rm power \,
  (kW) \times time\, (hr)} & = & 10^6~{\rm kW} \times 24~{\rm
  hr} \times 30~{\rm days}
\nonumber \\ & = & 7.2
\times 10^8~{\rm kWh} \, .\nonumber
\end{eqnarray}
Per capita, this is $7.2 \times 10^8/ 2 \times 10^6 = 360$~kWh per month.\\

{\bf \S~Electric circuit.} Any path along which electrons can flow is a
{\it circuit}. For a continuos flow of  electrons, there must be a
complete circuit with no gaps. A gap is usually provided by an
electric switch that can be opened or closed to either cutoff or allow
energy flow. Most circuits have more than one device that receives
electric energy. These devices are commonly connected in a circuit in
one of two ways: {\it series} or {\it parallel}. When connected in
series, they form a single pathway for electron flow between the
terminals of the battery, generator, or wall socket (which is simply
an extension of these terminals). When connected in parallel, they
form branches, each which is a separate path for the flow of
electrons. Both series and parallel connectors have their own
distinctive characteristics.

\begin{figure}[t]
    \postscript{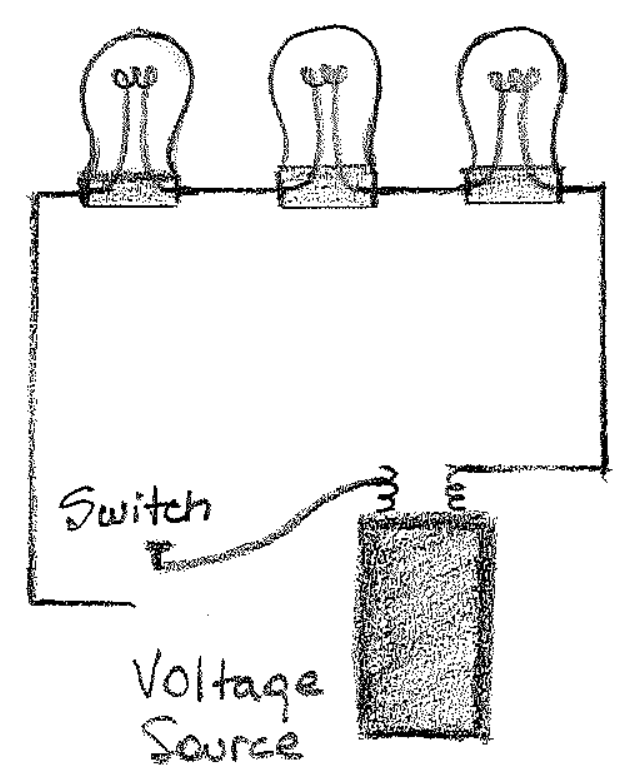}{0.9}
\caption{Series circuit.
 \label{fig:11}}
\end{figure}

\begin{figure}[t]
    \postscript{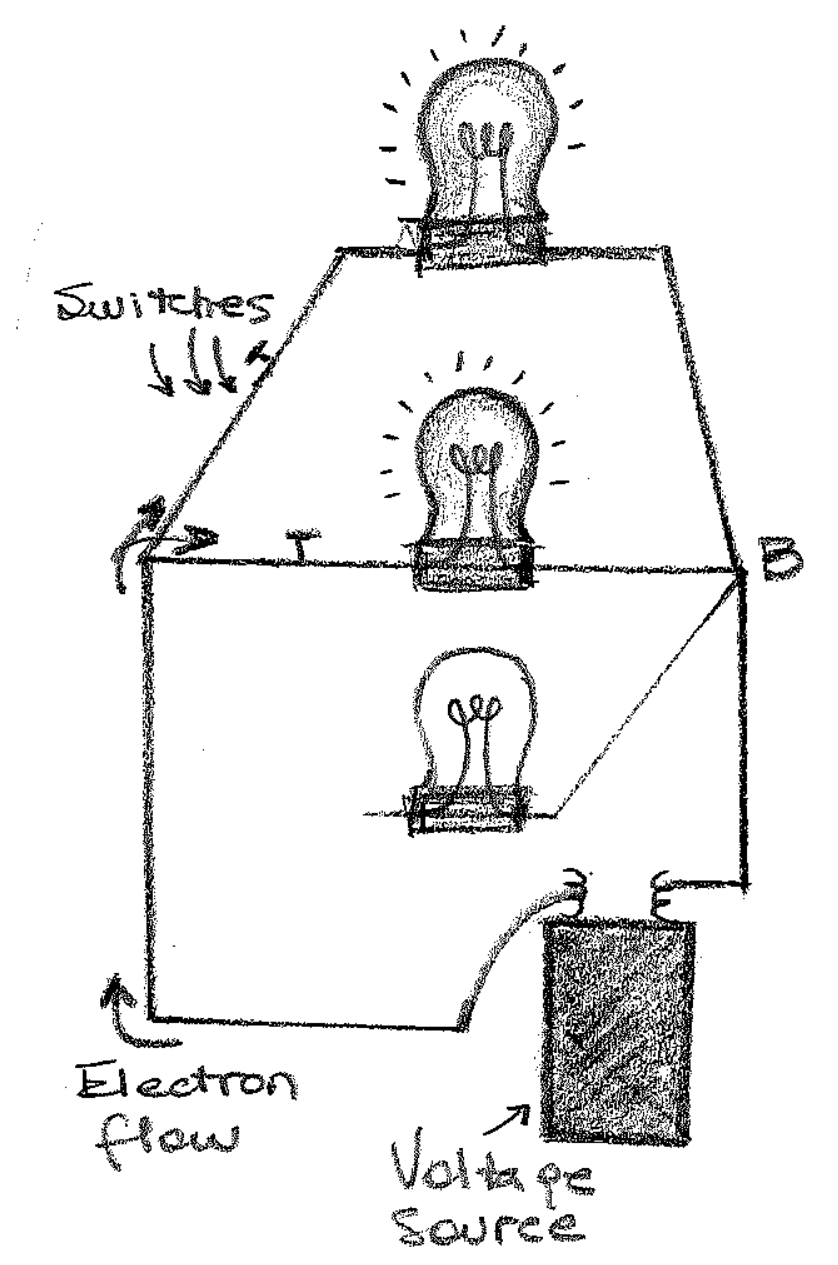}{0.9}
\caption{Parallel circuit.
 \label{fig:12}}
\end{figure}
A simple example of a {\it series} circuit is shown in
Fig.~\ref{fig:11}. All devices, lamps in this case, are connected end to end,
forming a single path for electrons. The same current exists almost
immediately in all three lamps, and also in the battery, when the
switch is closed. The greater the current in a lamp, the brighter it
glows. Electrons do not ``pile up'' in any lamp but flow through each
lamp, simultaneously.  Some electrons move away from the negative
terminal of the battery, some move toward the positive terminal, and
some move through the filament of each lamp. Eventually, the electrons
may move all the way around the circuit. This is the only path of the
electrons through the circuit. A break anywhere in the path results in
an open circuit, and the flow of electrons ceases. Burning out one of
the lamp filaments or simply opening the switch could cause such a
break.

A simple example of a {\it parallel} circuit is shown in
Fig.~\ref{fig:12}.  Three lamps are connected to the same two points:
$A$ and $B$. Electrical devices directly connected to the same two
points of an electric circuit are said to be connected in
parallel. The pathway from one terminal of the battery to the other is
completed if only {\it one} lamp is lit. The
circuit shown in Fig.~\ref{fig:12} branches into three separate pathways from $A$ to $B$. A break
in any one path does not interrupt the flow of charge in the other
paths. Each device
operates independently of the other devices.\\

{\bf \S~How do voltage and current relate?}  The
relationship among voltage and current is summarized by 
Ohm's law, which states
that, {\it at a constant temperature,  the electrical current
flowing 
through a wire between
two points is directly proportional to the voltage across the two
points}~\cite{Ohm}.  Introducing the constant of proportionality, the resistance
$R$, one arrives at the usual expression that describes
this relationship
\begin{equation}
i = \frac{V}{R} \,,
\end{equation}
where $i$ is the current through the wire in units of amperes,
$V$ is the voltage measured across the wire in units of volts,
and $R$ is the resistance of the wire in units of ohms ($\Omega$). 

If resistors are connected in series, each resistor has the same
current $i$. Each resistor has voltage $iR$, given by Ohm's law. Then,
for the circuit shown in Fig.~\ref{fig:11}, 
the total voltage drop across all three resistors is
\begin{eqnarray}
V_{\rm total} & = & i R_1 + i R_2 + i R_3 \nonumber \\
& = & i (R_1+ R_2 + R_3) \, .
\end{eqnarray}  
When we look at all three resistors together as one unit, we see that
they have the same $i~{\rm vs.}~V$ relationship as one resistor, whose
value  is
the sum of the resistances. So we can treat these resistors as just
one equivalent resistance,   
\begin{equation}
R_{\rm eq} = R_1 + R_2 + R_3 \, ,
\end{equation}
as long as we are not interested in the
individual voltages. Their effect on
the rest of the circuit is the same,
whether lumped together or not.

Resistors in parallel carry the same voltage. All of the resistors in
Fig.~\ref{fig:12} have voltage $V$. The current flowing through each resistor
could definitely be different. Even though they have the same voltage,
the resistances could be different. If we view the three resistors as
one unit, with a current $i$ going in, and a voltage $V$, this unit
has the following $i$ vs. $V$ relationship: 
\begin{eqnarray}
i & = & i_1 + i_2 + i_3  \nonumber \\
& = & V \left(\frac{1}{R_1} +
\frac{1}{R_2} + \frac{1}{R_3} \right) \, . 
\end{eqnarray}
Thus, to the outside world, the parallel resistors look like one
satisfying
\begin{equation}
\frac{1}{R_{\rm eq} }= \frac{1}{R_1} + \frac{1}{R_2} + \frac{1}{R_3} \,,
\end{equation}
for the equivalent resistance $R_{\rm eq}$.\\

\begin{figure}[t]
    \postscript{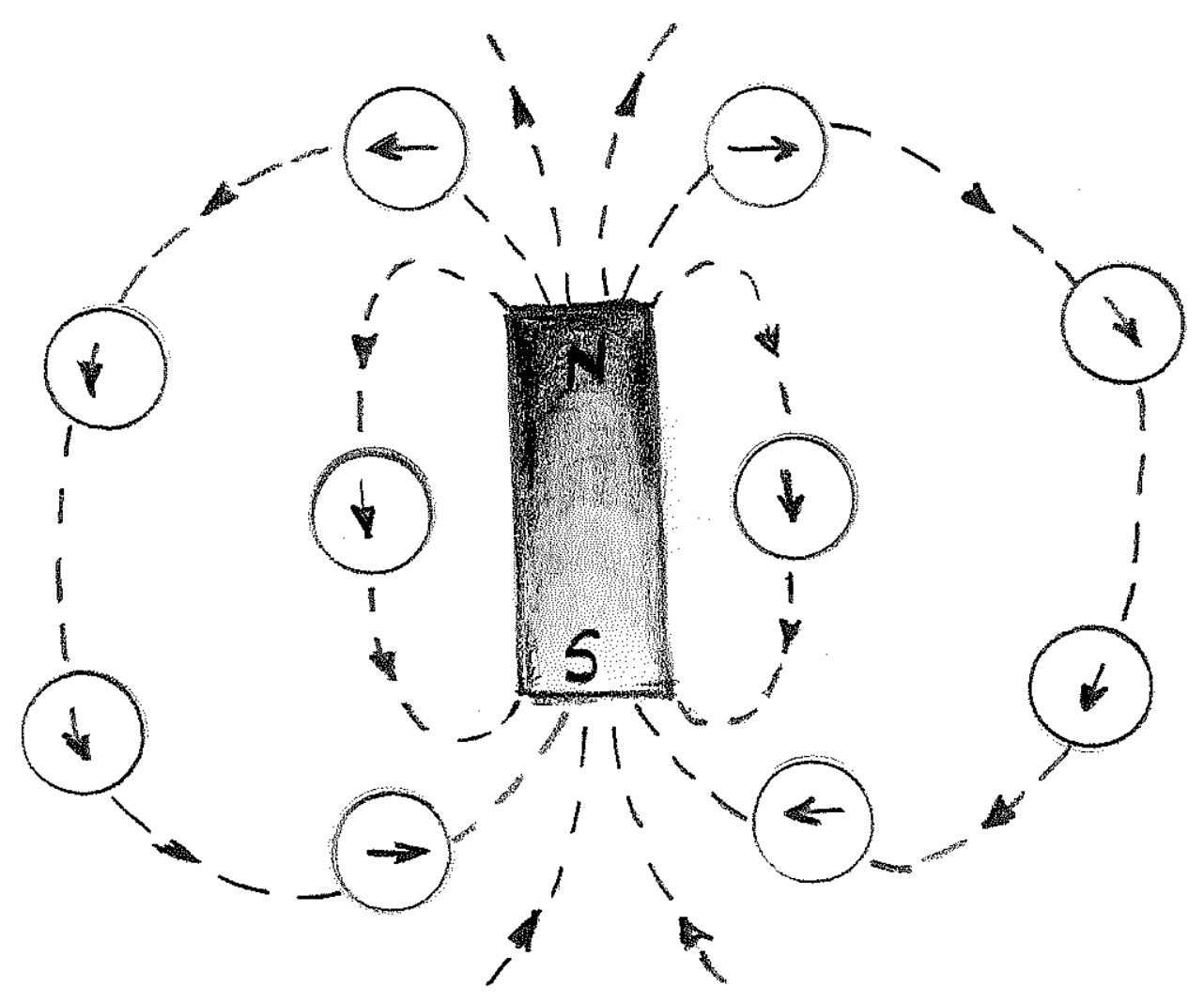}{0.9}
\caption{Magnetic field produced by a bar magnet.
 \label{fig:14}}
\end{figure}

{\bf \S~Magnetic field.}~We have seen that a charged object produces
an electric field $\vec E$ at all points in space. In a similar
fashion, a bar magnet is a source of a magnetic field $\vec B$. This
can be readily demonstrated by moving a compass near the magnet.  The
compass needle will line up along the direction of the magnetic field
produced by the magnet, which is shown in Fig.~\ref{fig:14}.

\begin{figure}[t]
    \postscript{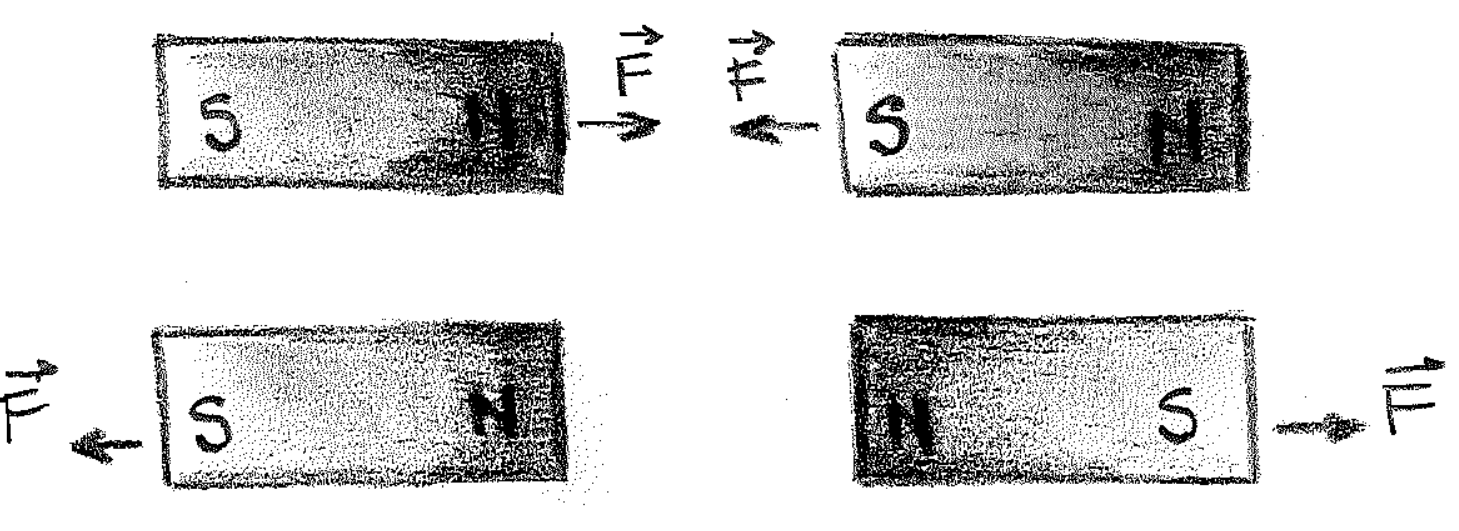}{0.9}
\caption{Magnets attracting and repelling.
 \label{fig:15}}
\end{figure}
\begin{figure}[t]
    \postscript{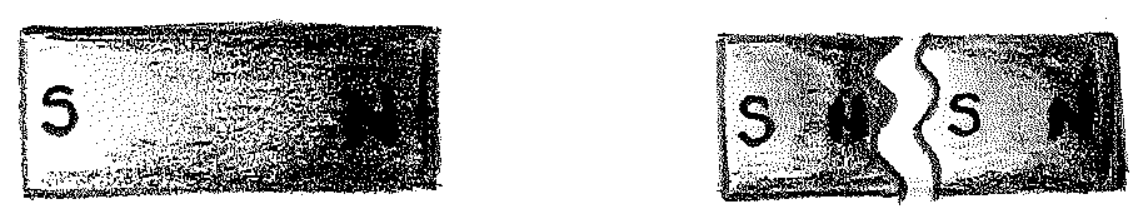}{0.9}
\caption{Magnetic monopoles do not exist in isolation.
 \label{fig:16}}
\end{figure}

Notice that the bar magnet consists of two poles, which are designated
as the north N and the south S. Magnetic fields are strongest at
the poles. The magnetic field lines leave from the north pole and
enter the south pole. When holding two bar magnets close to each
other, the like poles will repel each other while the opposite poles
attract, see Fig.~\ref{fig:15}.

\begin{figure}[t]
    \postscript{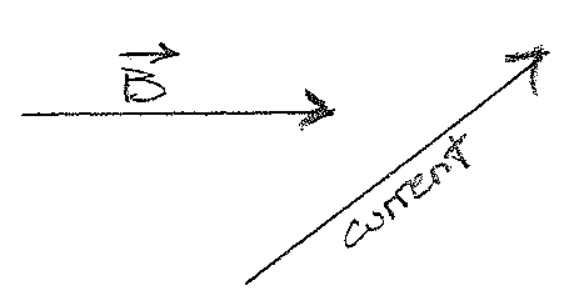}{0.9}
\caption{The direction of the force due to the magnetic field and the
  current is perpendicular to the plane of paper. 
 \label{fig:17}}
\end{figure}

\begin{figure}[t]
    \postscript{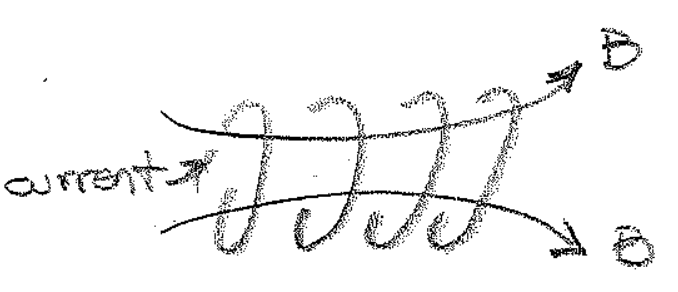}{0.9}
\caption{Magnetic field surrounding a current-carrying wire.
 \label{fig:18}}
\end{figure}

\begin{figure}[t]
    \postscript{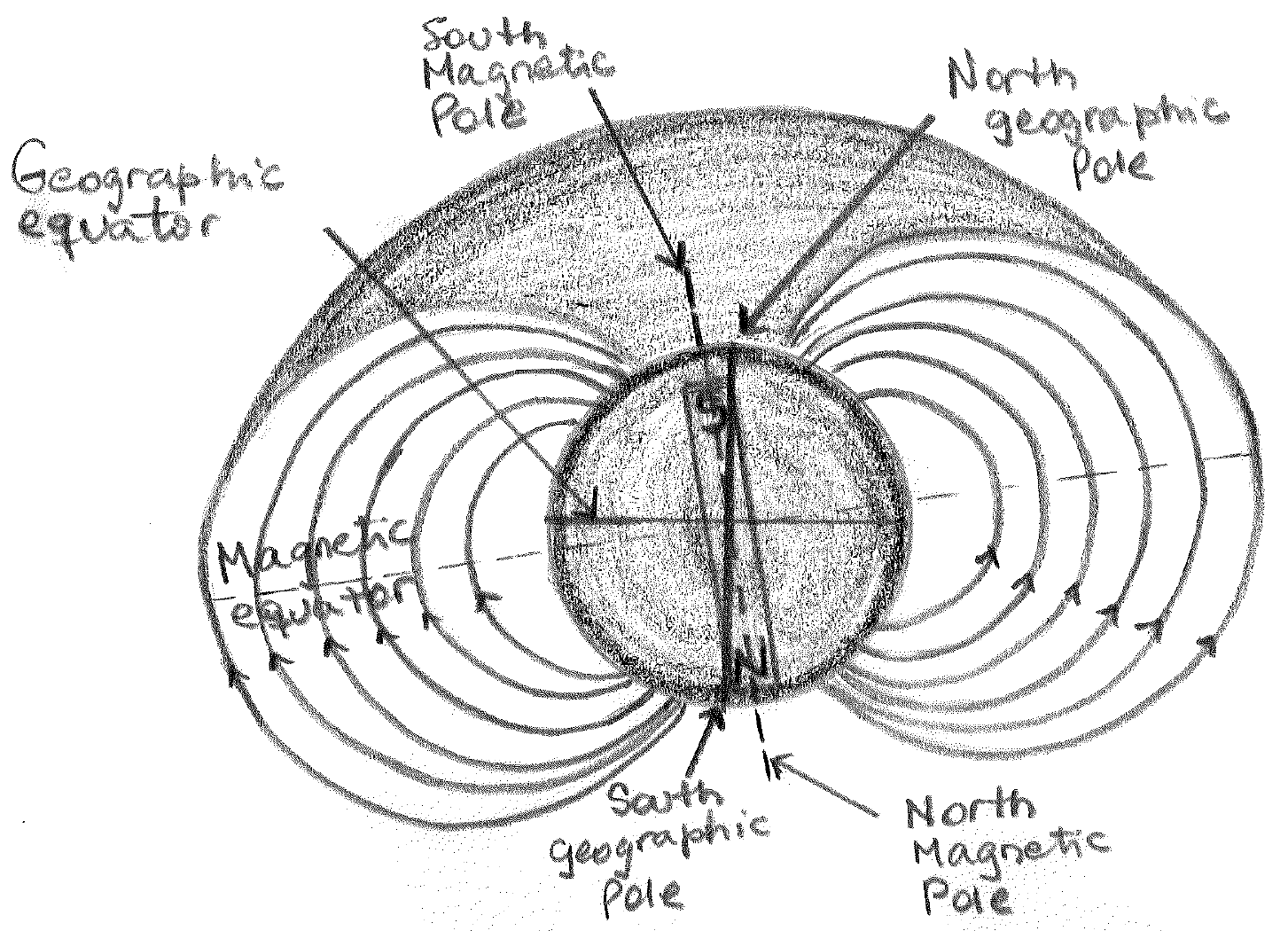}{0.9}
\caption{A pictorial representation of the Earth's magnetic field
  lines that is very useful in visualizing the strength and direction of
  the magnetic field. The $\vec B$ lines  describe the direction of
  the magnetic force on a north monopole at any given position.
  \label{fig:u23}}
\end{figure}

Unlike electric charges which can be isolated, the two magnetic poles
always come in a pair. When you break the bar magnet, two new bar
magnets are obtained, each with a north pole and a south pole, as
shown in Fig.~\ref{fig:16}. In other words, magnetic {\it monopoles}
have not been seen in isolation, although they are of theoretical
interest.

A magnetic field $\vec B$ exerts a force on a {\it moving} charge. The
direction of the force is perpendicular to both the field and the
motion of the charge. The magnitude of the
force depends on the velocity of the charge $v$, the magnetic field
strength $B$, and the angle between the direction of $\vec v$ and
$\vec B$,
\begin{equation}
F_B = |q| \,  v \,  B \, \sin \theta \,,
\end{equation}
where $|q|$ is the absolute value of the charge. The magnetic force
$F_B$ vanishes when $\vec v$ is parallel to $\vec B$. However, when
$\vec  v$ makes an angle $\theta$ with $\vec B$, the direction of $\vec F_B$
is perpendicular to the plane formed by $\vec v$ and $\vec B$, and the
magnitude of $\vec F$ is proportional to $\sin \theta$.  The unit of magnetic field is the tesla (T)
\begin{eqnarray}
{\rm tesla} & = & \frac{{\rm newton}}{{\rm (coulomb)} {\rm
    (meter/second)}} \nonumber \\
& = &  \frac{{\rm newton}}{{\rm (ampere)} {\rm (meter)}} \, .
\end{eqnarray}

We have just seen that a charged particle moving through a magnetic
field experiences a magnetic force $\vec F_B$. Since electric current
consists of a collection of charged particles in motion, when a
current-carrying wire is placed in a magnetic field, it will also
experience a magnetic force,  see Fig.~\ref{fig:17}. Moreover, Oersted noticed that an
electric current flowing through a wire can cause a compass needle to
deflect perpendicular to the wire, showing that a current also creates a
magnetic field $\vec B$~\cite{Oersted}. The lines describing the $\vec
B$ direction surround
the current, as shown in Fig.~\ref{fig:18}~\cite{Ampere}. 

In closing this section, it is important to note that the
configuration of the Earth's magnetic field (shown in Fig.~\ref{fig:u23}) 
is very much like the one that would be
achieved by burying a gigantic bar magnet deep in the interior of the
Earth. When we speak then of
a compass magnet having a north pole and a south pole, we should say
more properly that it has a ``north-seeking'' pole and a
``south-seeking'' pole. By this we mean that one pole of the magnet
seeks, or points to, the north geographic pole of the Earth. Because
the north pole of a magnet is attracted toward the north geographic
pole of the Earth, we conclude that the Earth's south magnetic pole is
located near the north geographic pole, and the Earth's north magnetic
pole is located near the south geographic pole.\\

\section{How Light Works}
\label{sec6}

\begin{figure}[t]
    \postscript{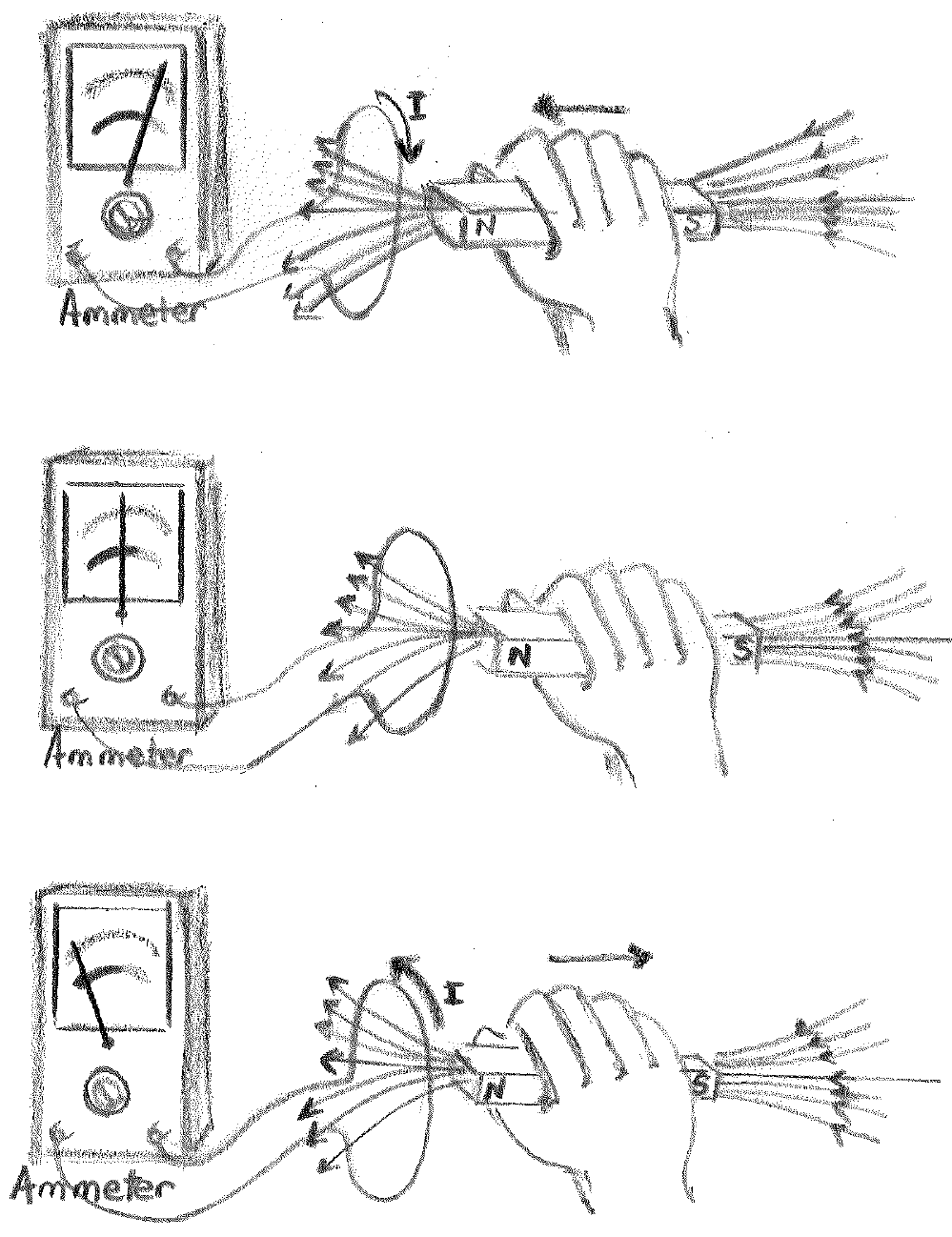}{0.9}
\caption{In the top panel a magnet is moved toward a loop of wire
  connected to an instrument for measuring current in amperes (i.e. an
  ammeter). The ammeter deflects as shown, indicating that a current
  is induced in the loop.  In the middle panel the magnet is held
  stationary and there is no induced current in the loop, even when
  the magnet is inside the loop. In the bottom panel the magnet is
  moved away from the loop. The ammeter deflects in the opposite
  direction, indicating that the induced current is opposite that
  shown in the top panel. Changing the direction of the magnet's
  motion changes the direction of the current induced by that
  motion. Note that the circulation of the charges producing the current has
  associated an electric field.
  \label{fig:u24}}
\end{figure}

\begin{figure}[t]
    \postscript{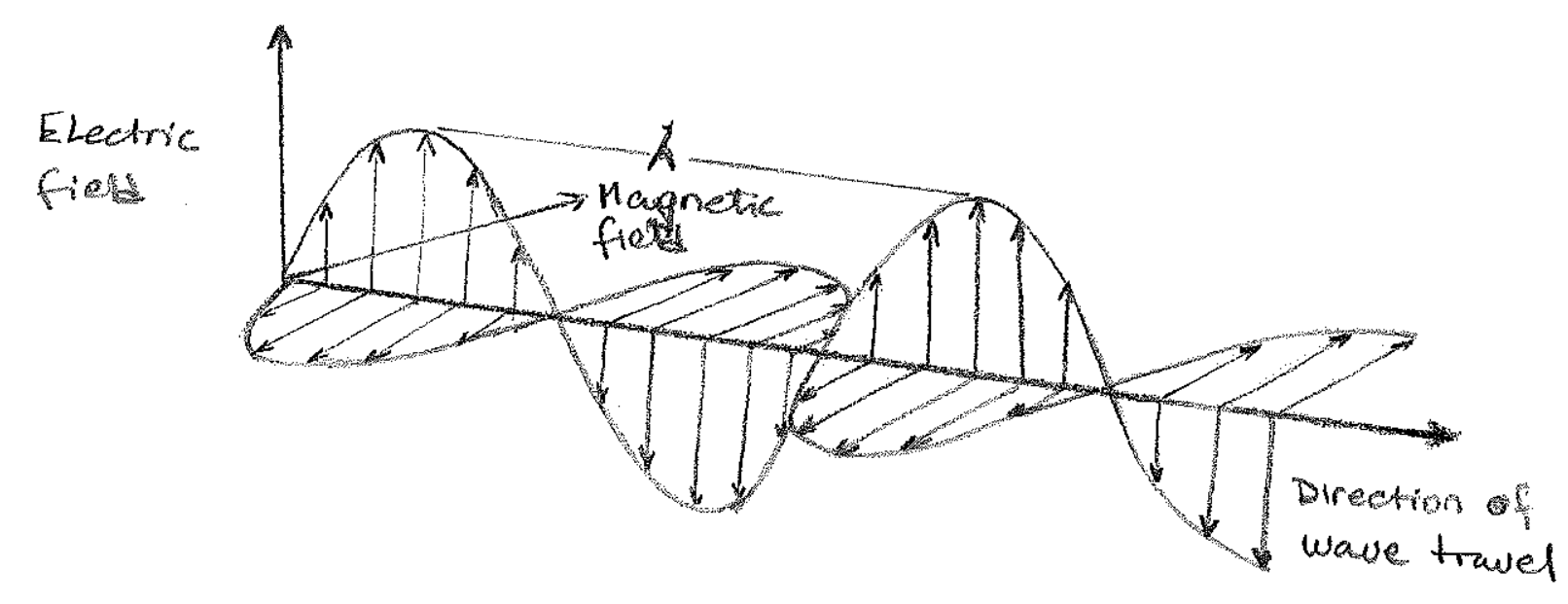}{0.9}
\caption{Electromagnetic waves transport energy through empty space,
  stored in the propagating electric and magnetic fields. A single
  frequency electromagnetic wave exhibits a sinusoidal variation of
  electric and magnetic fields in space. The magnetic
  field variation is perpendicular to the electric field. 
 \label{fig:19}}
\end{figure}

{\bf \S~Electromagnetic waves.}  In the mid-1800's, Faraday observed
(see Fig.~\ref{fig:u24}) that a changing magnetic field creates an
electric field~\cite{Faraday}.  Exhilarated by this discovery, Maxwell
hypothesized that a changing electric field creates a magnetic field,
and putting all this together he predicted that if you began changing
$\vec E$ and $\vec B$ in any region in space, a wave of these changing
fields propagates at the speed of light $c \simeq 3 \times 10^8~{\rm
  m/s}$, outward from the region where the change first took place;
see Fig.~\ref{fig:19}~\cite{Maxwell:1865zz}.  For example, moving an electron causes a
change in $\vec E$, which causes a change in $\vec B$, etc. These
changing fields zip over to a second electron, which was jiggled by
the electric field which arrives at a microscope time later. It was
not until 1888 that Maxwell's prediction passed an important test when
Hertz generated and detected certain types of electromagnetic waves in
the laboratory~\cite{Hertz}. He performed a series of experiments that
not only confirmed the existence of electromagnetic waves, but also
verified that they travel at the speed of light.

Light itself consists of electric and magnetic fields of this kind. But
what about photons? Good question. We will deal with this soon. But
meanwhile, we note the very important fact that we have here: a means
of transporting energy through empty space, without
transporting matter.  The {\it electromagnetic waves} (that's what they
are called) propagate any time an electron is jiggled.\\ 

{\bf \S~Ray optics.} We just learned that light is a wave.  Unlike
particles, waves behave in funny ways, {\it e.g.}, they bend around
corners (think of sound coming through a doorway).  However, the
smaller the wavelength $\lambda$ is, the weaker these funny effects
are, and so for light (tiny $\lambda$), no one noticed the ``wave
nature'' at all, for a long time. The wavelength of light is about 100
times smaller than the diameter of a human hair! This means that for
most physics phenomena of everyday life, we can safely ignore the wave
nature of light, because light waves travel through and around
obstacles whose transverse dimensions are much greater than the
wavelength, and the wave nature of light is not readily
discerned. Under these circumstances the behavior of light can be adequately
described by rays obeying a set of geometrical rules. This model of
light is called ray optics. Strictly speaking, ray optics is the limit
of wave optics when the wavelength is infinitesimally small.

To study the more {\it classical} aspects of how light travels we will
take
into account the following considerations:
\begin{itemize}
\item We will ignore time oscillations/variations ($10^{14}~{\rm Hz}$
  is too fast to notice, generally!)
\item We will assume light travels through a transparent medium in
  straight line (at 186,282~miles per second, super fast).\footnote{
    Fermat's principle states that when a light ray travels between
    any two points, its path is one that requires smallest time
    interval~\cite{Fermat}. An obvious consequence of this principle is that 
paths of light rays traveling in a homogeneous medium are straight lines, because a straight line is shortest distance between two points.}
\item Light can  change directions in 3 main ways:
\begin{itemize}
\item Bouncing off objects (reflection). 
\item Entering objects (e.g. glass) and bending (refraction).
\item Getting caught, and heating the object (absorption).
\end{itemize}
\end{itemize}
In other words, we  consider  that light travels in the form of rays. The rays are
emitted by light sources and can be observed when they reach an
optical detector. We further assume that the optical rays propagate in
optical media. To keep things simple, we will assume that the media
are lossless, or transparent. 

Any transparent medium (such as air, water, glass, $\cdots$) that lets
light through will be characterized completely by a number $n$, called the {\it index of refraction}. $n$ is determined by how fast light travels through
the material. Light only travels at $c$, the ``speed of light'', in
vacuum. In materials, it is always slowed down.  The bigger the $n$, the
slower the light travels. The  index of refraction is defined as the ratio of the
speed of light in vacuum to that in the medium,
\begin{equation}
{\rm index \ of \ refraction} = \frac{\rm speed \ of \ light \ (in \ vacuum)}{{\rm
  speed \  of \ light \ (in \ medium)}} \, .
\end{equation}

\begin{figure}[t]
    \postscript{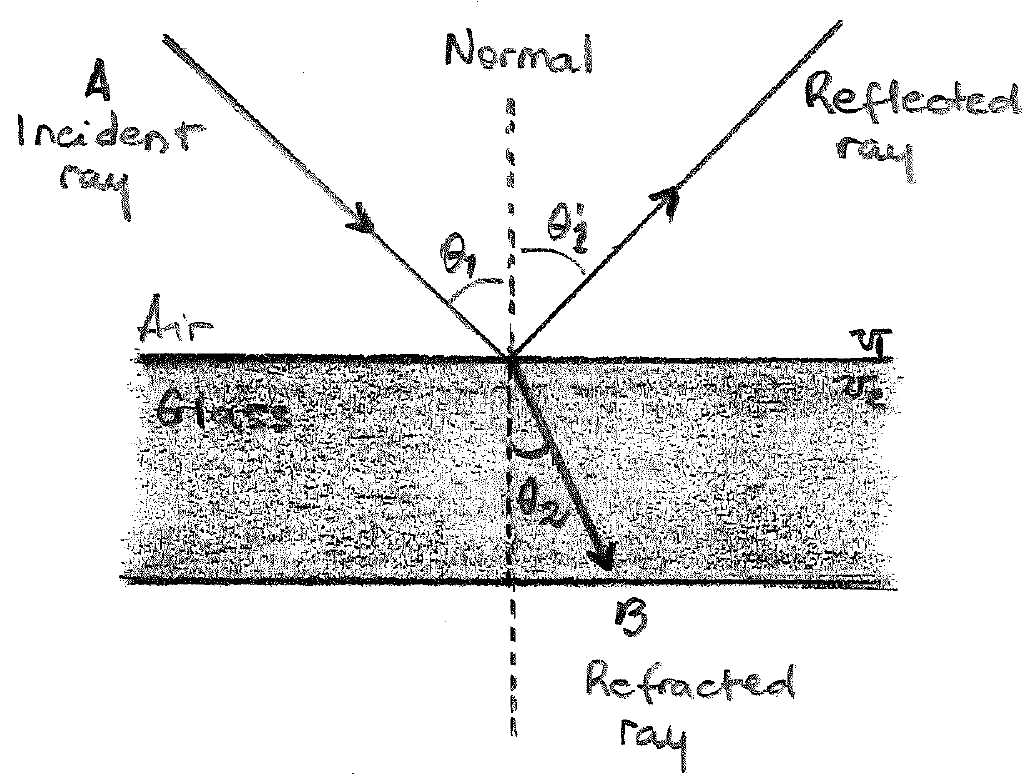}{0.9}
\caption{A ray obliquely incident on an air-glass interface. The
  refracted ray is bent toward the normal because $v_2 < v_1$.. All
  rays and the normal lie in the same plane.  \label{fig:20}}
\end{figure}

When a ray of light traveling through a transparent medium encounters
a boundary leading into another transparent medium, as shown in
Fig.~\ref{fig:20}, part of the energy is reflected and part enters the second
medium. The ray that enters the second medium is bent at the boundary
and is said to be refracted. The incident ray, the reflected ray, and
the refracted ray all lie in the same plane. Experiments and theory
show that the angle of reflection $\theta'_1$ equals the angle of
incidence $\theta_1$,
\begin{equation}
\theta'_1 = \theta_1 \, .
\end{equation}
This relationship is called the law of reflection.
The angle of refraction depends on the properties of the two media and on
the angle of incidence through the relationship
\begin{equation}
\frac{\sin \theta_2}{\sin \theta_1} = \frac{v_2}{v_1} = {\rm constant} \,,
\label{Snell-ley1}
\end{equation}
where $v_1$ is the speed of light in the first medium and $v_2$ is the
speed of light in the second medium. If we replace the $v_2/v_1$ term with the
 ratio of the refractive indexes $n_1/n_2$ we can express (\ref{Snell-ley1}) in an
 alternative form:
\begin{equation}
n_1 \ \sin \theta_1 = n_2 \ \sin \theta_2 \, .
\label{Snell-ley2}
\end{equation}
The experimental discovery of this relationship, known as law of
refraction, is usually credited to Snell~\cite{Snell}.

The path of a light ray through a refracting surface is
reversible. For example, the ray shown in Fig.~\ref{fig:20} travels from
point $A$ to point $B$. If the ray originated at $B$, it would travel to the
left along line $BA$ to reach point $A$, and the reflected part would
point downward and to the left in the glass.

Light rays can pass through several boundaries. For example, you might
have a sheet of glass: a light ray will enter (going from small $n_1$
to larger $n_2$) and then exit (large $n_2$ to small $n_1$). At each
boundary, Snell's law will hold. At the left boundary we have $n_1
\sin \theta_{\rm in} = n_2 \sin \theta_2$ (light bends toward the normal --
convince yourself of the equation and the physics). At the right
boundary we have $n_2 \sin \theta_3 = n_1 \sin \theta_{\rm out}$ 
(light bends away from the normal -- again, convince
yourself). 

But geometry tells us (if the walls are parallel) that $\theta_2 = \theta_3$
(do you see why?), which means $\sin \theta_2 = \sin \theta_3$.
So $n_1 \sin \theta_{\rm in} = n_2 \sin \theta_2 = n_2 \sin \theta_3 =
n_1 \sin \theta_{\rm out}$
(can you follow all the steps required to write that last line down?)
That means (compare the far left with the far right of the equation)
that $\sin \theta_{\rm in}  = \sin \theta_{\rm out}$, which says
$\theta_{\rm in} = \theta_{\rm out}$.

What if you have glass with walls that are not parallel? This is the
idea behind lenses.  As light enters, it is bent, and rays come out
different depending on where and how they strike. The focal length of
a given optical system is a measure of how strongly the system
converges or diverges light. For an optical system in air, it is the
distance over which initially collimated (parallel) rays are brought
to a focus. A system with a shorter focal length has greater optical
power than one with a long focal length; that is, it bends the rays
more sharply, bringing them to a focus in a shorter distance.  The lens
geometry usually looks complicated (and it is!) but for thin lenses,
the result is relatively simple. For a
given object position, the focal length defines where the image will
appear:
\begin{equation}
\frac{1}{\rm object \ distance} +
\frac{1}{\rm image \ distance} = \frac{1}{\rm focal \ length} \, .
\label{thinlens}
\end{equation}
(\ref{thinlens}) is known as the thin lens equation.

\begin{figure}[tbp] 
\postscript{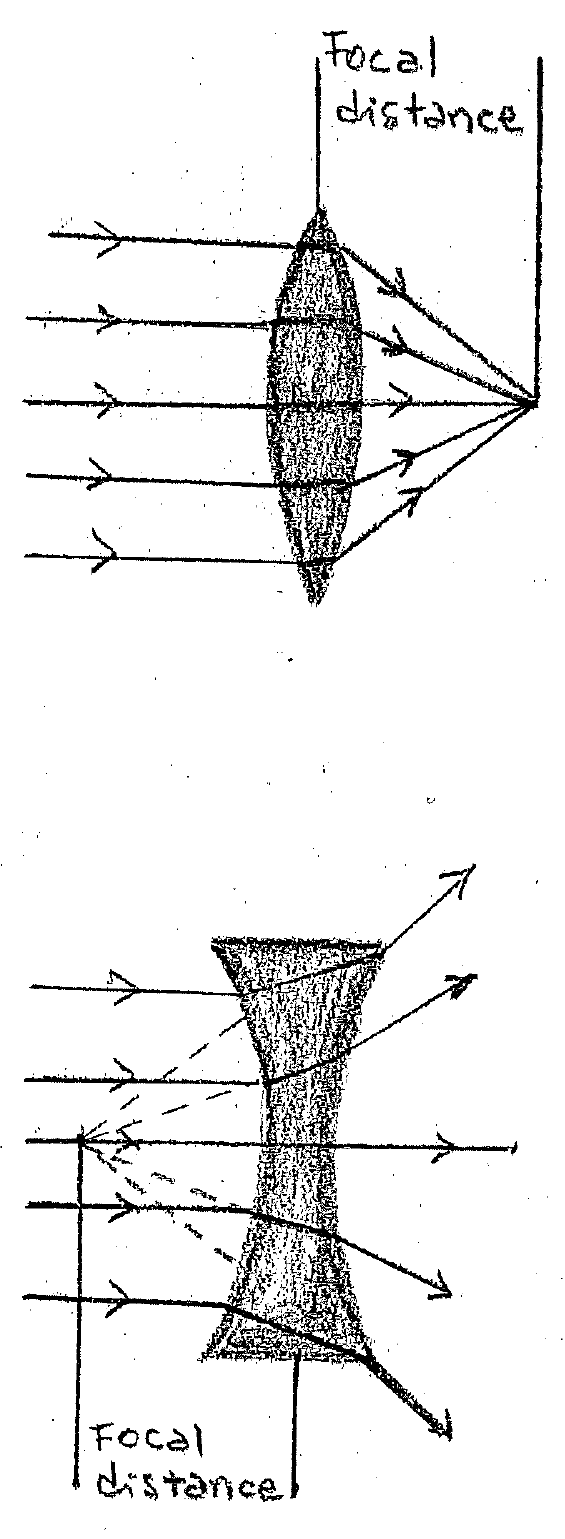}{0.8} 
\caption{The focii of converging (top) and diverging (bottom) lens.}
\label{fig:focii}
\end{figure}

Convex lenses (lenses which bulge outwards instead of curving inwards)
have the property of converging parallel light rays to a single
point, see Fig.~\ref{fig:focii}. This point is called the focus of the lens, and the distance of
the focus from the center of the lens is called the focal
length. Light rays are parallel if the distance to the object is very
large, effectively infinity compared to the size of the lens. If
(\ref{thinlens}) yields a negative image distance, then the image is a
{\it virtual image} on the same side of the lens as the object. If it
yields a negative focal length, then the lens is a concave (curving
inwards) {\it diverging lens} rather than a convex converging
lens, see Fig.~\ref{fig:focii}. The linear magnification relationship,
\begin{equation}
{\rm magnification} = - \frac{{\rm image \ distance}}{{\rm object
      \ distance}} \,,
\label{Magnification}
\end{equation}
allows one to predict the size of the image. The sign of the
maginification indicates whether the image is a real and inverted (negative) or a
virtual and upright (positive).

Now, we are in the position to answer two questions that have been
intriguing people since the time of Ptolemy. How do you know where
objects are? How do you see them?  You deduce the location (distance
and direction) in complicated ways, but it arises from the angle and
intensity (wavelength-weighted power emitted by a light source) of the
little {\it bundle} of light rays that make it into your eye. The eye
is an adaptive optical system. Unlike most optical systems, the
crystalline lens of the eye changes its shape to focus light from
objects over a great range of distances.\\

{\bf \S~True colors shinning through.} It is a common experience to
use a piece of shaped glass, a triangular cut of glass called a prism,
to produce a rainbow of color from sun light. This is basically a
refractive phenomenon and a simple extension of Fermat's least time
principle can be used to describe it. A narrow beam of white light
incident at a non-normal angle on one surface of the glass is
refracted; the beam changes direction. The spread of color appears
because the different colors in the light have different speeds in the
glass, with the red being faster than the blue and all colors slower
than for light in air. Hence, the red is bent less than the blue. The
separated rays then emerge from the other interface of the glass
spread in a familiar rainbow pattern. This spread of color can be seen
by placing a piece of paper after the second interface. It was Newton
who introduce the idea that white light was a complex phenomenon
composed of an internal structure -- the colors~\cite{Newton:1704}.
Prior to Newton's interpretation, the understanding was that the
different colors in the prism came from the glass and was not an
intrinsic property of the light. To show otherwise Newton placed a
prism in the path of a narrow beam of sunlight.  As expected, the beam
was spread over a band of angles. He then inserted a second prism and
allowed the spread beam to enter it. When arranged carefully, he found
that the second prism was able to reconstitute the original beam in
the original direction. He labeled the different colors with a
continuously varying parameter that had the units of a time, now
identified with the reciprocal of the frequency. The length $\lambda$
and time ${\cal T}$ characterizing a given color are connected by the
speed of light in the medium according to $\lambda/{\cal T} = c/n$
where $n$ is the index of refraction.

\section{Structure and Properties of Matter}
\label{sec:7}

{\bf \S~The basic building blocks of matter.} In the early fifth
century BC, the Greek philosopher Democritus proposed that matter
consists of small indivisible particles that he named atomos (meaning
uncuttable). Democritus' vision did not gain much favor with his
contemporaries and almost $2,300$~years passed before Dalton
reintroduced the idea of atoms in the early 1800s~\cite{Dalton}.
Despite the fact that Democritus' conception of atoms does not reflect our modern
understanding of atoms, the idea that matter is composed of
indivisible particles remains a simple but powerful idea.

We now know that all matter is composed
of {\it molecules}, and these molecules consist of one or more {\it
  atoms}. The molecules retain characteristics of the substance
(e.g. water molecules); they separate whereas atoms do not (e.g., hydrogen and
oxygen separately have nothing to do with water), unless the substance
is one of the {\it elements}. The molecules of an element contain only
one kind of atom (e.g., hydrogen, oxygen, uranium, iron). There are
about 100 elements, and therefore 100 different kinds of
atoms. Substances which are not elements are called compounds; at
present we know of about one million compounds.

Although Dalton conceived atoms to be indivisible, subsequent
experiments by Thomson~\cite{Thomson:1897cm} and Rutherford~\cite{Rutherford:1911zz} revealed a more complex
structure to the atom. Between 1900 and 1932, we had essentially
answered the question: {\it ``What are atoms themselves made off?''}
Atoms are made from three smaller subatomic particles: the {\it electron},
the {\it proton}, and the {\it neutron}. The atom, however, remains the smallest
division of matter with distinct chemical properties.

\begin{figure}[t]
    \postscript{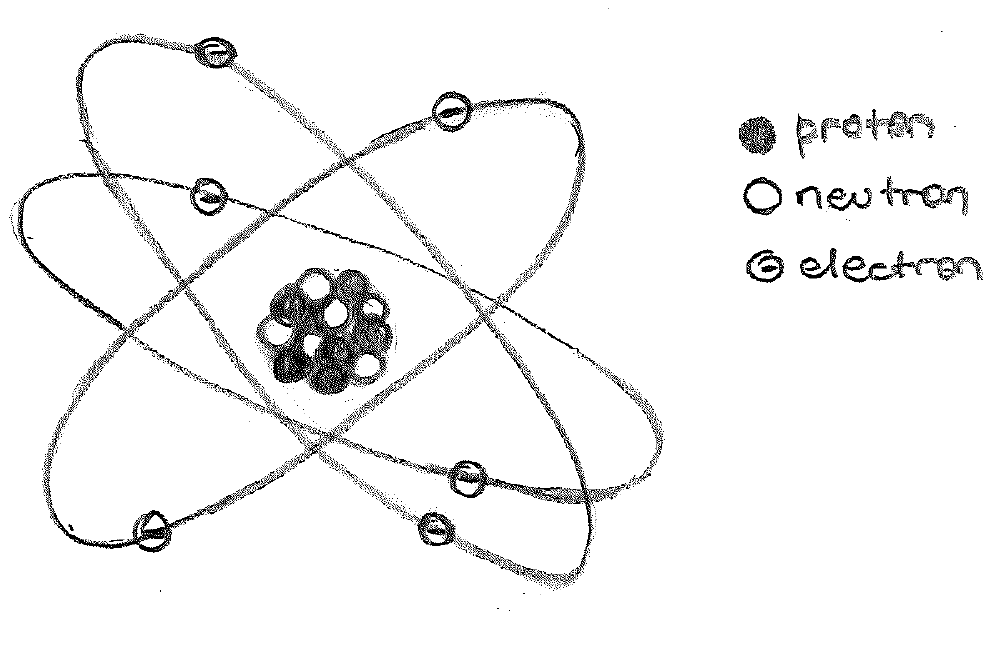}{0.9}
\caption{The early-20-th-century Bohr's
  model~\cite{Bohr:1913zba,Bohr:1913bca} 
of the carbon-12 atom, with a
  central nucleus and orbiting electrons, much like a solar system
  with orbiting planets. The $^{12}$C atom 
  consists of  6 protons and 6 neutrons that are found in the nucleus at the
  center of the atom, and 6 electrons that are outside the nucleus.
  \label{fig:21}}
\end{figure}

The 99.97\% of the mass of the atom is concentrated in a very small
{\it nucleus} at its center, consisting of two kinds of heavy
particles: protons and neutrons. The other 0.03\% of the
mass consists of the very light electrons, which buzz around in
fixed orbits very far from the nucleus, see Fig.~\ref{fig:21}. The
nucleus of the atom has a radius of approximately $5 \times
10^{-15}~{\rm m}$. The remainder of the atom, which has radius of
approximately $10^{-10}~{\rm m}$, is mostly empty space in which the
electrons buzz around. The scale of the atom is such that if the
nucleus were the size of a golfball, the electron orbits would be at a
distance of half a mile.

Now, besides mass, the electron and proton have something called
electric charge, or charge for short. Although the electron is $2,000$
times lighter than the proton, it has the same charge as the proton
or, more precisely, it has an equal and opposite charge.\footnote{The
  neutron is {\it neutral}; it has no charge.} This is because, as we
  have seen in Sec.~\ref{sec5},  electric
  charges come in two kinds: {\it positive} (or plus, $+$) and {\it
    negative} (or minus, $-$). Like charges repel one another, whereas
  opposite charges attract, and these forces become much stronger when
  the charges are closed together.  The characteristic properties of
  electrons, protons, and neutrons are summarized in
  Table~\ref{table1}.\footnote{The
  elementary (pointlike) particle model accepted today views quarks
  and leptons as the basic constituents of ordinary matter. By
  pointlike, we understand that quarks and leptons show no evidence of
  internal structure at the current limit of our resolution, which is
  about $2 \times 10^{-20}$~m. Leptons include electrons (along with muons, taus, and neutrinos).
Quarks are the fundamental particles that make up protons and
neutrons. They come in six different types, or {\it flavors}: up,
down, charm, strange, top, and bottom. For details, see e.g.,~\cite{Weinberg:2004kv}.} 

\begin{table}
\caption{Mass and charge of subatomic particles. \label{table1}}
\begin{tabular}{lccc}
\hline
\hline
~~~Particle~~~ & Mass (g) & Coulomb charge & ~Charge unit~\\
\hline
electron & ~~$9.10939 \times 10^{-28}$~~ & ~~$-1.6022 \times 10^{-19}$~~ &
$-1$ \\
proton & $1.67262 \times 10^{-24}$ & $+1.6022 \times 10^{-19}$  & $+1$ \\
neutron & $1.67493 \times 10^{-24}$ & 0 & 0 \\
\hline
\hline
\end{tabular}
\end{table}

\begin{table}
\caption{Stable isotopes of hydrogen, helium, and
  carbon. \label{table2}}
\begin{tabular}{lccc}
\hline 
\hline
Element & ~~~\# of protons~~~ & ~~~\# of neutrons~~~ & ~~~\# of electrons~~~
\\
\hline
hydrogen & 1 & 0,1,2 & 1 \\
helium & 2 & 1,2 & 2 \\
carbon & 6 & 6, 7 & 6 \\
\hline
\hline
\end{tabular}
\end{table}

The electrons are held in orbit by the electrical attraction of the
protons. The whole atom is neutral, because there are always equal
numbers  of electrons and protons. The reason the protons in the
nucleus do not fly appart (due to their strong electrical repulsion)
is that at these small distances, a much stronger attractive force (a
hundred times as strong) comes into play between neutrons and protons:
it is called the {\it nuclear force}.

We can now address the following question: {\it ``Why is an atom of carbon
different from an atom of hydrogen or helium?''} One possible
explanation is that carbon,  hydrogen, and helium have different numbers of
electrons, protons, or neutrons. Table~\ref{table2} provides the
relevant numbers.\footnote{$^3$H has a half-life of 12.32 years.}  Note that atoms of helium and carbon each have
two possibilities for the numbers of neutrons, and that it is even
possible for a hydrogen atom to exist without a neutron. Clearly the
number of neutrons is not crucial to determining if an atom is carbon,
hydrogen, or helium. Although hydrogen, helium, and carbon have
different numbers of electrons, the number is not critical to an
element's identity. For example, it is possible to strip an electron
away from helium forming a helium ion, with a charge of $+1$ that has
the same number of electrons as hydrogen. What makes an atom carbon is
the presence of 6 protons, whereas every atom of hydrogen has 1 proton
and every atom of helium has 2 protons. 

\begin{table*}
\caption{List of the elements with their symbols, atomic numbers, and
  average atomic masses. Approximate atomic masses for radioactive
  elements are given in parentheses. \label{elements}}
\postscript{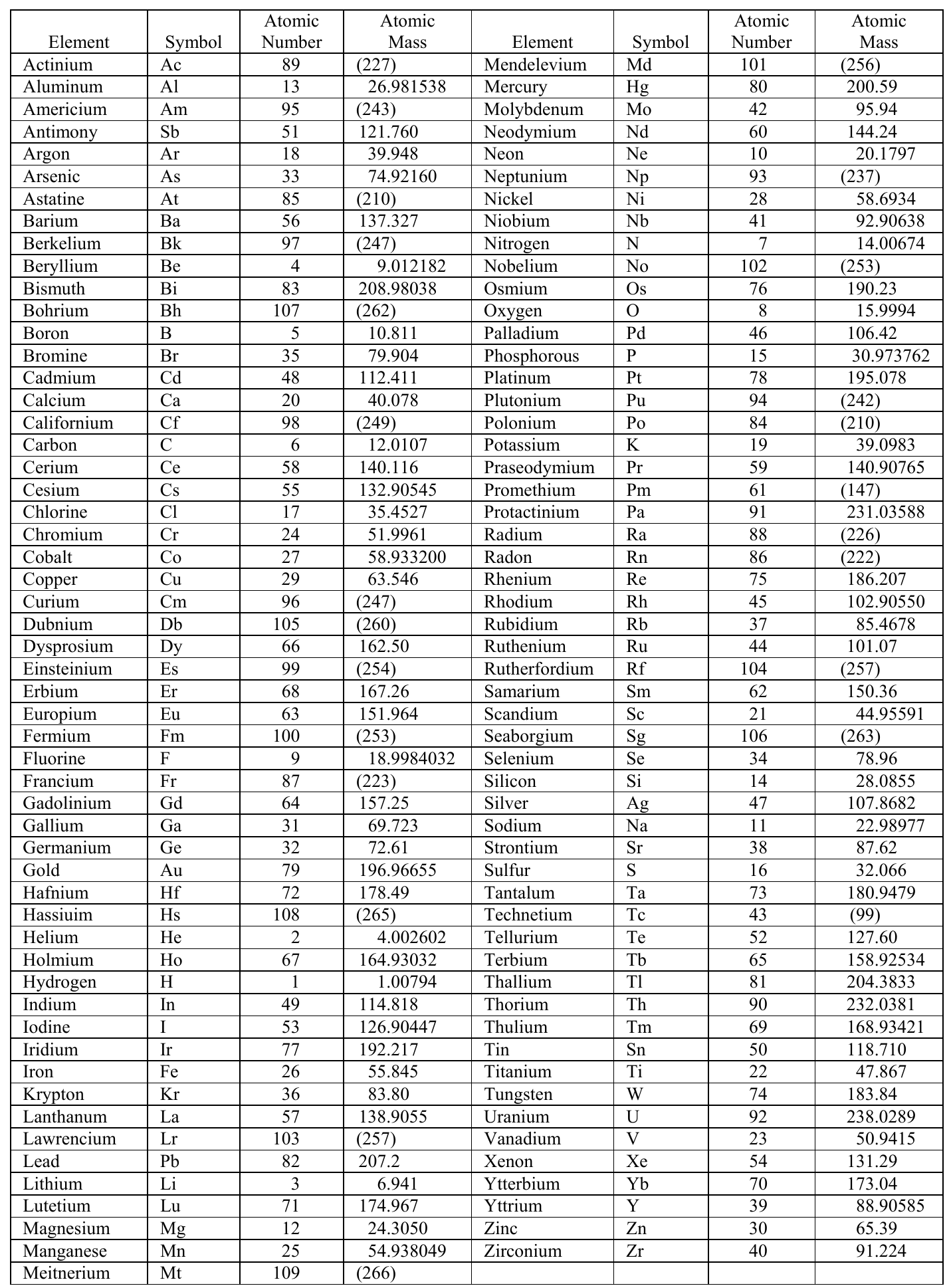}{0.9}
\end{table*}

We describe the atomic structure in terms of: {\it (i)}~the atomic
number $Z$, which equals the number of protons inside the nucleus;
{\it (ii)} the baryon number $A$, which equals the number of nucleons
inside the nucleus.\footnote{A nucleon is either a proton or a
  neutron.} The nuclei of all atoms of a particular element contain
the same number of protons but often contain different numbers of
neutrons. Nuclei that are related in this way are called isotopes.
Now, a bit of nomenclature. For an isotope with baryon number $A$ of
an element with symbol $X$, it is common writing {$_Z^A$X}. For
example, when we write $^{238}_{92}$U we mean the isotope of uranium
which has a total
of 238 neutrons plus protons, and 92 protons. (How many neutrons?)\\

{\bf \S~The microworld $\leftrightharpoons$ macroworld 
  connection.}~Individual atoms weigh very little, typically about
$10^{-24}~{\rm g}$ to $10^{-22}~{\rm g}$. This amount is so small that
there is no easy way to measure the mass of a single atom. To assign
masses to atoms it is necessary to assign a mass to one atom and
report the masses of other atoms relative to that absolute standard.
By agreement, atomic mass is stated in terms of atomic mass units (u),
where 1~u is defined as $1/12$ of the mass of an atom of carbon-12.
The atomic mass of carbon-12, therefore, is exactly 12~u. The atomic
mass of carbon-13 is 13.00335~u because the mass of an atom of
carbon-13 is 1.0836125 times greater than the mass of an atom of
carbon-12. Because carbon exists in several isotopes, the atomic mass of an
``average'' carbon atom is neither 12.0~u nor 13.00335~u. Instead it is
12.0107~u, a value that is closer to 12.0~u because
98.90\% of all carbon atoms are carbon-12. The average atomic masses
for the various elements are given in Table.~\ref{elements}.

Although the atomic mass unit provides a scale for comparing the
relative masses of atoms, it is not a useful unit when working in the
laboratory because it is too small (approximately $10^{-24}$~g).
Additionally, the atomic mass unit applies to a single atom, whereas
we work with gazillions of atoms at a time. To get around this problem
we introduce another unit that is better suited for samples containing
enormous numbers of atoms. Basically, the idea is to define a unit
that represents a particular number of objects, just as we use a dozen
to represent a collection of 12 eggs and a baker's dozen to represent
13 cookies.

To scale up from the microscopic level to the macroscopic world, we use
a unit called {\it mole} that has the unit symbol {\it mol}. The {\it
  mole} is defined as the amount of a substance containing the same
number of objects as there are atoms in exactly 12~g of carbon-12.
This number has been determined experimentally to be~\cite{Deslattes} 
\begin{equation}
N_A = 6.0220943 \times 10^{23}~{\rm mol}^{-1} \,,
\end{equation}
and is known as Avogadro's number. For our purposes, we usually
round Avogadro's number to $6.022 \times 10^{23}$~particles/mol. A mole of zinc
contains $6.022 \times 10^{23}$ atoms of zinc and a mole of jellybeans contains
$6.022  \times 10^{23}$ jellybeans.

The advantage of defining a mole in this way is that an element's average atomic mass is identical to its molar mass. Why is the true? A single atom of carbon-12 has an atomic mass of exactly 12~u and a mole of carbon-12 atoms has a molar mass of exactly 12~g. A single atom of carbon-13 has an atomic mass of 13.00335~u because its mass is 1.0836125 times greater than the mass of an atom of carbon-12. A mole of carbon-13, therefore, will have a mass of 13.00335~g.

A straightforward calculation
shows that the two mass scales,
grams and atomic mass units, are related by  
\begin{equation}
1~{\rm g} = 6.022 \times 10^{23}~{\rm u} \, .
\label{ukg}
\end{equation}
The proton and neutron each have a mass of
  approximately 1~u, and the electron has a mass that is only a small
  fraction of an atomic mass unit. More specifically, $m_p =
  1.00725~{\rm u}$, $m_n = 1.00864~{\rm u}$, and $m_e =  0.0005486~{\rm u}$.\\

\begin{figure}[tbp]
\postscript{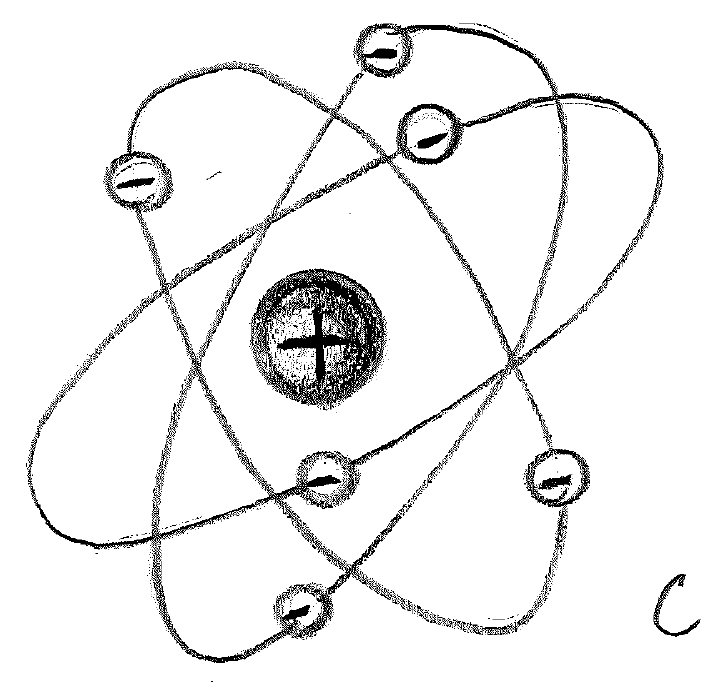}{0.7}
\postscript{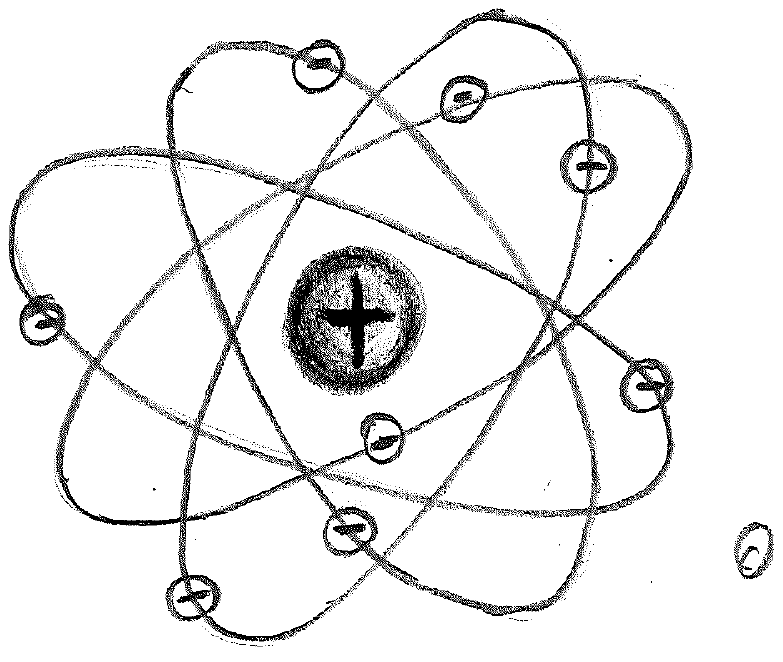}{0.7}
\caption{Simplified view of a carbon (up) and oxygen (down) atoms. The $^{12}$C nucleus
  consistes of 6 protons and 6 neutrons, while the $^{16}$O nucleus
  contains 8 protons and 8 neutrons.
  \label{fig:22}}
\end{figure}

\begin{figure}[t]
    \postscript{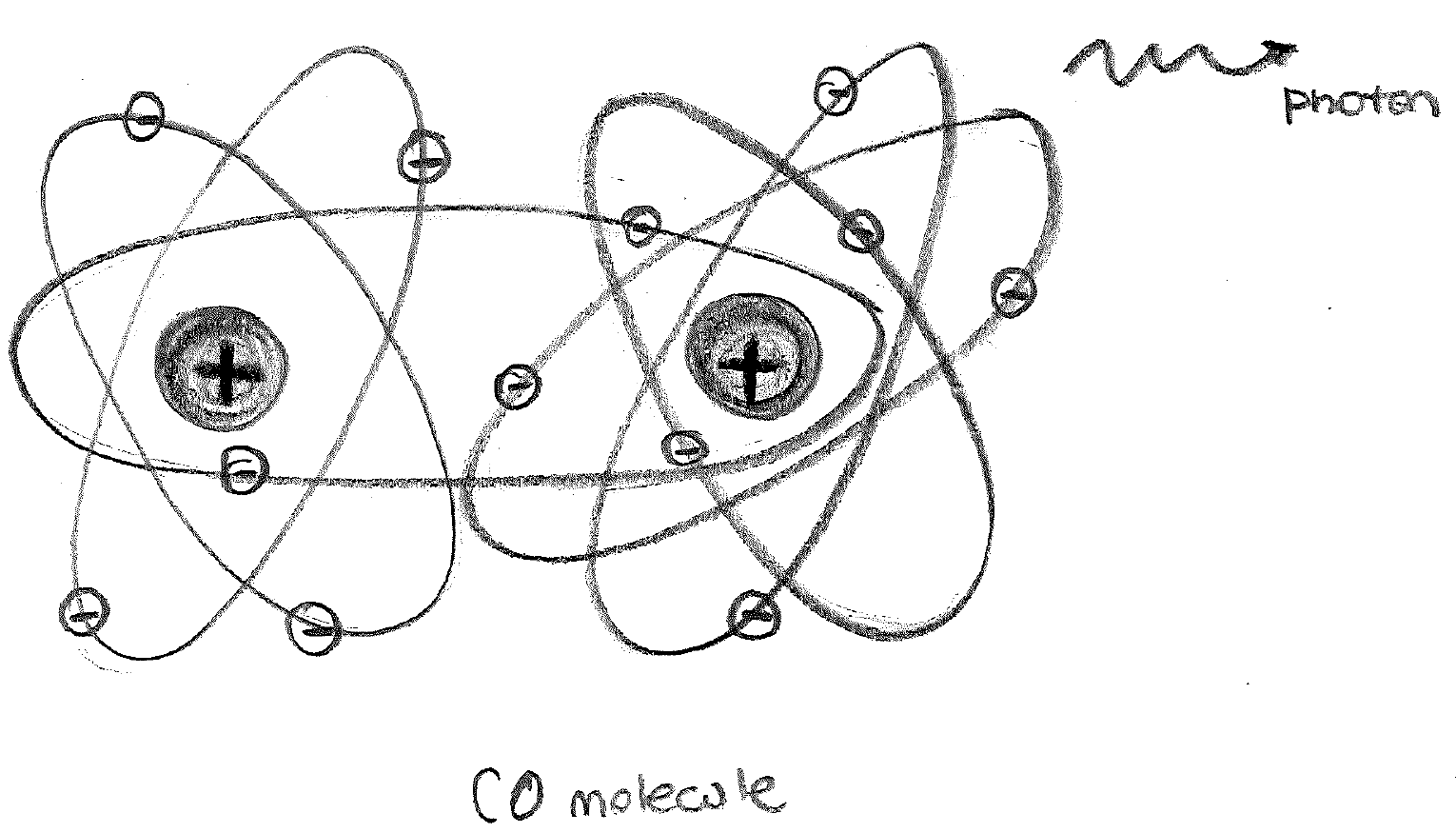}{0.9}
\caption{Structure of a carbon monoxide molecule.
 \label{fig:23}}
\end{figure}

{\bf \S~Photons.} Two atoms (such as C and O; see Fig.~\ref{fig:22})
combine to form a molecule (CO, or carbon monoxide; see
Fig.~\ref{fig:23}) as follows. In one way or another (which we will
discuss soon) the C and O atoms are driven up against one another. The
negative electrons repel each other, and the atoms fly appart. But
once in a while, a pair of atoms come together so hard that the
electrons are driven past each other, and a new situation comes being:
The negative electrons of the carbon get far enough past the negative
electrons of the oxygen, and begin to experience the {\it attractive
  force of the positive oxygen nucleus}. Same history for the oxygen
electrons. When this happens, the C and O combine into a stable CO
molecule.

But as they snap together, something very important happens: a small
packet of energy, called a photon, is emitted (sort of like when two
magnets snap together, a little heat is generated).\footnote{We have
  seen that on the macroscopic, everyday-world scale, light clearly
  exhibits the properties of a wave. It undergoes refraction at
  boundaries between different media, reflects at surfaces, and
  exhibits interference: all of these are wave phenomena. On the
  microscopic scale, however, the electromagnetic wave picture of the
  nature of light breaks down. Molecules, atoms, nuclei, and
  elementary particles do not ``see'' light as an electromagnetic
  wave; that is, they do not interact with light as they should
  interact with a simple set of time-varying electric and magnetic
  fields. Rather, they interact as if a beam of light were made up of
  discrete particles, each one carrying a specific amount of
  energy~\cite{Einstein:1905cc}.  These particles are called photons,
  and the discrete packet (a.k.a. quanta) of energy $= h \nu$, where
  $\nu$ is the photon frequency and $h= 6.626 \times 10^{-34}~{\rm m^2
    \, kg/s}$ is Planck's constant~\cite{Planck:1901tja}.} For each
molecule of CO formed, a photon is emitted and each of the photons has
the same energy. So every time 3 grams of C combines with 4 grams of
O, the same amount of energy (6 Cal) is released (in the form of a
million billion billion photons). The reason C and O always combine in
the proportions 3~g of C to 4~g of O is that 3~g of C contain the
same number of atoms as 4~g of O (convince yourself of this relation),
and the atoms just combine one-to-one.\\

{\bf \S~Burning.}~Now we come to a crucial consideration: the {\it
  chain reaction}. Let us suppose the following two considerations are
fulfilled: {\it (i)}~there are many C and O molecules together; {\it
  (ii)}~the C and O molecules are well interspersed. Then the photon
which originates during the combination of one pair of C and O has a
good chance of hitting another C (or O) and driving it with enough
force so that it will combine with an O (or C), and so on. Thus, the
combination will go on without outside energy, so long as there are
enough C and O atoms close together, and they are interspersed enough
[conditions {\it (i)} and {\it (ii)}]. This is what we call {\it
  burning}. Condition {\it (i)} can be translated to mean ``we need enough
fuel'' and condition {\it (ii)} ``give the fire some air!''

How do we {\it ignite} the reaction? We  must really agitate the C-O
mixture so that the C's really slam into the O's, overcoming the
repulsion of the outer electrons. Agitating a group of molecules to a
larger average velocity is the definition of raising the temperature
of the group. We can raise the temperature of the mixture by
introducing another source of photons (a light match), or by
physically agitating the C's (friction), e.g., as a rocket nose cone
burns when it rushes through the air.

A question which may have arisen at this point is, ``Where does the
photon come from?''  Unfortunately,   a complete revelation of these
secrets would require you to take a graduate physics course; at this
point we can simply provide the following explanation. The photon does
not exist in the atoms before it appears. But when it appears,
something else does disappear: {\it mass}. The mass of the CO molecule
is less than the sum of the masses of the C and O atoms. Einstein
discovered the relationship between the energy of the photon, $E$, and
the disappearing mass $m$,
\begin{equation}
E = mc^2 \,,
\label{energy-mass}
\end{equation}
with $c$ being the speed of light~\cite{Einstein:1905}. In the energy
units you have learned, the appearance of 2~Cal of energy in photons
is associated with the disappearance of $10^{-10}$~g of mass. Roughly
speaking,  $10^{-8}$-th of the mass of fuel burned
disappears.

By the way, if you calculate the masses of carbon-12 and carbon-13 by
adding up the masses of each isotope's electrons, neutrons, and
protons you will obtain a mass ratio of 1.08336, not 1.0836125. The
reason for this is that the masses given in Table~\ref{table1} are for
{\it free} electrons, protons, and neutrons; that is, for electrons,
protons, and neutrons that are not in an atom. When an atom forms,
some of the mass can be converted to energy according to
(\ref{energy-mass}), and the ``lost'' mass is the nuclear binding
energy that holds the nucleus together.\\

{\bf \S~Why the Sun is not burning chemical fuel.} The rate at which
solar energy reaches the Earth's upper atmosphere is 1.5
kilowatts/square meter ($=1.5~{\rm kW/m^2}$). Since $1 {\rm kW}
\approx 
1~{\rm Btu/second}$, this means an energy flux of 1.5~Btu/second per
$m^2$. We can use this number to obtain the total rate at which energy
is being radiated by the Sun.

\begin{figure}[t]
    \postscript{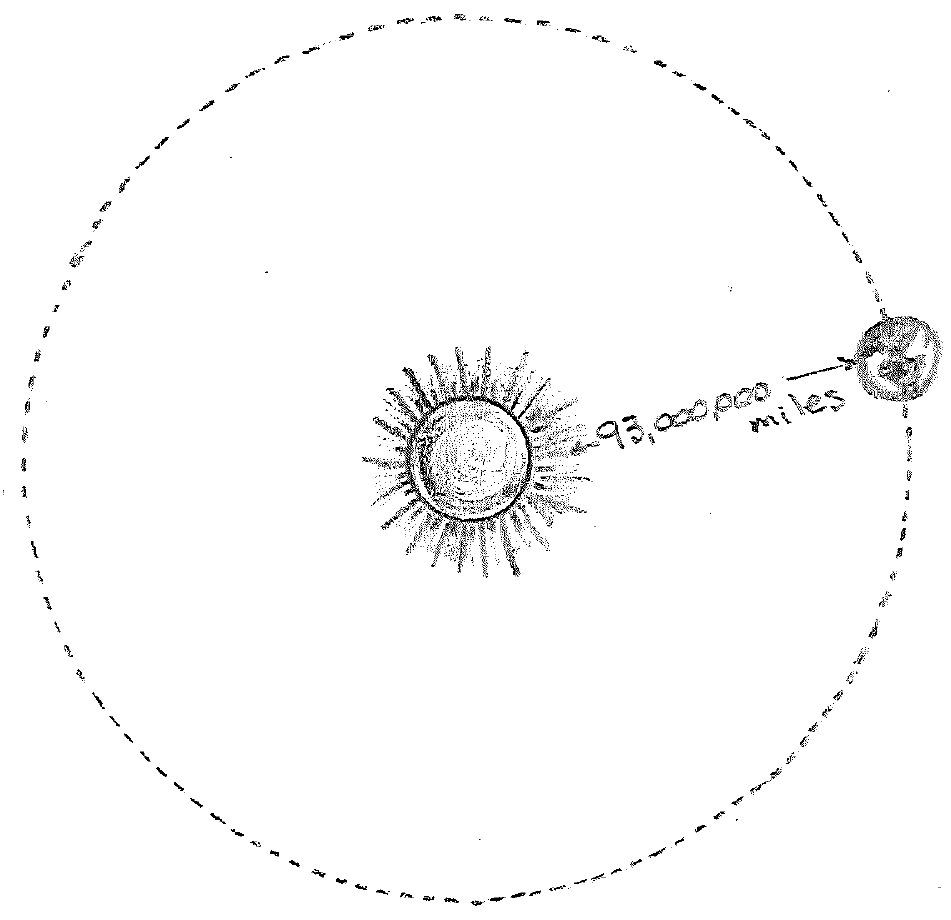}{0.9}
\caption{The intensity of the influence at any given radius $r$ is the source strength divided by the area of the sphere. Being strictly geometric in its origin, the inverse square law applies to diverse phenomena. 
  \label{fig:25}}
\end{figure}

Imagine a sphere drawn with the Sun at the center and with the Earth
at the surface; see Fig.~\ref{fig:25}. Since we know the rate at which radiation arrives at
1~m$^2$ of the sphere, we could obtain the total radiation emitted by
the Sun by multiplying 1.5~Btu/(m$^2$ s) by the area of
the sphere. So let's find that.

The distance from the Sun to the Earth is $93 \times 10^6$~miles. Now,
1 mile = 1.6~km = 1600~m, so
\begin{eqnarray}
{\rm the \ earth-sun \ distance} & = & 93 \times 10^6 \times 1600~{\rm m} 
\nonumber \\ & = & 150 \times 10^9~{\rm
  m} \, .
\end{eqnarray}
The area of the whole sphere is
\begin{equation}
A = 4 \pi r^2 = 4 \pi (150 \times 10^9)^2 \, .
\end{equation}
Now, $\pi \approx 3$, so $4 \pi \approx 12$, and $(150)^2 = 22500 =
2.25 \times 10^4$. Thus, the area of the sphere is
\begin{eqnarray}
A &  \approx & 12 \times 2.25 \times 10^4  \times 10^{18} \nonumber \\
  & \approx & 27 \times 10^{22}~{\rm square \ meters} \, .
\end{eqnarray}
The total radiation arriving at the sphere is
\begin{eqnarray}
L_\odot & \approx & 1.5~{\rm Btu/s/m}^2 \times 27 \times 10^{22}~{\rm m}^2 \nonumber
\\
& \approx & 40 \times 10^{22}~{\rm Btu/s} \nonumber \\
& \approx &  4 \times 10^{23}~{\rm Btu/s}
\, .
\label{Lodot}
\end{eqnarray}
This must be the rate at which energy is being radiated by the Sun. We
know that the Sun is not heating up, so $4 \times 10^{23}$~Btu/s must
also be the rate at which energy is being created at the Sun.

Now,  suppose the Sun were burning oil, or some other chemical with
energy content of the order of 20,000~Btu/lb. If the entire mass of
the Sun (which is $M_\odot = 4 \times 10^{30}~{\rm lb}$) were composed of such fuel,
and it were all to burn, then the {\it total} energy release would be
$4 \times 10^{30}~{\rm lb} \times 2 \times 10^4~{\rm Btu/lb} = 8 \times
10^{34}~{\rm Btu}$. Since the rate of energy release is $4 \times
10^{23}~{\rm Btu/s}$, we use the general formula
\begin{equation}
{\rm amount} = {\rm rate} \times {\rm time} 
\end{equation}
to obtain
\begin{equation}
8 \times 10^{34}~{\rm Btu} = 4 \times 10^{23}~{\rm Btu/s} \times {\rm
  time} \,,
\end{equation} 
yielding
\begin{equation}
{\rm time} = \frac{8 \times 10^{34}}{4 \times 10^{23}}~{\rm s} = 2 \times
  10^{11}~{\rm s}
\label{time-globito}
\end{equation}
as the time available for burning. Since a year has $3 \times 10^7$
seconds, the time in (\ref{time-globito}) is equivalent to
\begin{equation}
\frac{2 \times 10^{11}}{3 \times 10^7}~{\rm yr} = \frac{2}{3} \times
10^4~{\rm yr} =
\frac{2}{3} \times 10,000~{\rm yr} = 6,666~{\rm yr} \, .
\end{equation}
Way too short. To last $6 \times 10^9~{\rm yr}$ (or so), the fuel must
have an energy content of $10^6$ time greater than that of oil. But that is another story and shall be told another time.

\section{Radioactivity}
\label{sec8}

{\bf \S~Nuclear structure.}~The size and structure of nuclei were
first investigated in Rutherford's scattering
experiments~\cite{Rutherford:1911zz,Geiger}. In these experiments,
positively charged nuclei of helium atoms ($\alpha$ particles) were
directed towards a thin piece of metal foil. As the particles moved
through the foil, they often passed near a metal nucleus. Because of
the positive charge on both the incident particles and the nuclei,
particles were deflected from their straight-line paths by the Coulomb
repulsive force. In fact, some particles were even deflected backward,
through an angle of $180^\circ$ from the incident direction.  Those
particles were apparently moving directly toward a nucleus in a
head-on collision course.

\begin{figure}[t]
    \postscript{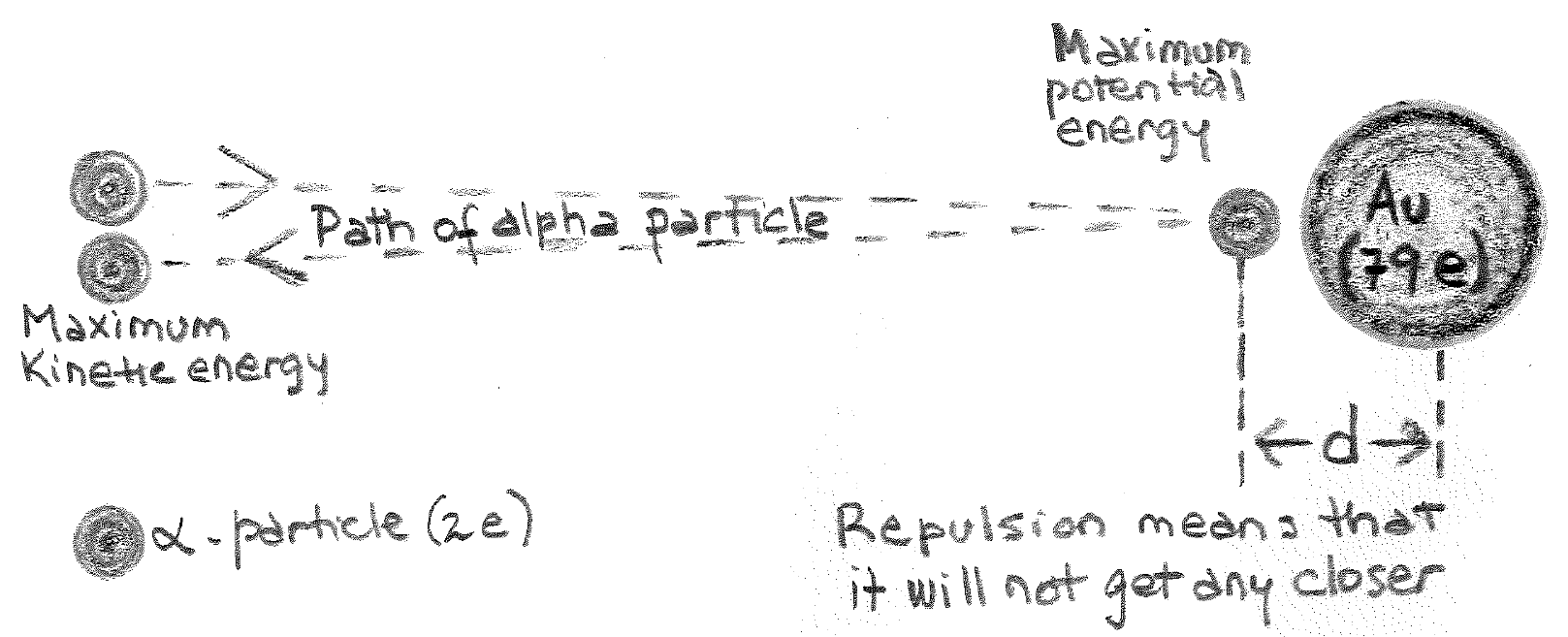}{0.9}
\caption{Rutherford's scattering experiment.
  \label{figu:31}}
\end{figure}

Using energy conservation it is straightforward to derive an expression for
the distance $d$ at which a particle approaching a nucleus is turned
around by Coulomb repulsion. Namely,  in a head-on collision, the kinetic
energy of the incoming $\alpha$ particle must be converted completely
to electrical potential energy when the particle stops at the point of
closest approach and turns around, see Fig.~\ref{figu:31}. If we equate the
initial kinetic energy of the $\alpha$ particle to the electrical
potential energy of the system ($\alpha$ particle plus target nucleus), we
have
\begin{equation}
\frac{1}{2} m_\alpha v_\alpha^2 = k_e \frac{q_1 q_2}{r} = k_e
\frac{(2e) (Ze)}{d} \, .
\end{equation}
Solving for $d$, the distance of closest approach, we get
\begin{equation}
d = \frac{4 k_e Z e^2}{m_\alpha v_\alpha^2} \, .
\label{Rutherforo}
\end{equation}
From (\ref{Rutherforo}) it follows  that $\alpha$ particles with
kinetic energy 
$K_\alpha = 1.12 \times 10^{-12}~{\rm J}$ would approach nuclei to
within a distance of 
$3.2 \times 10^{-14}~{\rm m}$ when the foil is made of gold.  Thus, the
radius of the gold nucleus must be less than $3.2 \times 10^{-14}~{\rm m}$. For silver
atoms, the distance of closest approach is found to be $2 \times
10^{-14}~{\rm m}$. From these results, Rutherford concluded that the
positive charge in an atom is concentrated in a small sphere with a
radius of approximately $10^{-14}~{\rm m}$, which he called the
nucleus. Because such small dimensions are common in particle physics,
a convenient length unit is the femtometer (fm), almost always called the
fermi, defined as $1~{\rm fm} = 10^{-15}~{\rm m}$.

Modern particle colliders are essentially a repeat of the Rutherford
scattering type experiment, but at a much higher energy.  We now know
from many such experiments that the nucleons are all made up of 
elementary particles called quarks, which glue together via the strong interaction. The CERN Large Hadron Collider has directly probed distance
scales well inside the proton, as short as $2 \times 10^{-5}~{\rm fm}$.

Some nuclei are unstable. An unstable nucleus tries to achieve a
balanced state by given off  neutrons or protons and this is done via
radioactive decay. It is this that we now turn to study.\\

{\bf \S~Radioactive decay.}  When
$^{238}$U (we many times omit the ``92'') undergoes radioactive decay,
it emits an $\alpha$-particle which consists of 2 protons and 2
neutrons~\cite{Becquerel}.  This leaves an atom with 90 protons (92
minus 2) and 234 nucleons (238 minus 4)., which is called
thorium-234. $^{234}$Th then decays again and we have a chain of
decays ending up with lead ($^{206}_{82}$Pb).

Beta decay takes place by the emission of an electron (or $\beta$-ray)
and a neutrino, from the nucleus at the same time that one of the
neutrons changes to a proton.\footnote{Neutrinos are very weird teeny,
  tiny, nearly massless particles that do not carry electric charge.}
So when a carbon-14 (6 protons, 8 neutrons) $\beta$-decays it changes
to nitrogen-14 (7 protons, 7 neutrons -- ordinary nitrogen!).

\begin{figure}[t]
    \postscript{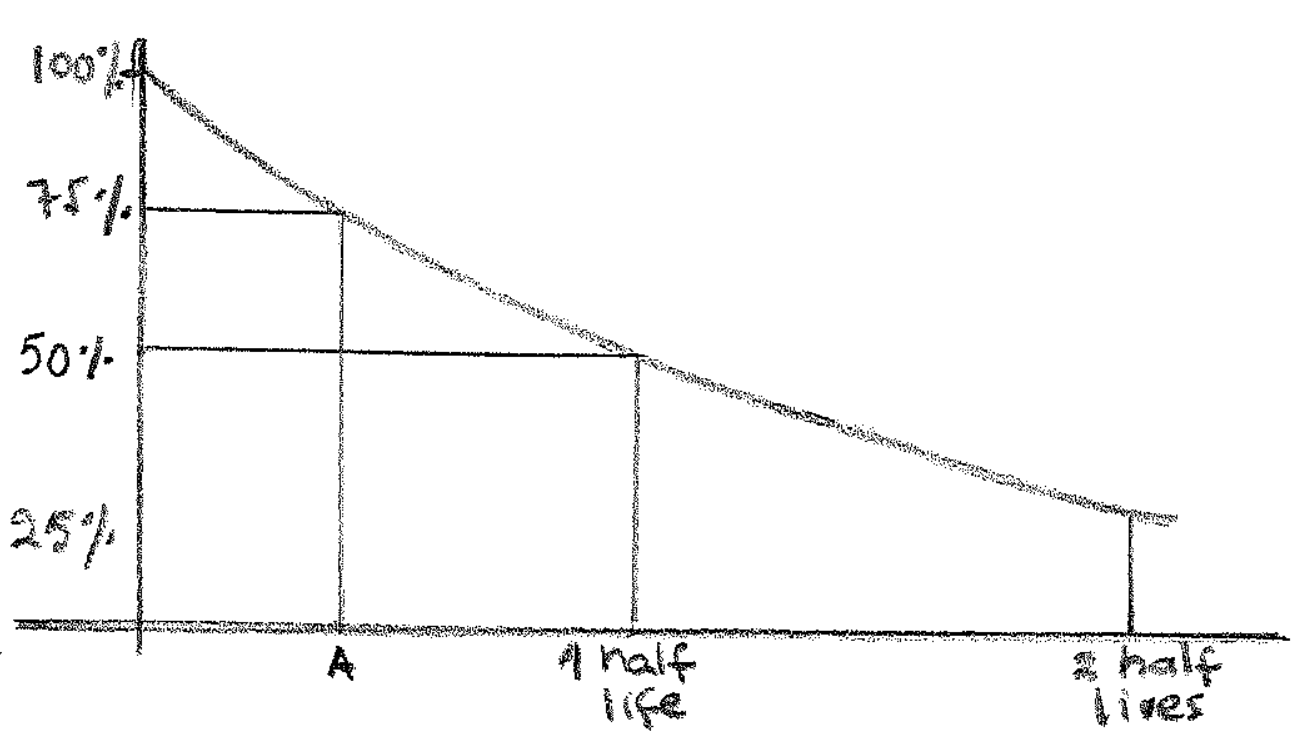}{0.9}
\caption{A plot of the exponential decay law for radioactive nuclei. The
vertical axis represents the percentage of radioactive nuclei at any
time and the horizontal axis is the time. After one half-life, half of
the radioactive atoms have decayed. After another half-life, half of
the remaining atoms have changed. This pattern continues, with half of
the atoms changing over each half-life time.
  \label{fig:24}}
\end{figure}

There is a law governing the way a substance undergoes radioactive
decay: a constant {\it fraction} of the atoms in the sample
disintegrate per second~\cite{Rutherford:1902}. The time it takes for half
of the sample to decay is called the {\it half-life}. For example, if
100 atoms are initially present and the half life is 10 years, then
after ten years, 50 atoms will remain undecayed, after another 10
years, 25 will be left, after 30 years, 12 will be left, etc.
Since the radioactivity level depends directly on the number of
undecayed atoms remaining, we can also say that after each half-life time
the radioactivity is cut in half.  

Knowing the half-life, we know how to draw the graph shown in
Fig.~\ref{fig:24}, which will allow us to find the radioactivity level
(or the amount of substance left) after any amount of time. So, for
example, to find out how much long substance must decay in order that
its radioactivity level drop by 25\%, we look at the graph for the
time which corresponds to 75\% activity. This is point $A$ on the time
axis, and is about 0.4 of a
half-life. \\

{\bf \S~Radioactivity carbon dating.}~Because of cosmic ray
bombardment, a tiny fraction (1 millionth of 1\%) of the carbon in the
atmosphere (in CO$_2$) is carbon-14. The CO$_2$ is breath by
plants. Hence all plants have a tiny bit of radioactive $^{14}$C in
them.

Through radioactive decay, the $^{14}$C in living things changes to
stable nitrogen, but because living plants breathe, the decayed
$^{14}$C is replenished, and there is a constant ratio of $^{14}$C to
$^{12}$C (the ordinary carbon). The equilibrium is such that there is a
radioactive level of 15 disintegrations per second for every gram of
the carbon mixture. When a plant dies, however, the replenishment
stops.

The percentage of $^{14}$C steadily decreases with a half-life of
$5,730$~years. Since we know the radioactivity of plants today, we are
able to determine the ages of ancient objects by measuring their
radioactivity. If we extract a small but precise quantity of carbon
from an ancient papyrus scroll, for example, and find it has 1/2 as much
radioactivity as the same amount of carbon extracted from a living
tree, then the papyrus must be $5,730$ years old. If it is 75\% as
radioactive, then 0.4 of one half-life, or $2,292$ years have elapsed
since the papyrus was alive.\\

{\bf \S~Age of the elements.} {\it (i)}~$^{238}$U: The half-life of
$^{238}$U is $4.5 \times 10^9$~years. Since this isotope is no
markedly less abundant  than the other heavy elements (bismuth,
mercury, gold, etc.), we conclude that these elements were formed not
much longer than $4.5 \times 10^9$~years ago (like maybe $5 \times 10^9$
or $6 \times 10^9$). {\it (ii)}~$^{235}$U: On the other hand,
$^{235}$U is only $\frac{1}{140}$ as abundant as $^{238}$U, and has a
half-life of $0.9 \times 10^9$~years. If $^{238}$U and $^{235}$U were
formed in roughly equal amounts, it must have taken about 7 half-lives
to get them to the present ratio (since $(\frac{1}{2})^7 =
\frac{1}{128}$ close to $\frac{1}{140}$). So we estimate that both
these elements were formed $7 \times 0.9 \times 10^9 = 6.3 \times
10^9$ years ago (6.3 billion).\\

{\bf \S~Age of the Earth.} {\it (i)}~{\it Age of the rocks.} When
$^{238}$U undergoes radioactive disintegration the final products of
the sequence of decays are an isotope of lead ($^{206}$Pb), 8 helium
atoms, and various electrons and neutrinos. When the $^{238}$U  became
encased in rock, the lead (and helium, to some extent) was locked into
close proximity to the $^{238}$U. As time passes the ratio
$^{206}$Pb/$^{238}$U increases. By knowing this ratio, and the
half-life of $^{238}$U ($4.5 \times 10^9$~years), one can estimate the
time which has passed since the $^{238}$U was encased in rock. The
same procedure can be used with other ``mother-daughter'' pairs
($^{232}$Th $\to$ $^{208}$Pb and $^{235}$U $\to$ $^{207}$Pb). Using
this method, the oldest rocks found on Earth have been dated 4 billion
years. More sophisticated methods eventually date the formation of the
earth's crust at $4.5 \times 10^{9}$~years. {\it (ii)}~{\it Age of the
  oceans.} The oceans have become salty as a result of minerals
being washed into them by rivers flowing to the sea. The evaporation
of water from the ocean (leaving behind the brine) and its subsequent
return to the rivers as fresh-water rain leads to an increase in the
salinity from year to year. From the knowledge of the total volume of
the oceans, the rate of fresh water flow to the sea, and the mineral
content of the river water, one may estimate that the salinity of the
oceans has been increasing at the rate of one billionth of a percent
per year ($10^{-9}$\% per year). Thus, to reach a concentration of
3\%, it must have taken about 3 billion years.\\

\section{Birth and Death of the Sun}

\begin{figure}[t]
    \postscript{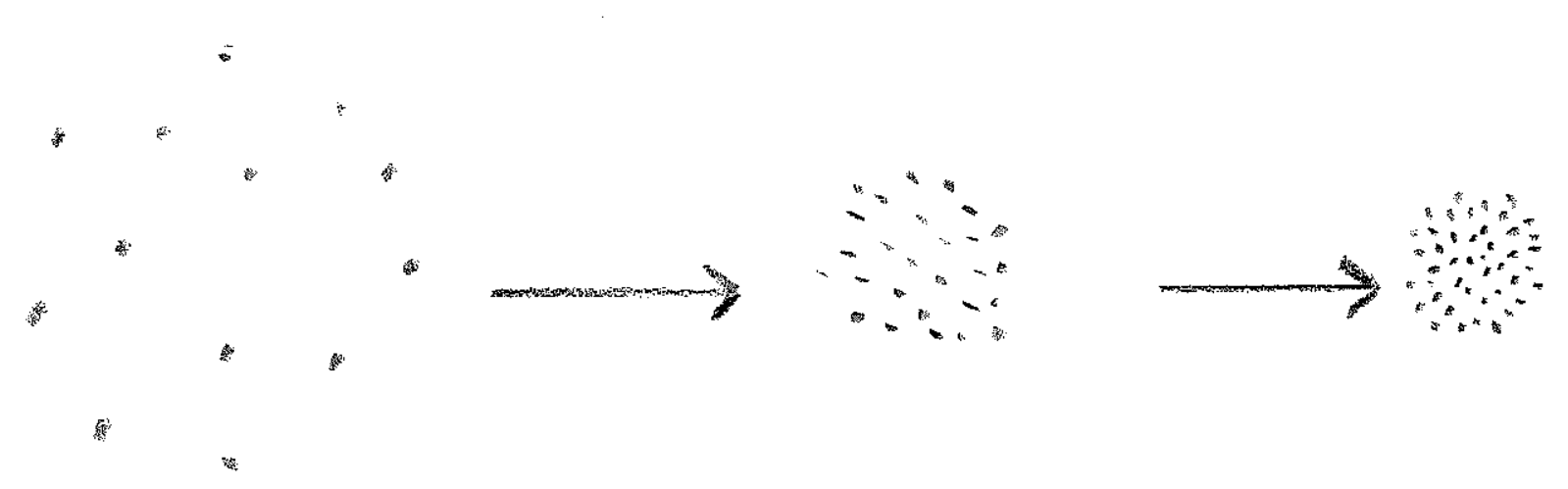}{0.9}
\caption{Contraction of a gas cloud. As the collapse due to
  gravitational attraction proceeds it speeds
  up. The gas cloud heats up, in the center most of all. Nuclear
  fusion starts. The contraction stops and a balance is established
  between pressure in the gas cloud and gravity.
 \label{fig:26}}
\end{figure}

{\bf \S~In the beginning $\bm{\cdots}$ !} there is a huge cloud of hydrogen
and helium, about a trillion miles in diameter (this is a million
times the sun's diameter). If the cloud has a mass greater than about
$10^{30}~{\rm lb}$, the attractive forces due to gravitation will be
sufficient to overcome the dispersive effects of the random motion of
the atoms in the cloud, and the cloud will begin to contract; see Fig.~\ref{fig:26}.

As the cloud contracts, one can think of all the atoms falling in
toward the center, and just like any other kind of falling, they pick
up speed. Their speed gets randomized through collisions, so that the
net effect is a large increase in the temperature (like the rise in
temperature of a falling brick when it strikes the ground).\\

{\bf \S~Formation of the plasma.} After about 10 million ($10^7$) years
of contraction, the hydrogen cloud shrunk to a diameter of 100 million
miles ($100 D_\odot$, or 100 times the diameter of the sun). This is
like shrinking the entire Lehman campus down to an inch. The
temperature of the cloud has gone from about $100^\circ$K
(representing an average speed of 1 mile per second for the hydrogen
atoms) to about $50,000^\circ$K (representing an average speed of 20
miles per second). The density of the cloud is about $\frac{1}{1000}$
of the density of air. At this speed and density the collisions are
frequent enough (each atom collides about a billion times/second) and
violent enough to ionize all the atoms in the cloud (that is, remove
the electrons from the atoms). So at this stage, the star-to-be
(called a protostar) consists of gas of positively charged protons and
helium nuclei, and negatively charged electrons in equal balance. This
kind of hot gas is called a {\it plasma}.\\

{\bf \S~Further contraction.} The plasma continues to contract under
the influence of the gravitational force, getting hotter and
denser. The temperature rises because the surface area of the star is
not large enough {\it in relation to its volume} to get rid of the
heat as it is produced. After another 10 million years the plasma has
shrunk by another factor of 100 to the size of the sun ($D_\odot = 1$
million miles), the density near the center has risen to 10 million
$^\circ$K (corresponding to an average speed of 280 miles per second
for the protons). At this stage the plasma is contracting so quickly
that is about one hour from a total collapse into a point. But in the
nick of time, we get $\cdots$\\

\begin{figure}[t]
    \postscript{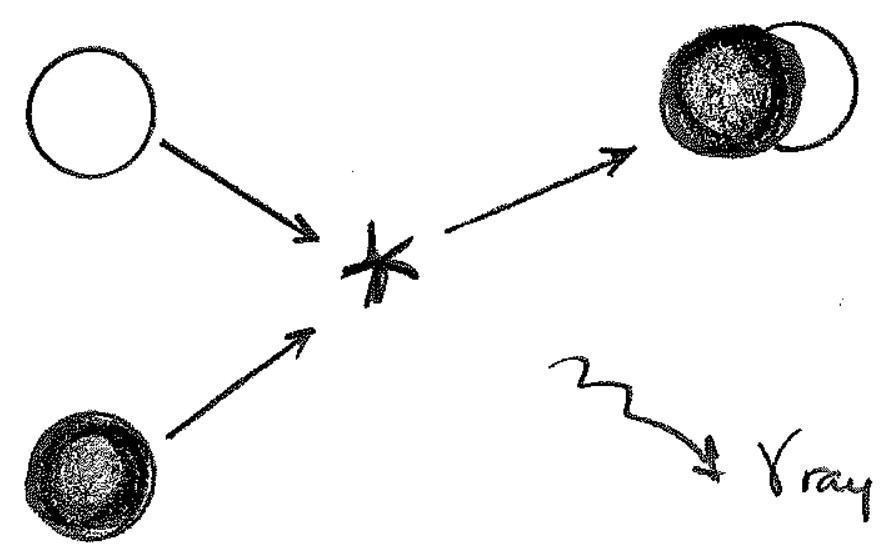}{0.6}
\caption{$p + n \to D + \gamma$.
\label{fig:27}}
\end{figure}

{\bf \S~Nuclear ignition.} We learned in Sec.~\ref{sec8} that nuclei
stay together in spite of the strong repulsive electric forces between
the protons. This is because there is a much stronger attractive
nuclear force which acts like a glue between protons and protons (or
$p$'s and $n$'s, or $n$'s and $p$'s): namely, when they get close
enough to touch, the attraction due to the nuclear force overwhelms
any electrical repulsion which may exist (like between $p$ and
$p$)~\cite{Gamow:1928zztop}. It is also a property of elementary
particles (like $p$'s and $n$'s) that they never sit still -- they are
always jiggling and moving about. This tends to greatly weaken the
binding due to the glue. In the case of neutrons touching protons, the
glue holds, the binding is stable, and we get a deuteron; see
Fig.~\ref{fig:27}. If the Sun were to contain large numbers of
neutrons, it would burn up immediately into deuterium.

For the case of protons touching other protons, the glue is slightly
weakened by the presence of the electrical repulsion, and the net
attractive force is just insufficient to overcome the jiggling and bind
the $pp$ to form $^2$He. So in our star, the protons will collide,
hang around each other a little, but no binding, no $\gamma$-ray, no
``burning.''\\

{\bf \S~However, enters the weak force!} There is another short range
interaction between the nucleons which is extremely weak, and is generally completely
masked by the strong force. This is called the weak
interaction~\cite{Fermi:1934hr}. If two protons did bind, you would
never know about the weak force. But they don't, so instead, once in
every 10 billion collisions, the protons stick together to form a
deuteron. But a deuteron has a charge $+1$, and the two protons have
charge $+2$, so we need something to carry off the extra positive
charge: the carrier is a positron, a positive electron $e^+$. In
addition to the positron,  a neutrino,
carries off a little of the energy and leaves the Sun.

\begin{figure}[t]
    \postscript{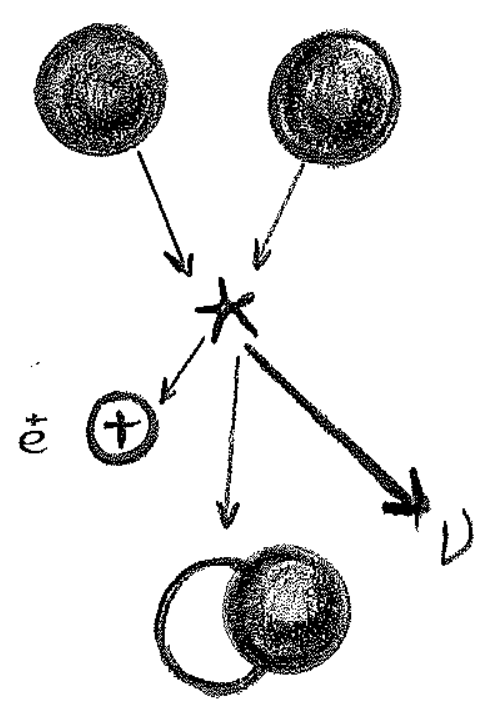}{0.6}
\caption{$p + p \to D + e^+ + \nu$.
\label{fig:28}}
\end{figure}

\begin{figure}[t]
    \postscript{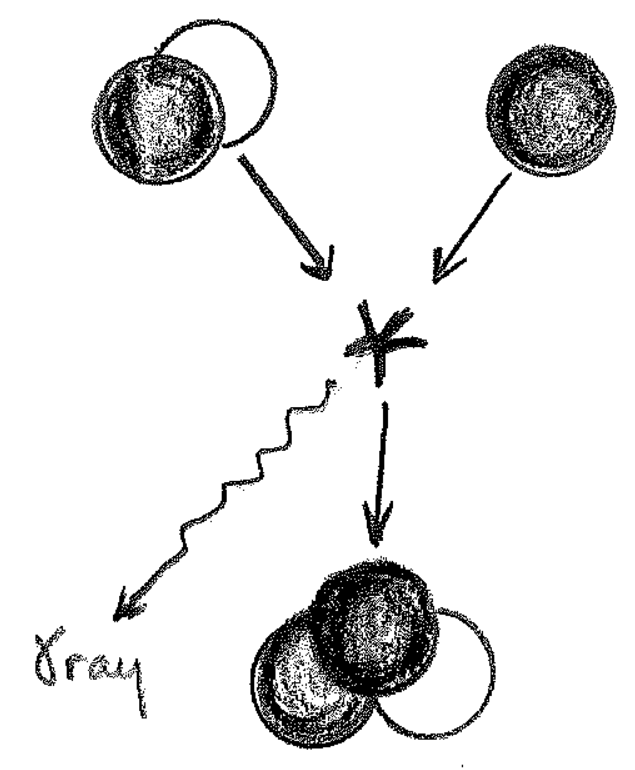}{0.7}
\caption{$D + p \to {^3{\rm He}} + \gamma$.
\label{fig:29}}
\end{figure}

The reaction shown in Fig.~\ref{fig:28} is very unlikely for two
reasons: {\it (i)}~At 10 million $^\circ$K, the protons get close
enough to touch only very rarely: the electric repulsion is not fully
overcome at this temperature. {\it (ii)}~Even when they do touch the
fusion reaction is so weak that it only occurs in a million
times. Altogether, to see how unlikely the whole thing is, if you had
a lb of hydrogen at the center of the sun, it would take 10
billion years for half of it to burn into deuterium.\\

\begin{figure}[t]
    \postscript{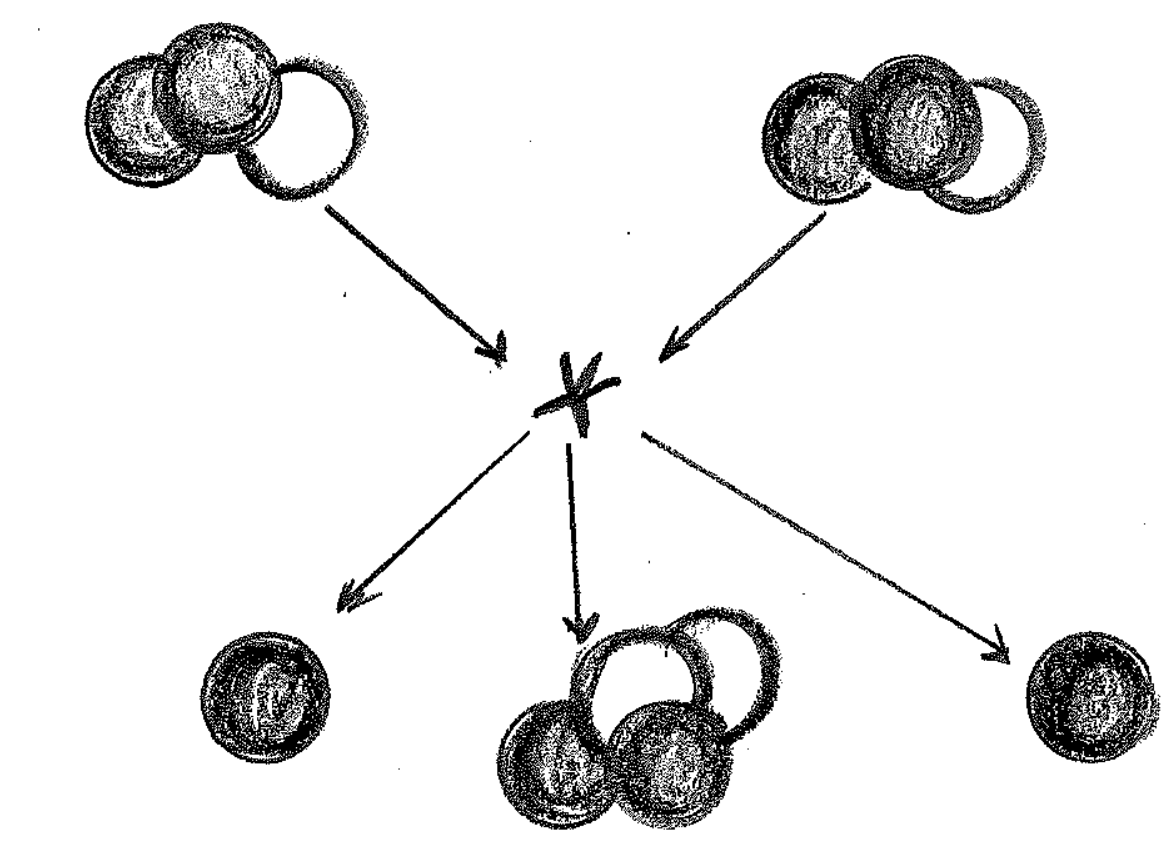}{0.9}
\caption{${^3{\rm He}} + {^3{\rm He}} \to {^4{\rm He}} + 2 p$.
\label{fig:30}}
\end{figure}

{\bf \S~Cooking up the helium isotopes.} We left off with the process
$p^+ + p^+ \to D^+ + e^+ + \nu$. This is not the end of this sequence,
however. The deuterium, in its many collisions with protons, undergoes
rapid conversion to $^3$He (see Fig.~\ref{fig:29}), and the $^3$He's
collide to form He$^4$ and two protons. The last reaction, shown in
Fig.~\ref{fig:30}, proceeds slowly because the two $^3$He, each having
charge $+2$, have a hard time getting close to one another, in order
to undergo their nuclear reaction.

The net effect of this sequence, which is called the $pp$-cycle, is
for four protons to combine to form one $^4_2$He nucleus, plus two
positrons, two neutrinos, and two gamma rays.\footnote{The theory of
  the $pp$ cycle as the source of energy for the Sun was first worked
  out by Bethe~\cite{Bethe:1939bt}. Interestingly, the same set of
  nuclear reactions that supply the energy of the Sun's radiation also
  produce neutrinos that can be searched for in the
  laboratory~\cite{Bahcall:1964gx,Davis:1964hf}.} Now, nothing new
happens for a long time except for the accumulation of helium at the
center of the
star.\\

\begin{figure}[t]
    \postscript{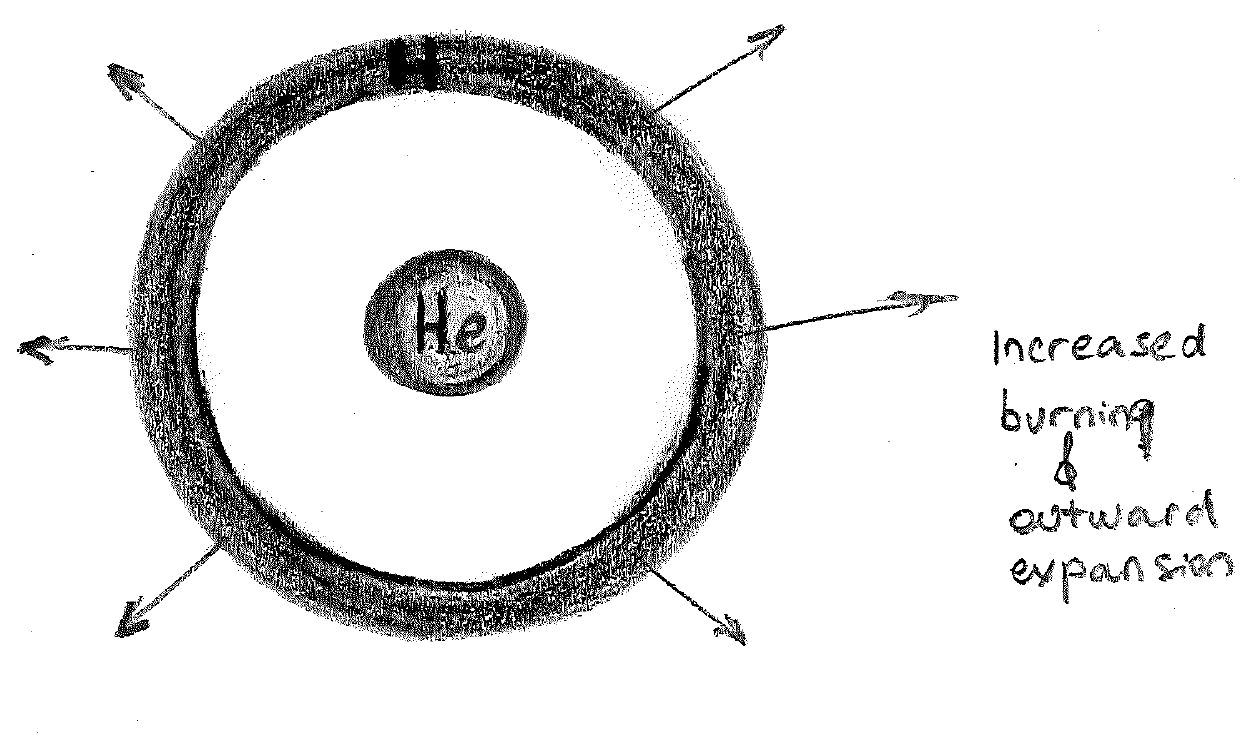}{0.9}
\caption{Outward expansion into the red giant phase.
 \label{fig:31}}
\end{figure}

\begin{figure*}[t]
    \postscript{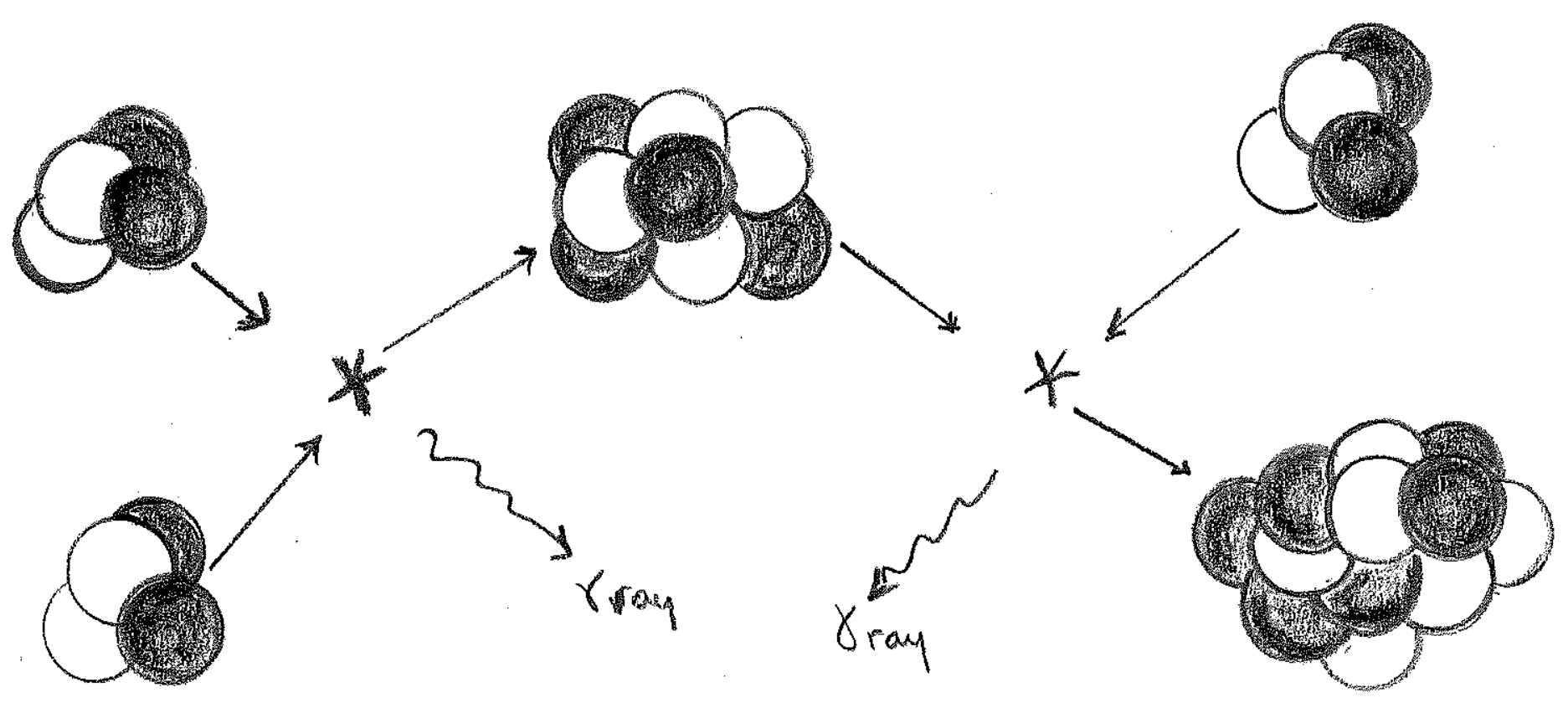}{0.9}
\caption{${^4{\rm He}} +  {^4{\rm He}} \to  {^8{\rm Be}} + \gamma$ and
  $ {^4{\rm He}} +  {^8{\rm Be}} \to  {^{12}{\rm C}} + \gamma$.
\label{fig:32}}
\end{figure*}

{\bf \S~Red giant phase.} As the helium accumulates, it starts
undergoing gravitational contraction, heating up in the process (just
like the hydrogen did before it). This heating and contracting
proceeds at a rather rapid rate (over a period of tens of millions of
years) leading to a great increase in temperature of the hydrogen
surrounding the helium core, and a consequent acceleration of its
burning. It also expands outward as shown in Fig.~\ref{fig:31}.

The end result is that the star becomes huge, the core becomes small
and dense ($20,000$ miles across, 1 ton per in$^3$). The large size of
the star allows the surface to remain ``cool'' (i.e., only red hot) so
that the star at this stage is visible as a ``red giant.'' An example
is Betelgeuse (pronounced ``beetle juice'') which is visible as
Orion's left shoulder in the night spring sky, looking south.\\

{\bf \S~Burning of the helium.}~When the temperature of the core
reaches 100 million $^\circ$K, the helium begins to burn, undergoing
the reaction 
\begin{equation}
^4{\rm He}~ + ~^4{\rm He} \to ~^8{\rm Be}~~~(\textnormal{beryllium-8}) \, .
\end{equation}
Now, there is another of those ``just so'' accidents. $^8$Be is not
stable; although it can exist for a short time, the nucleons out of
which it is made do not attract each other strongly enough, and in
less than $10^{-12}~{\rm s}$ it breaks up into separate helium nuclei again:
\begin{equation}
^8{\rm Be}~ \to ^4{\rm He}~ + ^4{\rm He} \, .
\end{equation}
However, at the temperature and density of the helium core, each
helium nucleus undergoes about $10^{12}$ collisions per second. So it
is not unlikely that $^8$Be will be hit by another helium-4 before it
breaks apart; see Fig.~\ref{fig:32}. This leads to the formation of carbon, and a
$\gamma$-ray (which is energy).\\

{\bf \S~On to the white dwarf.} So now a carbon core begins to
form. The heat generated by these interactions and the subsequent 
collapse of the carbon core lead to a blow-off of most of the outer
gas layers. The star has become a hot carbon core, about $20,000$
miles across, with a density of 10 tons per in$^{3}$. The electron gas
exerts enough outward pressure to keep the carbon core from
contracting much in diameter, if the star is less than one and a half
times as massive as the Sun. But the inward pressures still generate a
great deal of heat, and the carbon star glows with a white heat. This
is  a white dwarf. After a few million years, the dwarf cools some,
becomes yellow, then red, then cools completely and the fire goes
out. The star has met its death as a massive, dense lump of coal.\\

{\bf \S~Heavy stars.} An entirely different end awaits the 5\% of
stars whose masses exceed 1.5 times the mass of the
sun~\cite{Chandrasekhar:1935zz}. For such masses, the inward
gravitational pressure of the carbon core can generate a temperature
sufficiently large (600 million $^\circ$K) to ignite the carbon. The
carbon burns to form magnesium and other elements.

The pressure and temperature mount, nuclear reactions continue, until
finally the {\it iron} is reached. At this point, the process stops,
because iron is a very special element. Any reaction that takes place
involving an iron nucleus will {\it use up} energy. The iron, instead
of  providing more fuel to burn, puts the fire out. The center of the
star commences to collapse again, but this time, because of the
presence of the iron nuclei, the fire cannot be rekindled; it has gone
out for the last time, and the entire star commences its final collapse.  

The collapse is a catastrophic event. The materials of the collapsing
star pile up at the center, creating exceedingly high temperatures and
pressures. Finally, when all the nuclei are pressed against each other
, the star can be compressed no further. The collapsed star,
compressed like a giant spring, rebounds instantly in a great
explosion. About half the material within the star disperses into
space, the other half remains as a tiny core (about 10 miles in
diameter), containing a mass equal to that of the Sun. This is now a
{\it pulsar} or {\it neutron star} (for reason to be explained
soon). The entire episode (collapse and explosion) lasts a few
minutes.

The exploding star is called a supernova. The most famous supernova
was recorded by Chinese astronomers in A.D. 1054, and its remnants are
now visible as the Crab Nebula.

At the very high temperatures generated in the collapse and explosion,
some of the nuclei in the star are broken up, and many neutrons and
protons freed. These are captured by other nuclei, building up heavier
elements, such as silver, gold, and uranium. In this way the remaining
elements of the periodic table, extending beyond iron, are {\it
  perhaps} 
manufactured in the final moments of the star's life~\cite{Burbidge:1957vc}. Because the time
available for making these elements is so brief, they never become as
abundant as the elements up to including iron.

The core of the collapsed star then contracts until all the nuclei
are touching, and then stops. The forces are so great that all the
nuclei (iron, etc.) disintegrate into their constituents (neutrons and
protons); the protons combine with electrons to leave a dense core of
neutrons -- a nucleus about as large as
Boston~\cite{Oppenheimer:1939ne}. This object, a neutron star, rotates
madly about its axis, emitting energy at a billion times the rate at
which the sun does so. We see it in the sky as a pulsar.

If the mass of the neutron star is greater than about three solar
masses, then the star further contracts under gravity and eventually
collapses to the point of zero volume and infinite
density~\cite{Oppenheimer:1939ue}.  As the density increases, the
paths of light rays emitted from the star are bent and eventually
wrapped irrevocably around the star. The ``star'' with infinite
density is completely enclosed by a boundary known as the event
horizon, inside which the gravitational force of the star is so strong
that light cannot escape~\cite{Finkelstein:1958zz}. This is called a
{\it black hole}, because no light escapes the event horizon.

\section{Nuclear Processes}

\begin{figure}[t]
    \postscript{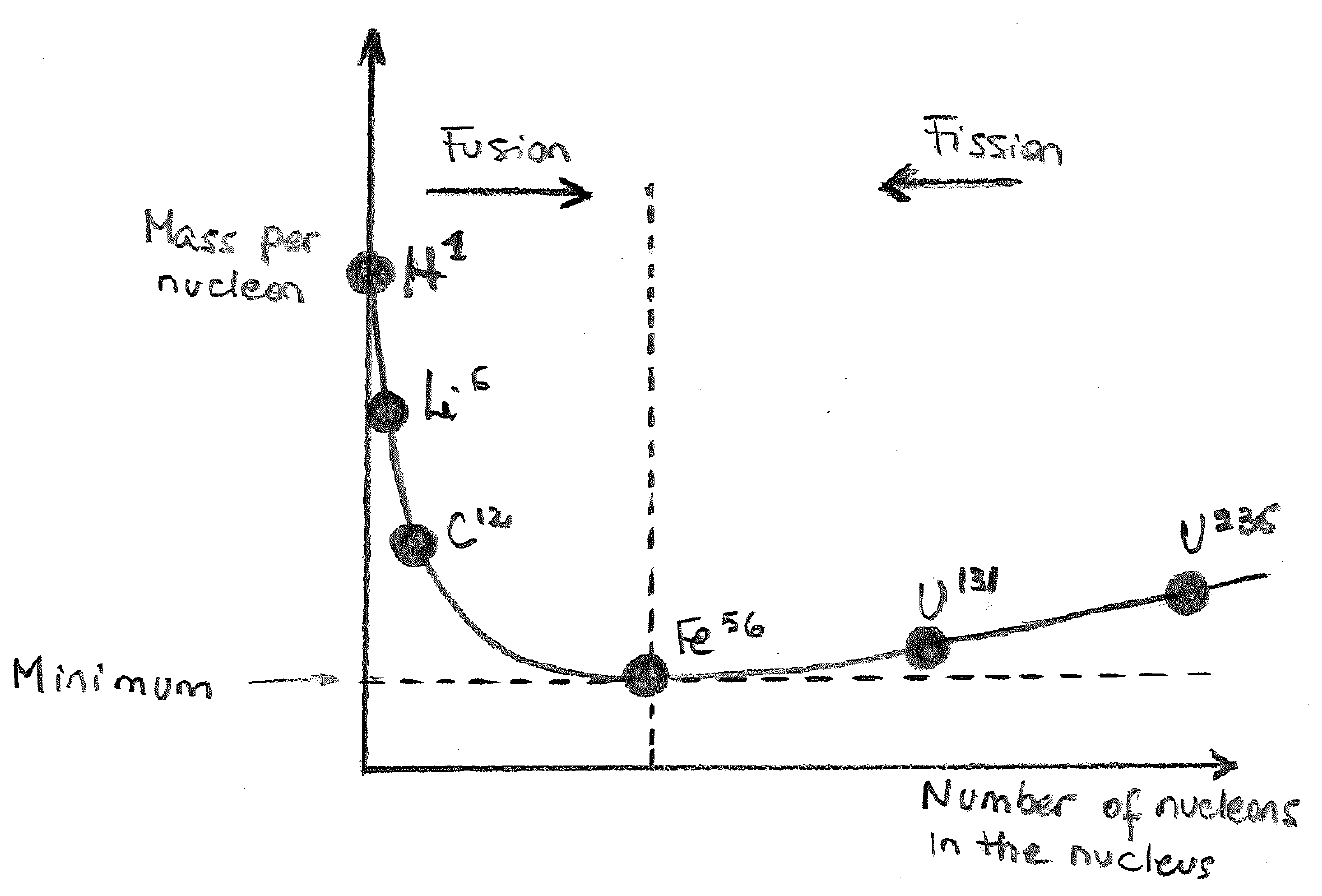}{0.9}
\caption{Schematic diagram of mass per nucleon as a function of the number of
nucleons in the nucleus. Note that we are only illustrating the
general trends. There are for instance a few light elements for which
the mass per nucleon increases with increasing number of nucleons in
the nucleus.
\label{fig:33}}
\end{figure}

{\bf \S~Nuclear masses and binding energies.}~Two distinct processes
involving the nuclei of atoms can be harnessed, in principle, for
energy production: fission (the splitting of a nucleus) and fusion
(the joining together of two nuclei). In Fig.~\ref{fig:33} we show the
mass per nucleon for the different elements. We see that we have a
minimum for $^{56}$Fe. This means that $^{56}$Fe is {\it relatively}
more tightly bound than the other nuclei, i.e., the average binding
energy {\it per nucleon} is greater for $^{56}$Fe than for other
nuclei.  This also means that for lighter elements (with less than 56
nucleons), the mass per nucleon decreases when combining nuclei to
form more heavier elements. Thus, for lighter elements, energy is
usually released in a fusion reaction. For elements heavier than iron
however, the mass per nucleus increases with increasing number of
nucleons and energy is liberated in a fission reaction. For any given
mass or volume of fuel, nuclear processes generate more energy than
can be produced through any other fuel-based approach.\\

\begin{figure*}[t]
    \postscript{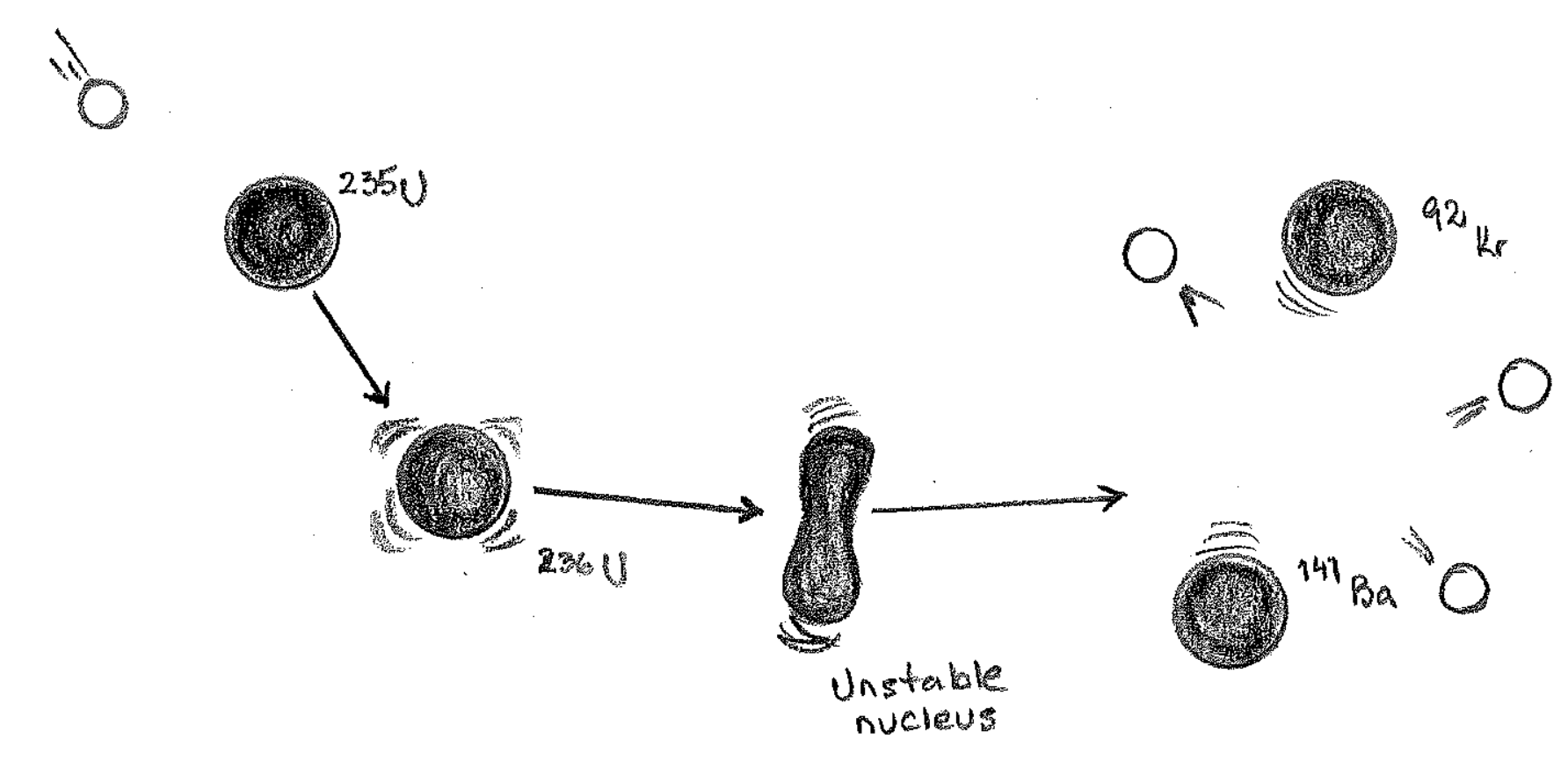}{0.9}
\caption{The stages in a nuclear fission event.
\label{fig:34}}
\end{figure*}

{\bf \S~Nuclear fission} The fission of a uranium or plutonium nucleus
liberates a large amount of energy, which comes from the changes in
the internal energy of the nuclei involved in the reaction. The
splitting of uranium-235 shown in Fig.~\ref{fig:34}, for example, releases over $3 \times
10^{-11}~{\rm J}$, i.e., 50 times more than the alpha decay of the
same nucleus.\footnote{The mass deficit for the reaction ${^{235}{\rm
      U}} + n \to {^{141}{\rm Ba}} +{^{92}{\rm Kr}} + 3 n$ is
  $140.91441~{\rm u} + 91.92616~{\rm u} + 2 \times 1.008665~{\rm u} -
  235.04394~{\rm u} = -0.18604~{\rm u}$. From (\ref{ukg}) we have that
  $1~{\rm u} =
  1.66058 \times 10^{-27}~{\rm kg}$, and so using (\ref{energy-mass})
  it is straightforward to see that the annihilation of $3.08934
  \times 10^{-28}$~kg of mass produces $2.78041 \times 10^{-11}~{\rm
    J}$ of energy.} Only the $^{235}$U part of uranium (0.7\%)
undergoes fission when irradiated with slow neutrons; the $^{238}$U
(99.3\%) will absorb fast neutrons (becoming $^{239}$U, but it will
not fission. The neutrons must be {\it slow} because they have to stay
around a nucleus a while in order to be {\it captured}. They don't
really knock the $^{235}$U appart during fission; it is really more
like they enter the nucleus to form $^{236}$U, which is very unstable
and undergoes fission. During each fission, about 3 neutrons (on the
average) are produced. The possibility of a chain reaction is then
present, except that these fission neutrons are fast. As a
consequence, not only do they fly past $^{235}$U without causing
further fission, but they also get captured by the $^{238}$U. So we
need some way of slowing down the neutrons produced in fission
(moderating them) and then letting them hit more uranium.

\begin{figure}[t]
    \postscript{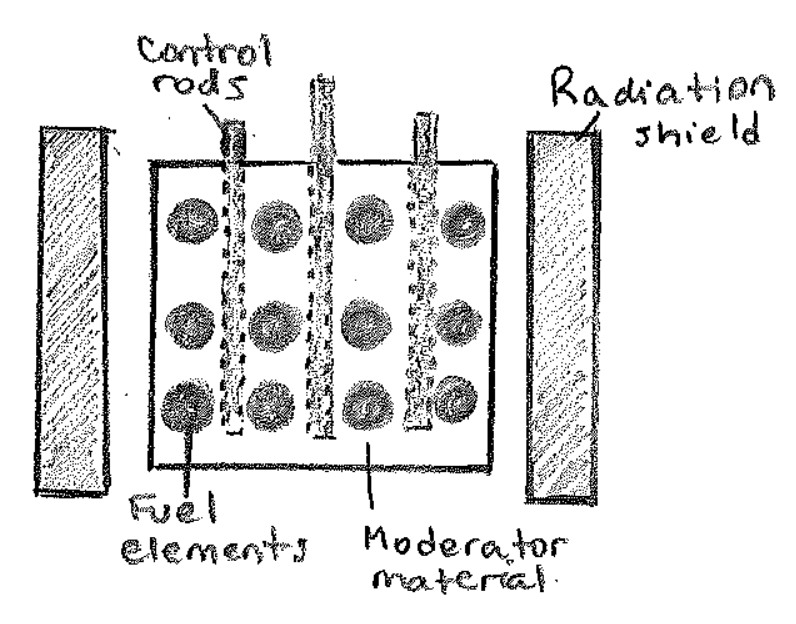}{0.9}
\caption{A cross section of a reactor core showing the control rods, fuel
elements, and moderating material surrounded by a radiation shield.
\label{fig:35}}
\end{figure}

The {\it modus operandi} is as follows. Arrange a series of uranium rods
sitting in the moderator, as shown in Fig.~\ref{fig:35}. You must make sure that: 
\begin{itemize}
\item moderator
moderates (slow the neutrons);
\item moderator does not absorb
too many neutrons; 
\item moderator does not become radioactive,
because it is also used for cooling.
\end{itemize}
For the moderator to slow down neutrons, it must consist of light
atoms (neutrons would just bounce off heavy atoms without losing
speed). Hydrogen (as part of water) is excellent except that it
absorbs neutrons,
\begin{equation}
{\rm hydrogen} + {\rm neutron} = {\rm deuterium} \, .
\end{equation}
So to use ordinary water, one must eliminate some of the absorption of
neutrons by $^{238}$U by making the uranium rod richer in $^{235}$U
(``enrichment''). This is a very expensive process. Carbon (graphite)
does not absorb neutrons, but it is also pretty heavy so it doesn't
moderate as well. So, if you use graphite, you must enrich the
uranium. Heavy water (D$_2$O) is excellent: it moderates almost as
well as ordinary water and it doesn't absorb neutrons; {\it reactors
  using heavy water do not need to have the uranium enriched} -- they
use the natural mixture of $^{238}$U and $^{235}$U. Neither water,
graphite,
or heavy water become radioactive when neutrons pass through them.\\

{\bf \S~The bomb and all that...} In December 1942, Fermi constructed
the first nuclear reactor. It used natural uranium (not enriched), a
graphite moderator (to slow neutrons) and boron control rods (to
control the reaction by absorbing
neutrons)~\cite{Anderson1,Anderson2}. The stage was then set
for building the uranium bomb.

For the bomb, one has {\it no moderator, no control rods}, but pure
$^{235}$U (very expensive). Two pieces of $^{235}$U are brought
together  through a small explosion or a spring, and are irradiated
with neutrons from a small radioactive source (in the bomb). Each
piece by itself is not enough to ``go critical'', but together they
can sustain a chain reaction using the small number of slow neutrons
that are emitted during fission, and the small probability that a fast
one will be captured by a $^{235}$U to induce fission. Unless you have
almost pure $^{235}$U, this won't work. (That is why we have
moderators in reactors, which do not use pure $^{235}$U.)

The first (test; code name Trinity) bomb was exploded in July
1945. The first $^{235}$U bomb was dropped on Hiroshima on August 6,
1945, killing $80,000$ people. The second bomb (Plutonium) was dropped
on Nagasaki two days later with similar results. Plutonium
($_{94}^{239}$Pu) also undergoes fusion with slow neutrons. It is {\it
  not} found naturally, but is made whenever $^{238}$U (the common
isotope) is radiated with fast neutrons:
\begin{eqnarray}
{_{92}^{238}{\rm U}} + n \to {_{92}^{239}{\rm U}}
\xrightarrow[\text{23~min}]{\beta^-\text{decay}}  {_{93}^{239}{\rm
    Np}} \xrightarrow[\text{2.35~days}]{\beta^-\text{decay}}
{{_{94}^{239}}{\rm Pu}} \, ,
\end{eqnarray}
where the times are respectively half-lives of uranium-239 and
neptunium-239. Plutonium is highly radioactive, but has a long
half-life of $24,100$ years, so it can be used. It is lethal in {\it
  microgram} quantities (much less than a speck), due to both its
radioactive and chemical properties.\footnote{A 5~kg mass of
  $^{239}$Pu contains about $12.5 \times 10^{24}$ atoms. With a
  half-life of $24,100$ years, about $11.5 \times 10^{12}$ of its
  atoms decay each second by emitting an alpha particle of $8.3 \times
  10^{-13}$~J. This amounts to 9.68 watts of power.} {\it Terrible
  stuff!}\\

\begin{figure}[t]
    \postscript{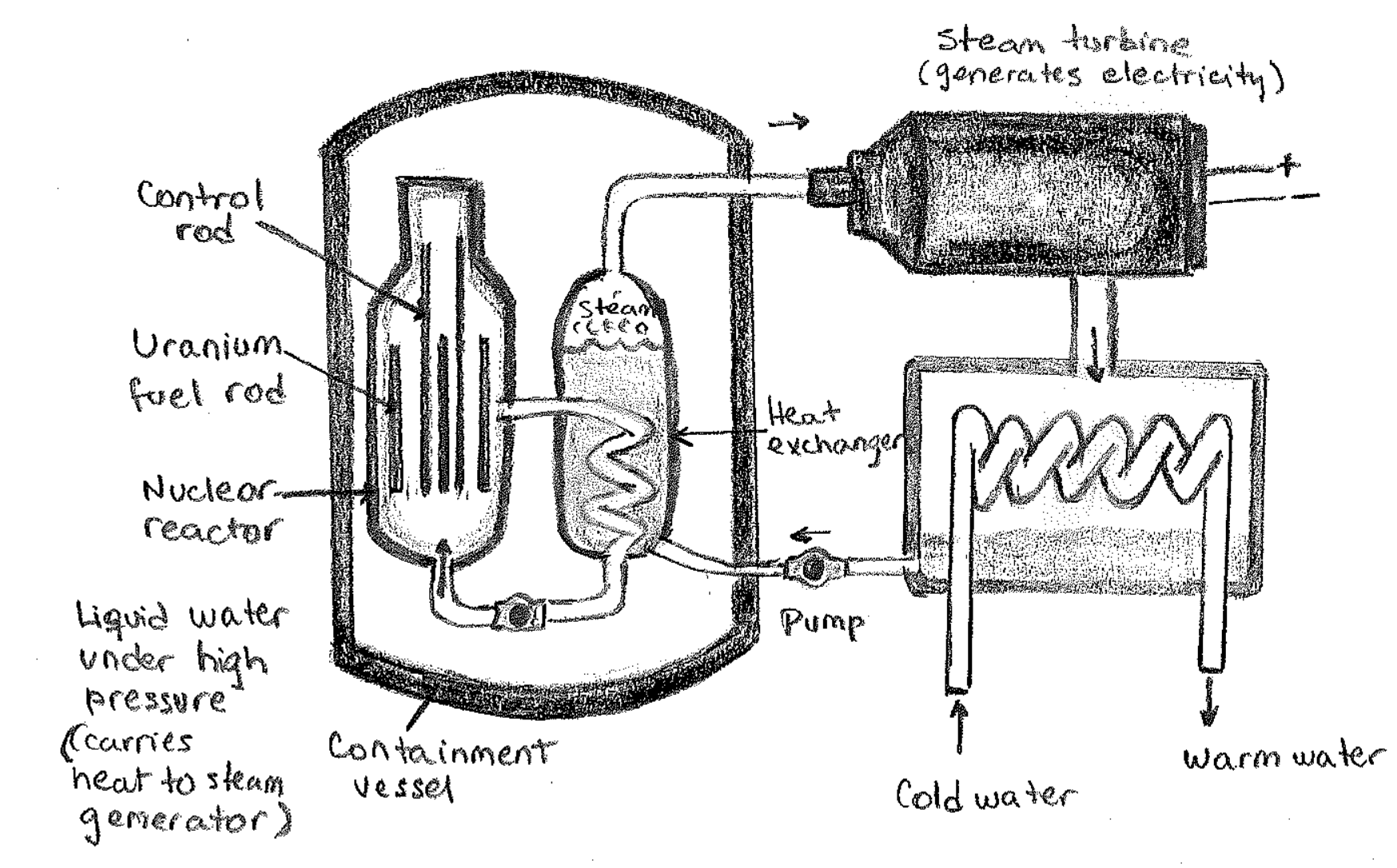}{0.9}
\caption{Main components of a pressurized-water reactor.
\label{fig:36}}
\end{figure}

{\bf \S~Nuclear power plants.}~There are 61 commercially operating
nuclear power plants with 99 nuclear reactors in 30 states of the
U.S. Though there are several types of reactor that convert the kinetic
energy of fission fragments to electrical energy, the most common type
in use in the U.S.  is the pressurized-water reactor shown in
Fig.~\ref{fig:36}. Its main parts are common to all reactor
designs. Fission events in the reactor core supply heat to the water
contained in the primary (closed) loop, which is maintained at high
pressure to keep it from boiling. This water also serves as the
moderator. The hot water is pumped through a heat exchanger, and the
heat is transferred to the water contained in the secondary loop. The
hot water in the secondary loop is converted to steam, which drives a
turbine-generator system to create electric power.  The secondary loop
has a condenser, where steam from the turbine is condensed by
cold water (from a lake or a river). Note that the water
in the secondary loop is isolated from the water in the primary loop
to prevent contamination of the secondary
water and steam by radioactive nuclei from the reactor core.\\

{\bf \S~Dangers of operating nuclear reactors.} The 1986 accident at
the Chernobyl reactor in the Ukraine rightfully focused attention on
reactor safety. Despite of their advantage (energy content of fuel
very high, negligible atmospheric pollution), there are some serious
potential hazards associated with the use of nuclear power. These may
be catalogued as follows:

{\it (1)}~Loss of coolant accident. Possible failure of a pump or pipe
  would prevent the water from cooling. Unless the emergency core
  cooling system went into effect, the core would melt down through
  the casing of the reactor in about 30 seconds, with the possible
  release of an enormous amount of radioactivity to the outside
  world. To control the power level, control roads are inserted into
  the reactor core (see Fig.~\ref{fig:35}). These roads are made of
  material such as cadmium  that absorb neutrons very efficiently.

{\it (2)} Radioactive waste disposal. After a certain amount of time,
the $^{235}$U concentration in the fuel rod becomes too low (it gets
used up in fission!) and the road must be replaced. Unfortunately, it
now contains a {\it large} amount of radioactive material, including
the products of the fission:
\begin{eqnarray}
n + {^{235}{\rm U}} \to {\rm fission} \to & \textnormal{cesium-137} &~~~
({^{137}{\rm Cs}}) \nonumber \\
& \textnormal{stronium-90} & ~~~({^{90}{\rm Sr}}) \nonumber \\
& \textnormal{krypton-85} & ~~~({^{85}{\rm Kr}}) \, .
\end{eqnarray}
The {$^{85}$Kr} escapes as a gas -- not too dangerous. But
{$^{137}$Cs} and {$^{90}$Sr} take about 60 years (two half-lives) to
lose 75\% of their activity, so they create a storage
problem. Strontium-90 is very dangerous because if release to the
natural surroundings, it finds its way into cow's milk in very high
concentrations, and from there into children's bones. The wastes, in
liquid, form are usually stored in stainless steel casks buried in
concrete vaults. They need constant surveillance because of leakage
and the possibility of corrosion. This method is regarded as a
temporary expedient.

Unfortunately, the problem does not die with the Cs and Sr, because
wastes also include some very long-lived materials, such as
plutonium-239. Since a large amount of $^{238}$U is present in every
commercial reactor, and fast neutrons are released during the fission
of $^{235}$U, some of these fast neutrons (the ones that aren't slowed
by the moderator) will react with the $^{238}$U to form plutonium {\it
  in the same fuel road}. So, a fair amount of plutonium is formed
every present day commercial reactor.  {$^{239}$Pu} undergoes fission
just like $^{235}$U, so more fuel for fission has been created in the
reactor. This is called converting. When the reactor makes more fuel
than it consumes, it is called a {\it breeder}.  Breeders were at
first found attractive because their fuel economy was better than
light water reactors, but interest declined after the 1960s as more
uranium reserves were found, and
new methods of uranium enrichment reduced fuel costs.\\

{\bf \S~Nuclear fusion.} The reactions that take place in
the Sun cannot take place on Earth. To hold the hydrogen cloud
together by gravitation (against the expanding forces) until the slow
interaction \mbox{$pp \to D + e^+ + \nu$} takes place, requires a mass of
material equal to the mass of the Sun. So if we would attempt to burn
``protons'' here, the ``fire'' would keep going out.

\begin{figure}[t]
    \postscript{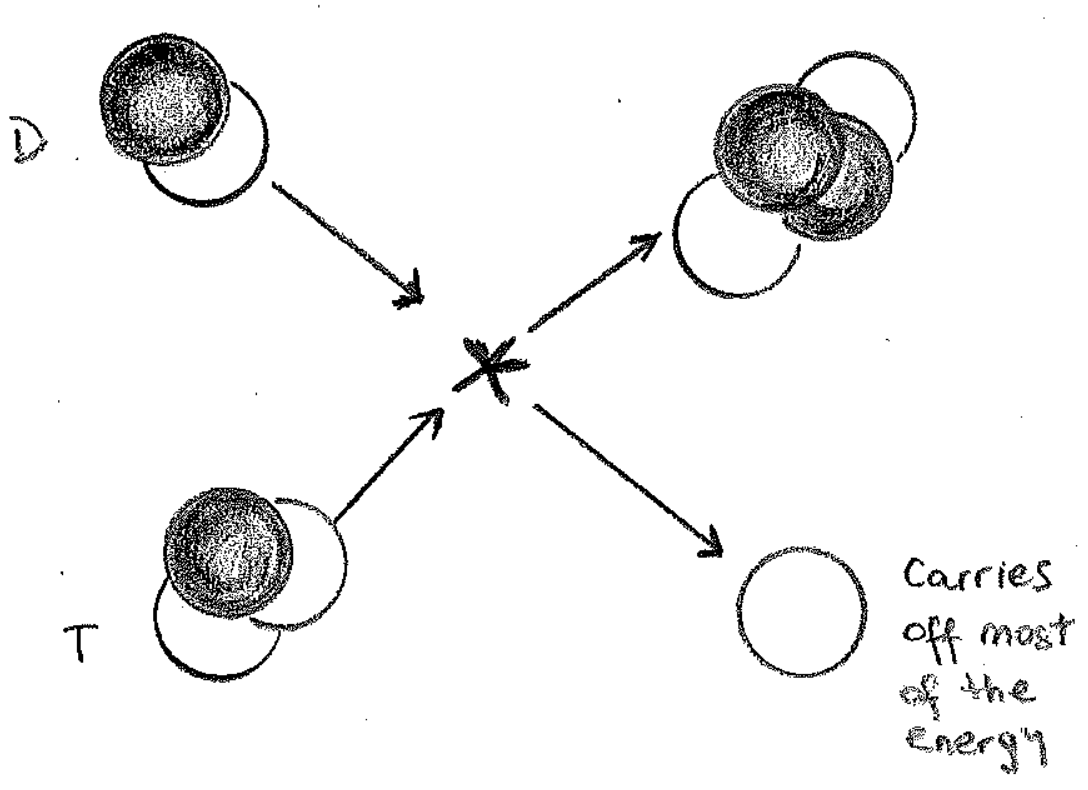}{0.9}
\caption{$D + T \to  {^4{\rm He}} + n$.
\label{fig:37}}
\end{figure}

However, there may be other nuclear fusion reactions which release
enormous amount of energy. These play no role in the nuclear chemistry
of the Sun because of the negligible amounts of the necessary
materials present there. We will discuss two of these. They involve
the isotopes of hydrogen, deuterium (D), and tritium (T). The
deuterium-tritium reaction, shown in Fig.~\ref{fig:37},  is
\begin{equation}
{\rm D+T} \to \, ^4{\rm He} + n + 6 \ {\rm billion \ Btu/oz \ of \ (D+T)}
\,  . 
\end{equation}
(Compare to an energy release of $1,000$~Btu/oz of oil burnt.)

The D-D reaction is more complicated, and it releases about 1/4 as much
energy. Nevertheless, in the very long run, it is very important (as we
will see). For now, we will concentrate on the D-T reaction.

First, why D-T? Well, to make a nuclear fusion reaction go, the nuclei
must be brought close enough together to touch. All nuclei are
positively charged, so that one must accelerate them to high speeds in
order to overcome the effects of the electric repulsion. Since the
positive charge of on a nucleus increases with its mass (more
protons), it is then clear why it is easiest to overcome the
electrical repulsion  for the lightest nuclei. We have seen this in
the case of the Sun's life cycle, where the fusion of helium (charge
+2 each) or carbon (+6 each) require a higher temperature than the
fusion of hydrogen. So, we want to fuse hydrogen, if we can. But we
have seen that the fusion of protons goes only via the weak
interaction, so we then may inquire if the fusion of deuterium or
tritium (same 1 + charge) can go via strong interaction. The answer is
yes!

What about availability? Deuterium is stable and very plentiful: one
out of every $6,000$ atoms of hydrogen in water is $^2$H
(a.k.a. deuterium). So, a gallon of H$_2$O yields about a spoonful of
heavy water D$_2$O. It is relatively easy to separate off. Tritium is
another problem. It is a radioactive gas, decaying with a half-life of
12 years (to $^3$He + electron + neutrino) so it doesn't occur
naturally. It can be produced in small quantities at ordinary nuclear
reactors through the bombardment of lithium with neutrons
\begin{equation}
n + {^6{\rm Li}}  \to {^4{\rm He}} + {\rm T} \, .
\end{equation}
To overcome the difficulty, the usual design of the future fusion
reactor involves the following general procedure: 
\begin{enumerate}
\item Start the reaction (in one way or another) using D and T.
\item Surround the reaction region with a liquid lithium blanket. This
  serves two processes:
\begin{enumerate}
\item The neutrons emerging from the D-T reaction will react with {$^6$Li}
  to generate tritium and
\item the liquid litium will absorb energy of the reaction, become hot
  and can be used to boil water, make steam, etc.
\end{enumerate}
\end{enumerate}
So two things are apparent:
\begin{enumerate}
\item One must deal with technology of liquid lithium (similar to that
  of liquid sodium) and
\item the limiting fuel resource is lithium. At present, the world
  supply of lithium is equivalent (in eventual Btu) to that of all the
  fossil fuels (a few hundred years). By that time, one will have
  learned to run the D-D reaction, where the resources are almost
  infinite (sea-water!).
\end{enumerate}

The basic fact is, that to
overcome the electrical repulsion, the D and T must approach one
another at a relative speed of about 750 miles per second. As we have
seen in the case of the Sun, this involves heating the D-mixture to
temperatures of tens of millions of degrees K (about 50 million, to be
exact). Such high temperatures are attainable on Earth in the
proximity of a nuclear explosion. Indeed, by surrounding a $^{235}$U
bomb with D-T mixture, you can ignite all at once, and it explodes:
Presto, a hydrogen bomb! But this is not a controllable source of
power.

\begin{figure}[t]
    \postscript{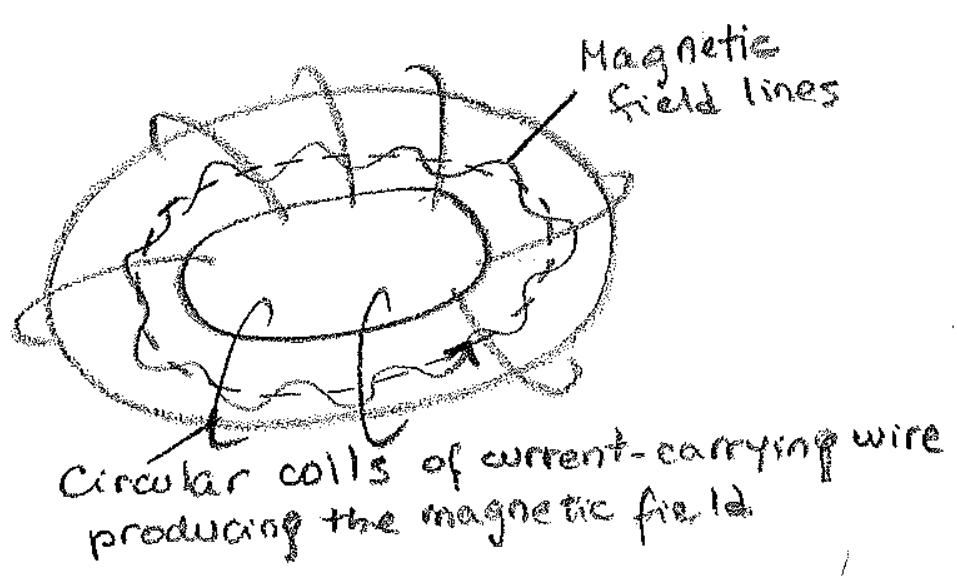}{0.9}
\caption{The toroidal geometry of plasma confinement. The ionized atoms circulate around the ring, trapped by the magnetic field lines. The coils produce a magnetic field along the axis of the toroid (dashed line). Another field component is produced by a current along the axis in the plasma. The two components of the field produce the helical field lines shown.
  \label{fig:38}}
\end{figure}

Given this, the basic problem is the following: Once the D-T mixture
is heated to $50 \times 10^{6\circ}$K, {\it how you get the reaction
  going before the mixture is cooled through contact with the
  environment?} There are two basic possibilities, and both are being
  actively pursed: Firstly, raise the temperature so quickly that the
  fuel mixture is all consumed {\it before} the heat
  escapes. Secondly, (and this at first sounds impossible), keep the
  mixture away from {\it all} contact with the outside world.   

The first process (look back!) is essentially what we mean by an
explosion. In the D-T case, this means setting off a series of tiny
hydrogen bombs; at present this is being researched under the heading
of laser-induced fusion. The scheme is basically very simple: a small
pellet (about 1/10 to 1~cm in diameter) of frozen D-T (at $4^\circ$K)
is injected into an explosion vessel. A massive laser pulse is
focussed onto the pellet, heating it ``instantaneously'' to 100
million$^\circ$K, causing the D-T reaction to consume the pellet,
giving off the fast neutrons which heat the lithium blanket and make
more tritium. Then another pellet is dropped, the process repeated and
so on. Each explosion is equivalent to the detonation of a small stick
of dynamite.

Let's just list some of the problems:
\begin{itemize}
\item The projected power required from the laser is about 100 times
  that envisaged in the next generation of giant lasers. At this
  point, the Russians are in a distinct lead in this field. 
\item Through computer studies, it looks like there will be a problem
  in heating the pellet fast enough and evenly enough to cause the
  reaction to go to completion before the D-T evaporates.
\item There may be a cost problem: the combined cost of pellet and
  energy for laser may be excessive.
\end{itemize}
However, the combination of military applicability (lasers for
antimissile defense, the fusion process itself as a simulation for
weapon testing) and the payoff in the event of success provide a
distinct impetus for the program.

The second way of preventing the cooling of the mixture before the
burning takes place depends on our old friend, the magnetic field. In
essence, one can keep charged particles out of touch with their
environment by getting them to circle around magnetic field lines. The
idea is one we have already dealt with: if you create a strong
magnetic field, charged particles will be confined to describe tight
circles about the field lines.

If the field is strong enough, the plasma will be kept from all
contact with the wall of the container. One of the most promising
setups at present was first developed in Russia and goes under the
name of Tokamac~\cite{Tamm}. In this machine the magnetic field takes
a toroidal, or doughnut, shape, and the plasma is squeezed into very
tight circles about the field lines; see Fig.~\ref{fig:38}. This
serves not only as a containment device, but also as a heating agent
for the plasma.

The main problem with magnetically contained fusion is the following:
for ``break-even'', that is, to get more fusion energy out than you put
in, the plasma must be contained at {\it a high density} for a {\it
  long enough time}, at a temperature of about 40 million$^\circ$K.
On May 2011, the Experimental Advanced Superconducting Tokamak (EAST)
was able to sustain hydrogen plasma to about 50 million$^\circ$K for
30~s~\cite{Li}.  EAST is one of the precursors of the International
Thermonuclear Experimental Reactor (ITER), a full-scale nuclear fusion
power plant currently being built in France~\cite{Clery}.  Thereby the
ITER machine aims to demonstrate the principle of producing more
energy from the fusion process than is used to initiate it, something
that has not yet been achieved in any fusion reactor.

\section{Spacetime}

{\bf \S~Foundations of special relativity.} In the 19th century, it
was thought that just as water waves must have a medium to move across
(water), and audible sound waves require a medium to move through
(e.g., air), so also light waves require a medium, which was called
the ``luminiferous'' (i.e. light-bearing) ``\ae ther.''  If this were
the case, as the Earth moves in its orbit around the Sun, the flow of
the \ae ther across the Earth's surface could produce a detectable
``\ae ther wind.'' Unless for some reason the \ae ther were always
stationary with respect to the Earth, the speed of a beam of light
emitted from a source on Earth would depend on the magnitude of the
\ae ther wind and on the direction of the beam with respect to it. In
1881, Michelson designed an experiment to measure the speed of light in
different directions in order to measure the speed of the \ae ther
relative to Earth, thus establishing its existence~\cite{Michelson}. The
Michelson-Morley experiment became what might be regarded as the most
famous failed experiment to date and is generally considered to be the
first strong evidence against the existence of the luminiferous
\ae ther~\cite{Michelson:1887zz}.

\begin{figure*}[t]
    \postscript{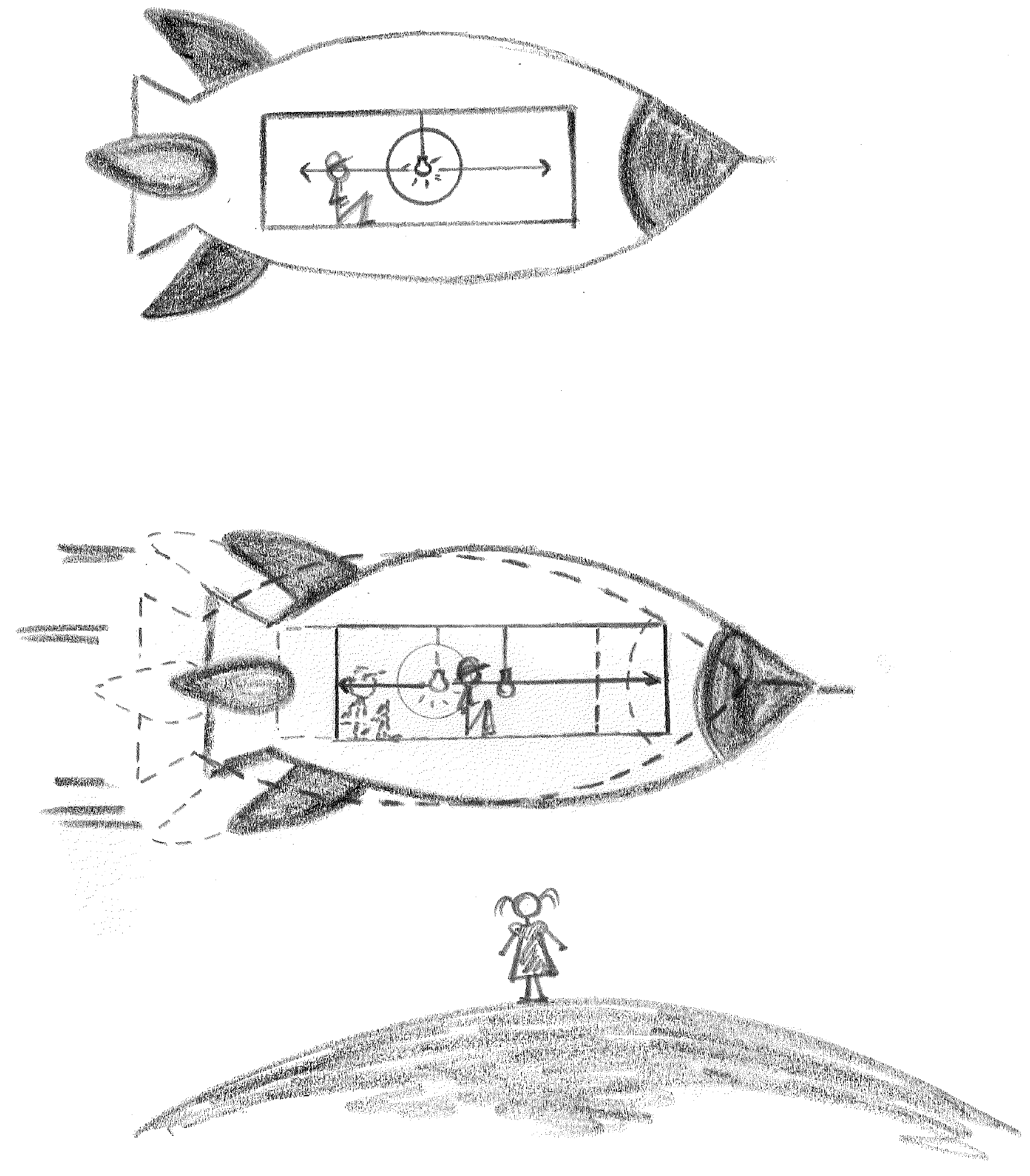}{0.8}
\caption{From Harry's viewpoint the light from the source travels
  equal distances to both ends of the spacecraft, and hence strikes
  both ends simultaneously.  The events of striking the front and the
  end of the spacecraft are not simultaneous in Sally's reference
  frame. Because of the rocket's motion,  light that strikes the back
  end doesn't have as far
  to go and strikes sooner than light that strikes the front end.
\label{fig:39}}
\end{figure*}

To explain nature's apparent conspiracy to hide the \ae ther
drift, in 1905  Einstein advanced the principle of relativity based
on the following two postulates~\cite{Einstein:1905ve}:
\begin{enumerate}
\item {\it All laws of nature are the same in all
  uniformly moving reference frames.}
\item {\it The speed of light in free space has the same measured value for
  all observers, regardless of the motion of the source or the motion
  of the observer, i.e. the speed of light is a constant.}
\end{enumerate}
The first postulate recollects the idea that all motion is relative,
and not to any stationary hitching post in the universe, but to
arbitrary reference frames. This implies that a spacecraft cannot
measure its speed with respect to empty space, but only with respect
to other objects. In other words, if Harry's rocket-ship drifts past
Sally's rocket-ship  in empty space, spaceman  and spacewoman will
each observe the relative motion, and from this observation, each will
be unable to determine who is moving and who is not, if either. The
second postulate introduces the idea that in empty space the speed of
light is the same in all reference frames. Einstein's postulates
describe in simple and clear terms the Michelson-Morley
experiment, which cannot be explained
otherwise.\\

\begin{figure*}[tbp] 
\postscript{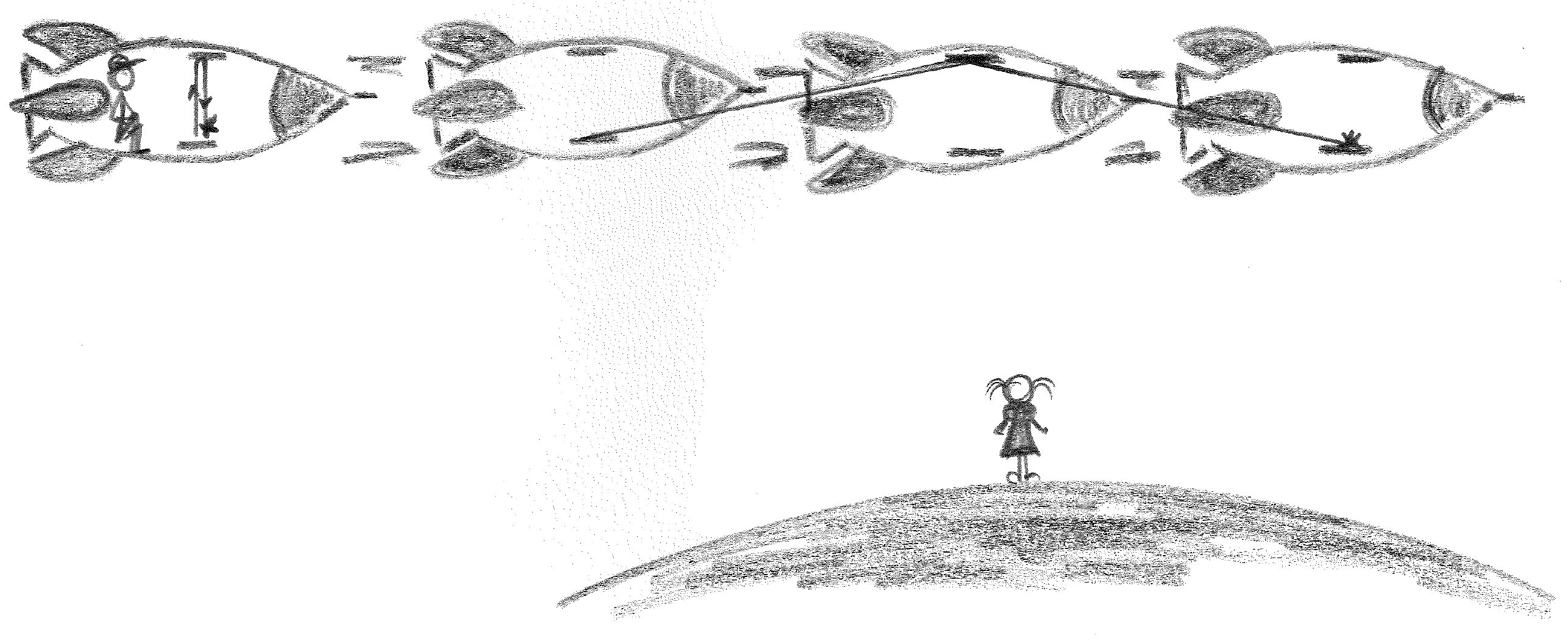}{0.9} 
\caption{Harry and the mirror are in a spaceship at rest in the $S'$ frame. The time it takes for the light pulse to reach the mirror and return is measured by Harry to be $2d'/c$.  In the frame $S$, the spaceship is moving to the right with speed $v$. For Sally, the time it takes for the light to reach the mirror and return is longer than $2d'/c$.}
\label{fig:40}
\end{figure*}

\begin{figure}[tbp] 
\postscript{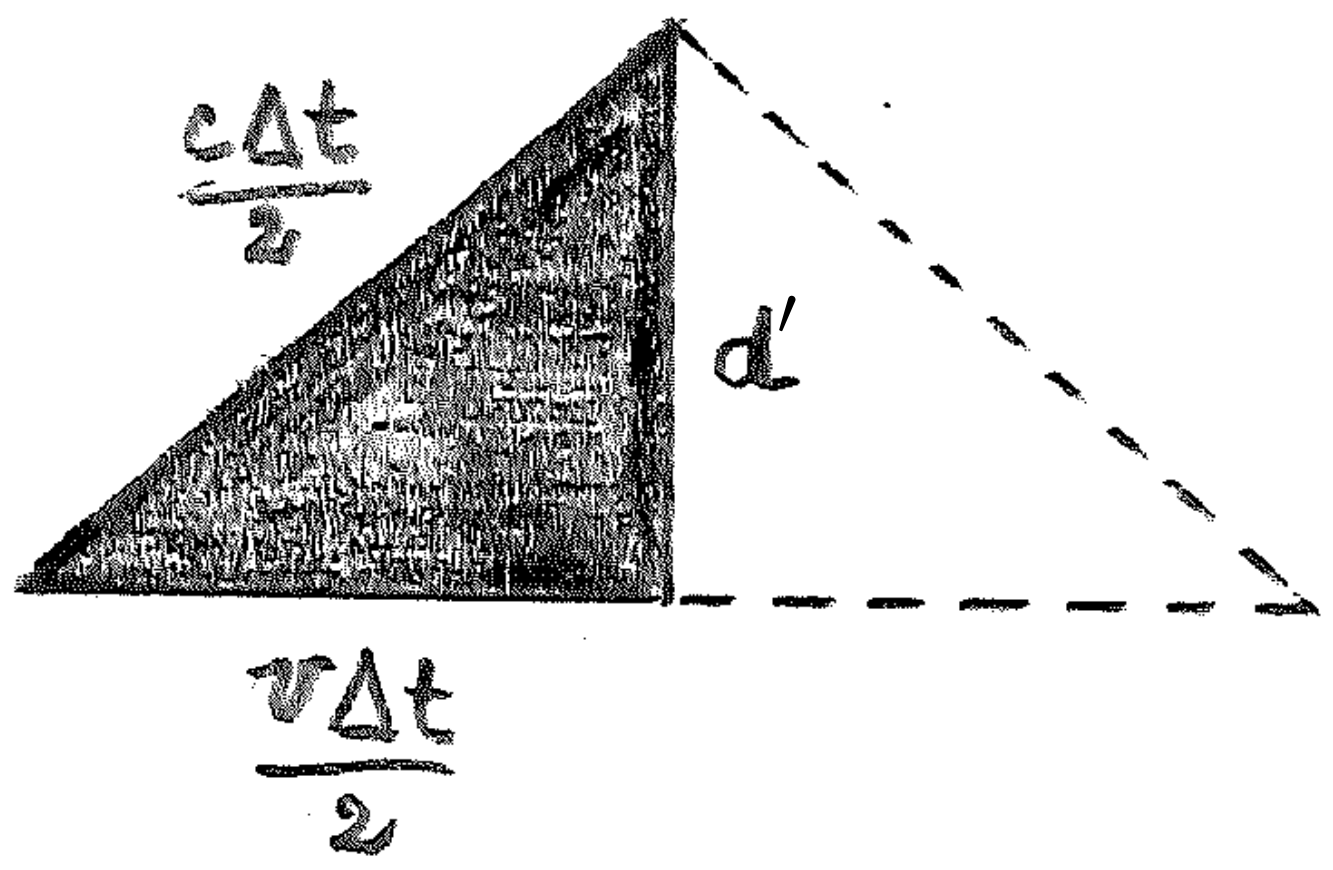}{0.8} 
\caption{A right triangle for computing the time $\Delta t$ in the $S$ frame.}
\label{fig:41}
\end{figure}

{\bf \S~Relativity of simultaneity.}~We experience time as a
succession of events with certain regularities. The days pass one
after another. We and others grow older according to the number of
days we have lived through. Time was traditionally measured in days,
which added up to months, and then to years. Slowly other types of
regularities came to be used to measure time: spring motion, pendulum
motion, the flow of sand, etc. These days we use electronic
oscillators (very accurate), atomic vibrations (incredibly accurate),
and other things.  All of these things could be used as clocks. In
some way, the flow of time governs them all.

Once we accept Einstein's 2nd postulate it turns out that our
commonsense notion of time needs to be profoundly modified in several
ways. The first modification is to the notion of simultaneity.  We
expect, based on everyday experience, that two events that are
simultaneous for observer $S$ will also be simultaneous for observer
$S'$, even if observers $S$ and $S'$ are moving with respect to one
another. This is certainly true in Newtonian mechanics. In other
words, two events being simultaneous is an invariant.  Now let's check
if this invariance carries over to the theory of special relativity.
This time, observer $S'$ (Harry) in a rocket-ship takes a picture with a
flashbulb, and the flash goes out in all directions.

We will focus on just the east and west directions, shown in
Fig.~\ref{fig:39}. Harry is
sitting in the middle of the spaceship. The vessel is not moving in his
reference frame. The camera  is at rest in his frame,
as are the front and back of the spacecraft. The flash, which travels at the
speed $c$ in both directions, reaches the front of the spacecraft and the rear
of the spacecraft at the same time since it started in the middle. Nothing
strange about that. Those two arrival events are simultaneous in his
frame.

But what about in Sally's frame, where the rocket is moving at the
velocity $v$? Sally also sees the light flashes moving
forwards and backwards at the speed $c$. But there is a time lapse
between the time when the flash is generated at the middle of the rocket
and the time when the flash hits the front of the rocket. During that
admittedly short time the front of the rocket has moved to the east, and
so has the rear of the rocket. So the flash takes longer to reach the
front of the rocket than it does to reach the rear of the rocket. The
arrival time at the front is later than the arrival time at the
back. Two events that are simultaneous in Harry's frame  happen at different
times in Sally's frame. This is called the relativity of simultaneity, since
whether two events are simultaneous depends on the
observer. Simultaneity is not an invariant.

But really, if two events that happen at the same time in one frame
happen at different times in another frame, how can we make sense of
motion at all?  We defined velocity as length traveled divided by
elapsed time. The elapsed time did not depend on which frame we were
talking about, and it seems like this is the only way we can have a
sensible theory of motion. To answer this question we have to think
carefully about how we actually do measurements of position and
time. Einstein pointed out that when we speak of the time that
something happens we must be present at the place that it happens and
we must have a clock to record the time. Up until now it was enough
for us to have a measuring stick for each reference frame, a rigid
body that defined units of a coordinate system. But we could all
depend on just one clock, a master timepiece that was used by all
observers. Now what we need is a measuring stick with clocks all along
it, maybe a little clock every centimeter or so, so that when
something happens we can record both the time and the place. All the
clocks on our measuring stick in one frame must be synchronized with
one another. Any such fancy measuring stick then defines the
observations in one reference frame. Then, the
description of an event 1 in frame $S$ is given by a statement that
looks like this: event $1$ happens at point $x_1$ at a time $t_1$ according to
the measuring-stick-plus-clocks of frame $S$. Note that space and
time are somehow intertwined together, so we can take this a little
further and speak of an event happening at a point in spacetime. \\

{\bf \S~Time dilation.} Having established our
measuring system we now turn to derive predictions of the principle of
special relativity using some thought experiments devised by
Einstein. Einstein's thought experiments involve an idealized clock in
which a light wave is bouncing back and forth between two mirrors. The
clock ``ticks'' when the light wave makes a round trip from mirror $A$
to mirror $B$ and back, that is the time that passes as the light
travels from one mirror to the other and returns is the unit of
time.  Let's go back to frame $S$, where Sally is standing on a planet with
measuring-stick-plus-clocks. She is watching Harry that goes by in his
rocket as before.

Assume the mirrors $A$ and $B$ are separated by a distance $d'$
in Harry's rest frame. In that frame a light wave will take a time
\begin{equation}
\Delta t' = 2 d'/c
\label{qqqqw}
\end{equation}
 for the round trip $A \to B \to A$. This is the proper time interval between two consecutive ticks of the clock. Let $\Delta t$ be the interval between two consecutive ticks of the clock in Sally's frame, in which the mirrors move with velocity $v$, as shown in Fig.~\ref{fig:40}. It is noted that when the light wave is bounced back at the mirror $B$, the latter has already move a distance $ v\Delta t/2$, as shown in 
Fig.~\ref{fig:41}. Since light has velocity $c$ in all directions
\begin{equation}
{d'}^2 + \left(v \, \frac{\Delta t }{2} \right)^2 = \left(\frac{c \Delta t}{2} \right)^2\,,
\end{equation}
or
\begin{equation}
\Delta t = \frac{2d'}{\sqrt{c^2 - v^2}} = \frac{\Delta t'}{\sqrt{1 - v^2/c^2}} \, .
\label{t-dilation}
\end{equation}
Hence the ticking of the clock in Harry's frame, which moves with
velocity $v$ in a direction perpendicular to the separation of the
mirrors, is slower by a factor 
\begin{equation}
\gamma = \sqrt{\frac{1}{1- v^2/c^2}} \,, 
\label{Lorentz-factor}
\end{equation}
usually refer to as the Lorentz factor~\cite{Lorentz:1904}.

A number of experiments support the time dilation predicted by special
relativity. One such experiments, conducted by Rossi and Hall in 1941,
revealed that muons decay more slowly while falling~\cite{Rossi:1941zz}.
Muons are sub-atomic particles generated when cosmic rays strike the
upper levels of our atmosphere. They have a half-life of about 2.2
microseconds ($\mu$s) meaning that every 2.2~$\mu$s, their population
will reduce by half. By observing the concentration of muons at both
the summit and base of Mount Washington, Rossi and Hall were able to
measure what proportion of them have decayed and to compare this result
with the predictions of special relativity. The experiment was carried
out using detectors that only count muons traveling within a certain
speed range, $0.9950c < v < 0.9954c$. Mount Washington has a height
difference of 1.9~km between the summit and the base. Flying 1.9~km
through the atmosphere at the above speed takes about $6.4~\mu {\rm
  s}$. Based on the stated half-life, we should thus expect that only
13\% of the original concentration of muons should arrive. However, it
is observed that about 82\% of the muons arrive below. This percentage
would correspond to a half-life of 22~$\mu$s, i.e. ten times greater
than the original. A factor of ten, however, corresponds to what
(\ref{Lorentz-factor}) would give for a speed of $0.995c$.\\

\begin{figure*}[tbp] 
\postscript{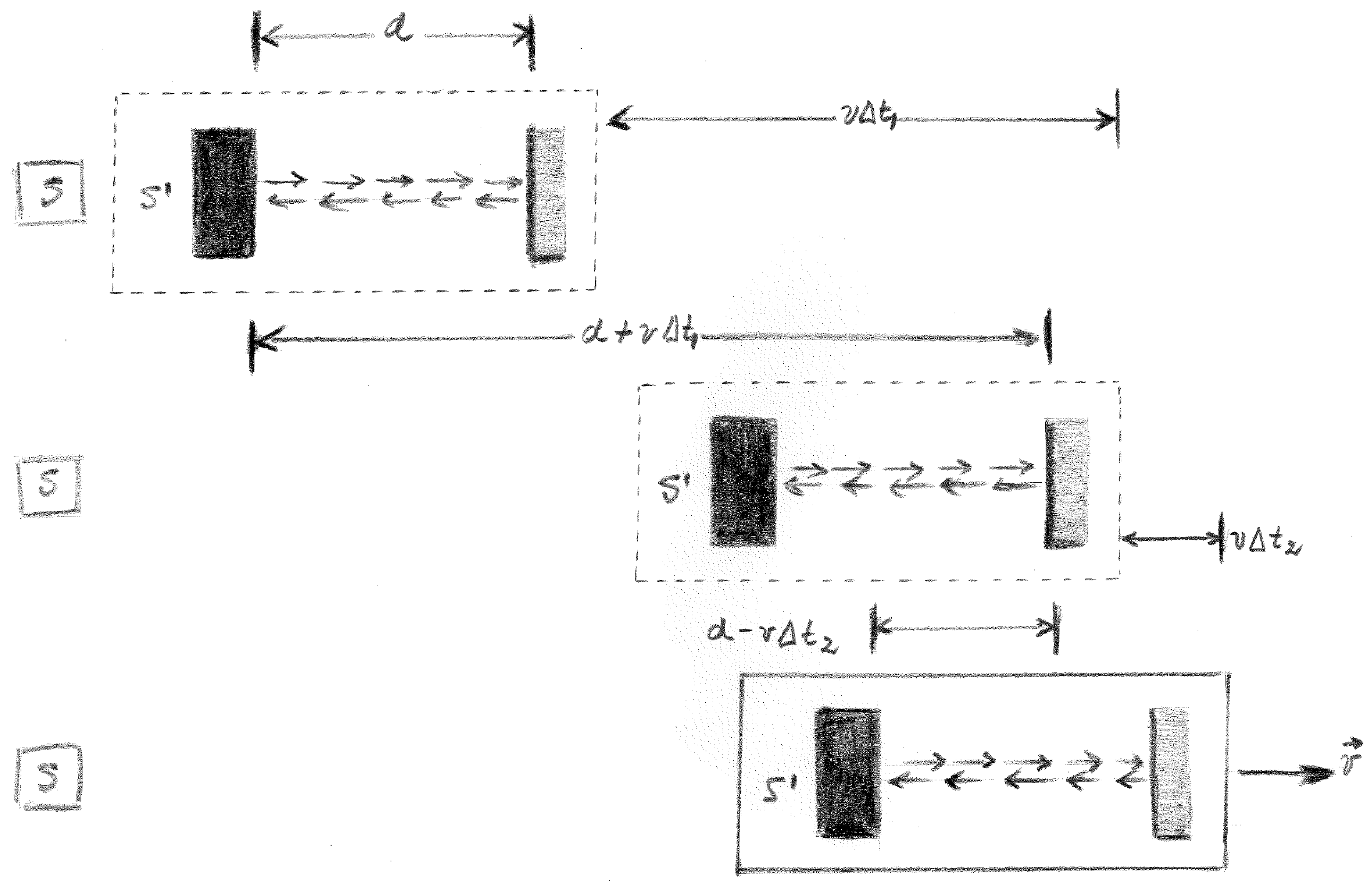}{0.9} 
\caption{The clock carried by Harry emits its flash from mirror $A$
  (on the left of the clock) towards
  $B$ in the direction of motion. The top, middle, and bottom panels
  show representative snapshots of the sequence of events.}
\label{fig:42}
\end{figure*}

{\bf \S~Length contraction.} Now, suppose that Harry rotates the clock
by $90^\circ$ before being set in motion, so that  it has velocity
$v$ parallel to the separation between the mirrors, see
Fig.~\ref{fig:42}. According to Sally (in reference frame $S$) the
length of the clock (distance between the mirors)
is $d$. As we will  see, this length is different from the length $d'$
measured by Harry in $S',$ relative to whom the clock is at rest.

The flash of light emitted from mirror $A$ reaches the mirror $B$ at
time $\Delta t_1$ later. In this time interval, the light travels a
distance $c \Delta t_1$, which is equal to the length $d$ of the clock
plus the additional distance $v \Delta t_1$ that the mirror moves
forward in this interval; namely,
\begin{equation}
d + v \, \Delta t_1 = c \, \Delta t_1 ~~~~~~~~ 
{\rm or} ~~~~~~~~
\Delta t_1 = \frac{d}{c-v} \, .
\end{equation}
Assuming the light wave, after bouncing at $B$, takes time $\Delta t_2$ to reach $A$ again, we have by the same reasoning
\begin{equation}
d - v  \Delta t_2 = c \Delta t_2 \,,
~~~~~~~~ 
{\rm or}
~~~~~~~~ 
\Delta t_2 = \frac{d}{c +v} \, .
\end{equation}
Hence the interval between two consecutive ticks in the moving frame is
\begin{eqnarray}
\Delta t  & = & \Delta t_1 +  \Delta t_2 = \frac{2d}{c (1 - v^2/c^2)} \nonumber \\
& = & \left(\frac{d}{d'} \right) \frac{\Delta t'}{1 - v^2/c^2} \, ,
\label{starbuckG}
\end{eqnarray}
where we have used (\ref{qqqqw}). Substituting (\ref{t-dilation}) into (\ref{starbuckG}) we have
\begin{equation}
d = \left( 1 - \frac{v^2}{c^2} \right)^{1/2} d' \, .
\label{l-contraction}
\end{equation}
Such a length contraction was independently proposed by Lorentz and FitzGerald  to explain the result of Michelson's experiment~\cite{FitzGerald,Lorentz:1892}.

Direct confirmation of length contraction is hard to achieve in
practice since the dimensions of the observed particles are
vanishingly small. However, there are indirect confirmations; for
example, the behavior of colliding heavy nuclei can only be explained if
their increased density due to Lorentz contraction is considered. The
implications of special relativity have been widely tested over the
past century. All of the experiments support Einstein's theory and are
in complete disagreement with non-relativistic predictions.
Consequently, both time dilation and length contraction must be
considered when conducting experiments in particle colliders.\\

{\bf \S~What is gravity?}~It turns out that Newton's universal law of
gravity is not compatible with special relativity. In particular,
having learned special relativity we now know that it should not be
possible to send messages faster than the speed of light. However,
Newton's relation (\ref{GF}) would allow us to do so using
gravity. The point is that Newton stated that the gravitational force
depends on the separation between the objects at a given instant of
time. 

To further convince yourself that (\ref{GF}) is not compatible with
special relativity consider the following situation. The Earth is
about eight light-minutes from the Sun. This means that, at the speed
of light, a message would take eight minutes to travel from the Sun to
the Earth. However, suppose that, unbeknownst to us, some aliens are
about to move the Sun. Then, based on our understanding of relativity,
we would expect it to take eight minutes for us to find out! But
Newton would have expected us to find out instantly because the force
on the Earth would shift (changing e.g., the tides).

Now, it is important to understand how Maxwell's light waves get
around this sort of problem. That is to say, what if the Sun were a
positive electric charge, the Earth were a big negative electric
charge, and they were held together by an electromagnetic field? We
know that Maxwell's light waves are consistent with relativity, so we
can investigate what would these waves tell us happens when the aliens
move the Sun.

The point is that the positive charge does not act directly on the
negative charge. Instead, the positive charge sets up an electric
field which tells the negative charge how to move.  When the positive
charge is moved, the electric field around it must change, but it
turns out that the field does not change everywhere at the same time.
Instead, the movement of the charge modifies the field only where the
charge actually is. This makes a ``{\it ripple}'' in the field which then
moves outward at the speed of light. 

Thus, the basic way that Maxwell's electromagnetic waves get around
the problem of instant reaction is by having a field that will carry
the message to the other charge (or, say, to the planet) at a finite
speed.  What we see is that the field concept is the essential link
that allows us to understand electric and magnetic forces in
relativity.

Something like this must happen for gravity as well. Let's try to
introduce a gravitational field by breaking Newton's law of gravity up
into two parts. The idea will again be than an object should produce a
gravitational field $\vec g$  in the spacetime around it, and that this
gravitational field should then tell the other objects how to move
through spacetime. Any information about the object causing the
gravity should not reach the other objects directly, but should only
be communicated through the field. We rewrite (\ref{GF}) to indicate
the force on $m$ produced by $M$ as follows:
\begin{equation}
\vec F_g = \vec g \ m \,,
\end{equation}
where $\vec g = GM/r^2$. Einstein used this idea to develop the
general theory of relativity, in which the gravitational field is
related to the geometry of
spacetime~\cite{Einstein:1915by,Einstein:1915bz,Einstein:1915ca,Einstein:1916vd}. A
major forecast of general relativity is that when two massive objects
crash into each other there should be a release of gravitational
waves, which transport energy as gravitational
radiation~\cite{Einstein:1916cc}.

The first direct detection of gravitational waves was made on 14
September 2015 by the LIGO and Virgo
collaborations~\cite{Abbott:2016blz}. The observed gravitational waves
were emitted during the final moments of the merger of a pair of black
holes. Previously gravitational waves had only been inferred
indirectly, via their effect on the timing of pulsars in binary star
systems~\cite{Taylor:1982zz}. More recently, the LIGO and Virgo
detectors observed {\it ``ripples''} in the geometry of spacetime 
originating in the violent collision of two distant neutron
stars~\cite{TheLIGOScientific:2017qsa}. When these two objects
combined, they spiraled around each other rapidly before smashing into
one another, creating a gigantic fireball of light also visible to
telescopes on Earth~\cite{GBM:2017lvd}. The substantial ejecta masses
inferred from observations at optical and infrared wavelengths suggest
that the accumulated nucleosynthesis from neutron star mergers could
account for all the gold, platinum, and many other heavy elements
around us~\cite{Kasen:2017sxr}.

\section{Across the Universe}

{\bf \S~Stars and galaxies.} A look at the night sky provides a strong
impression of a changeless universe. We know that clouds drift across
the Moon, the sky rotates around the polar star, and on longer times,
the Moon itself grows and shrinks and the Moon and planets move
against the background of stars.  Of course we know that these are
merely local phenomena caused by motions within our solar system. Far
beyond the planets, the stars appear motionless. In today's class  
we are going to see that this impression of changelessness is
illusory.

According to the ancient cosmological belief, the stars, except for a
few that appeared to move (the planets), where fixed on a sphere
beyond the last planet.  The universe was self
contained and we, here on Earth, were at its center. Our view of the
universe dramatically changed after Galileo's first telescopic
observations: we no longer place ourselves at the center and we view
the universe as vastly larger~\cite{Galileo:1610,Galileo,Copernicus}. 

The astronomical distances are so large
that we specify them in terms of the time it takes the light to travel
a given distance. For example, 
\begin{equation}
{\rm one \ light \ second } = 3 \times 10^8 {\rm m} = 300,000~{\rm
  km} \,, 
\end{equation}
\begin{equation}
{\rm one \ light \ minute} = 1.8 \times 10^7~{\rm km} \,, 
\end{equation}
and
\begin{equation}
{\rm one \ light \ year} = 1~{\rm ly}  =  9.46 \times 10^{15}~{\rm m} 
  \approx  10^{13}~{\rm km}. 
\label{unoEQ}
\end{equation}
For specifying distances to the Sun and the Moon, we usually use
meters or kilometers, but we could specify them in terms of light. The
Earth-Moon distance is 384,000~km, which is 1.28 ls. The Earth-Sun
distance is $150,000,000~{\rm km}$; this is
equal to 8.3~lm. Far out in the solar system, Pluto is
about $6 \times 10^9~{\rm km}$ from the Sun, or $6 \times 10^{-4}$ ly.
The nearest star to us, Proxima Centauri, is about 4.2~ly away.
Therefore, the nearest star is 10,000 times farther from us that the
outer reach of the solar system.

On clear moonless nights, thousands of stars with varying degrees of
brightness can be seen, as well as the long cloudy strip known as the
Milky Way. Galileo first observed with his telescope that the Milky
Way is comprised of countless numbers of individual stars. A half
century later Wright suggested that the Milky Way was a flat disc of
stars extending to great distances in a plane, which we call the
Galaxy~\cite{Wright:1750}.

Our Galaxy has a diameter of 100,000~ly and a thickness of roughly
2,000~ly. It has a bulging central {\it nucleus} and spiral arms. Our Sun,
which seems to be just another star, is located half  way from the
Galactic center to the edge, some $26,000$~ly from the center. The Sun
orbits the Galactic center approximately once every 250 million years
or so, so its speed is
\begin{equation}
v  =  
\frac{2\pi\ \ \  26,000 \times 10^{13}~\rm km}{2.5 \times 10^8~{\rm yr}\ 
3.156 \times 10^7~{\rm s/yr} } 
   = 200~{\rm km/s} \,.
\end{equation}
The total mass of all the stars in the Galaxy can be estimated using
the orbital data of the Sun about the center of the Galaxy. To do so,
assume that most of the mass is concentrated near the center of the
Galaxy and that the Sun and the solar system (of total mass $m$) move
in a circular orbit around the center of the Galaxy (of total mass
$M$), 
\begin{equation}
\frac{GMm}{r^2} = m \frac{v^2}{r}\,\, ,
\label{dmeq}
\end{equation}
where we recall that $a = v^2/r$ is the centripetal acceleration. All in all,
\begin{equation}
M = \frac{r\,v^2}{G} \approx 2 \times 10^{41}~{\rm kg}\, \,.
\end{equation}
Assuming all the stars in the Galaxy are similar to our Sun ($M_\odot
\approx 
2 \times 10^{30}~{\rm kg}$), we conclude that there are roughly
$10^{11}$ stars in the Galaxy.

In addition to stars both within and outside the Milky Way, we can see
with a telescope many faint cloudy patches in the sky which were once
all referred to as {\it nebulae} (Latin for clouds).  A few of these,
such as those in the constellations of Andromeda and Orion, can
actually be discerned with the naked eye on a clear night. In the XVII
and XVIII centuries, astronomers found that these objects were getting
in the way of the search for comets. In 1781, in order to provide a
convenient list of objects not to look at while hunting for comets,
Messier published a celebrated catalogue~\cite{Messier:1781}. Nowadays astronomers
still refer to the 103 objects in this catalog by their Messier
numbers, e.g., the Andromeda Nebula is M31.

Even in Messier's time it was clear that these extended objects are
not all the same. Some are star clusters, groups of stars which are so
numerous that they appeared to be a cloud. Others are glowing clouds
of gas or dust and it is for these that we now mainly reserve the word
nebula. Most fascinating are those that belong to a third category:
they often have fairly regular elliptical shapes and seem to be a
great distance beyond the Galaxy. Kant seems to have been the first to
suggest that these latter might be circular discs, but appear
elliptical because we see them at an angle, and are faint because they
are so distant~\cite{Kant:1755}. At first it was not universally
accepted that these objects were extragalactic (i.e. outside our
Galaxy). In 1920, Sir Hubble's observations revealed that individual
stars could be resolved within these extragalactic objects and that
many contain spiral arms~\cite{Hubble:b}.  The distance to our nearest
spiral galaxy, Andromeda, is over 2 million ly, a distance 20 times
greater than the diameter of our Galaxy. It seemed logical that these
nebulae must be galaxies similar to ours. Today it is thought that
there are roughly $4 \times 10^{10}$ galaxies in the observable
universe -- that is, as many galaxies as there are stars in the
Galaxy.

Sir Hubble also observed a persistent redshift in the spectra of
known elements and that the shift was greater the greater the distance
of the galaxy from the Earth. It was Hubble himself who explained the
redshift as indicating that distant galaxies were radially moving
away from the Earth~\cite{Hubble:1929}. In every direction, these vast
accumulations of stars and interstellar matter were moving outward at
enormous speeds. He called this motion, recession. He showed that the
velocity of recession was greater at greater distances. Hubble's law
of cosmic expansion states that an observer {\it at any point in the
  universe} will observe distant galaxies receding from him/her with
{\it radial velocities $V$ proportional to their distance $d$ from the
  observer},
\begin{equation}
V = H_0 \, d
\end{equation}
where $H_0$ is the Hubble's proportionality constant. Hubble's initial
determination of $H_0$ was approximately 160~km/s per
million-light-years. Most recent observations indicate that $H_0
\approx 22.4$~km/s per
million-light-years~\cite{Riess:2011yx,Freedman:2012ny,Ade:2015xua,Abbott:2017xzu}.

Hubble's law is consistent with a general expansion of the space
between galaxies (or galactic clusters), and is not a particular
characteristic of the galaxies (clusters) themselves. This statement
means that the galaxies themselves are not changing in any way; only
the regions between them are expanding with time. If the expansion is
run backward (as can be done with mathematics), then it would appear
that, very long ago, all the matter of the universe was once compacted
into a relatively small volume from which it was hurled outward by
some titanic force. This idea is the basis for the  hot {\it Big Bang} model~\cite{Gamow:1946eb,Alpher:1948ve,Alpher:1953zz,Peebles:1991ch}.\\

{\bf \S~Are we alone?}  {\it Of course} we are not alone! We now know
that our Sun is but one of ten billions of stars in the Milky Way, and our
Galaxy is but one of ten billions of galaxies in the universe. Over the
last decade, the search for life in the universe has been transformed
from speculation to a data-driven science.

It is well known that all organisms living on Earth require
carbon-based chemistry in liquid water. Therefore, according to the
hot Big Bang model, life (as we know it) could not have appeared
earlier than $t \sim 10~{\rm Myr}$ after the Bang, since the entire Universe was
bathed in a thermal radiation background above the boiling temperature
of liquid water. After a while ($10 \lesssim t/{\rm Myr} \lesssim
17$) though,
the Universe cooled down to habitable comfortable temperatures: $273
\lesssim T/^\circ{\rm K} \lesssim 373$~\cite{Loeb:2013fna}.  

Phase diagrams show the preferred physical states of matter at
different temperatures and pressure. Within each phase, the material
is uniform with respect to its chemical composition and physical
state. At typical temperatures and pressures enforced by the earth
atmosphere water is a liquid, but it becomes solid (that is, ice) if
its temperature is lowered below $0^\circ$C and gaseous (that is,
water vapor) if its temperature is raised above $100^\circ$C. This
means that water would remain liquid only under the external pressure
in an atmosphere, which can be confined gravitationally on the surface
of a planet. To keep the atmosphere restrained against {\it
  evaporation} depends upon the strong surface gravity of a rocky
planet, with a mass above about that of the Earth~\cite{Schaller}.

Life needs of stars for at least two reasons. We have seen that stars
are required to synthesize heavy elements (such as carbon, oxygen,
$\cdots$, iron) out of which rocky planets and the molecules of life
are made. We have also seen that sars maintain a source of heat to
power the chemistry of life on the surface of their planets.

Each star is surrounded by an {\it habitable zone}. Since water is essential
for life as we know it, the search for biosignature gases naturally
focuses on planets located in the {\it habitable zone} of their host stars,
which is defined as the orbital range around the star within which surface liquid
water could be sustained.  To understand this we need to take a quick
side trip into how one estimates temperature. The total energy flux
$\mathscr{F}$ (energy per unit area per unit time) passing through a
region can be related to the effective temperature $T$ according to
\begin{equation}
\mathscr{F} = \sigma_{\rm SB} \ T^4 \,,
\label{SB1}
\end{equation}
where $\sigma_{\rm SB} \approx 5.67 \times 10^{-8}$~watt per meter
squared per kelvin to the fourth is the Stefan-Boltzmann
constant~\cite{Stefan:1879,Boltzmann:1884}. We have seen in Sec.~\ref{sec:7}
that the luminosity (energy per unit time) of a star is $L$, and that
$L$ and  the flux at a distance $r$ from the star are related by
\begin{equation}
\mathscr{F} = \frac{L}{4 \pi r^2} \,, 
\label{SB2}
\end{equation}
because the area of a sphere of radius $r$ is $A = 4\pi r^2$ and the
flux is the luminosity divided by the area. A quick estimate of the
effective temperature at a given radius from a star can proceed by combining (\ref{SB1})
and (\ref{SB2}), yielding
\begin{equation} 
\sigma_{\rm SB} \ T^4 = {\mathscr F} = \frac{L}{4 \pi r^2} \, .
\label{SB3}
\end{equation}
When we consider our solar system, then not only $\sigma_{\rm
  SB}$ and $4\pi$, but also the luminosity $L_\odot$, are constants with
distance. This tells us that
\begin{equation}
T^4 \propto \frac{1}{r^2} \Rightarrow T \propto r^{-1/2} \, .
\label{SB4}
\end{equation}
Therefore, if we calculate the effective temperature at any radius
(say, one astronomical unit = 1~AU = earth-sun distance), we can use
the proportionality in (\ref{SB4}) to calculate the temperature at any other
radius. For example, if the temperature is $300^\circ$K at 1~AU, then four
times farther away the temperature is $4^{-1/2} = 1/2$ times as low,
or $150^\circ$K. Similarly, the radius where the temperature is $600^\circ$K would
be given by $(600/300)^{- 2}  \times 1~{\rm AU}= 0.25~{\rm AU}$.

The habitable zone of the solar system looks like a ring (with
habitable  temperatures $273 \lesssim T/^\circ{\rm K}
\lesssim 373$) around the Sun. Rocky planets with an orbit within this
ring may have liquid water to support life. The habitable zone around
any  other star in the cosmos looks similar to the habitable zone in our
Solar System. The only difference is the size of the ring. If the star
is bigger than the Sun it has a wider zone, if the star is smaller it
has a narrower zone. It might seem that the bigger the star the
better. However, the biggest stars have relatively short lifespans, so
the life around them probably would not have enough time to
evolve. The habitable zones of small stars face a different
problem. Besides being narrow they are relatively close to the star. A
hypothetical planet in such a region would be tidally locked. That
means that one half of it would always face the star and be extremely
hot, while the opposite side would always be facing away and
freezing. Such conditions are not very favorable for
life~\cite{Seager}.

As of today, we only know of life on Earth. The Sun formed about 4.6
Gyr ago and has a lifetime comparable to the current age of the
Universe. Tiny zircons (zirconium silicate crystals) found in ancient
stream deposits indicate that Earth developed continents and water --
perhaps even oceans and environments in which microbial life could
emerge -- 4.3 billion to 4.4 billion years ago, remarkably soon after
our planet formed~\cite{Mojzsis,Dodd}. The presence of water on the young Earth was
confirmed when the zircons were analyzed for oxygen isotopes and the
telltale signature of rocks that have been touched by water was found:
an elevated ratio of oxygen-18 to oxygen-16. If we insist that life
near the Sun is typical and not premature, we can assume that the
evolution of an advanced civilization requires approximately 4~Gyr~\cite{Loeb:2016vps}.

The Fermi paradox is the discrepancy between the strong likelihood of
alien intelligent life emerging (under a wide variety of assumptions)
and the absence of any visible evidence for such emergence~\cite{Jones}. By
adopting the starting point of a first approximation to the answer of
this intriguing unlikeness, we can write the number of intelligent
civilizations in our galaxy at any given time capable of releasing
detectable signals of their existence into space using a quite simple
functional form,
\begin{equation}
 N = R_\star \ f_{\rm p}  \ n_{\rm e}  \ f_{\ell} \ f_{\rm i} \ f_{\rm c} \ L_\tau \,,
\label{Drake}
\end{equation}
where $R_\star$ is the average rate of star formation, $f_{\rm p}$ is
the fraction of stars with planetary systems, $n_{\rm e}$ is the
number of planets (per solar system) with a long-lasting ($\sim 4~{\rm
  Gyr}$) ecoshell, $f_\ell$ is the fraction of suitable planets on
which life actually appears, $f_{\rm i}$ is the fraction of living
species that develop intelligence, $f_{\rm c}$ is the fraction of
intelligent species with communications technology, and $L_\tau$ is
the length of time such civilizations release detectable signals into
space (i.e. the lifetime of the {\it communicative
  phase})~\cite{Drake}. We can further separate (\ref{Drake}) into its
astrophysical and biotechnological factors
\begin{equation}
N = \langle \zeta_{\rm astro} \rangle \, \langle \xi_{\rm biotec} \rangle \, L_\tau \, ,
\end{equation}
where $\langle \zeta_{\rm astro} \rangle = R_\star \, f_{\rm p} \,
n_{\rm e}$ represents the production rate of habitable planets with
long-lasting ecoshell (determined through astrophysics) and $\langle
\xi_{\rm biotec} \rangle = f_\ell \, f_{\rm i} \, f_{\rm c}$
represents the product of all chemical, biological and technological
factors leading to the development of a technological
civilization. $\langle \cdots \rangle$ indicates average over all the
multiple manners civilizations can arise, grow, and develop such
technology, starting at any time since the formation of our Galaxy in
any location inside it. This averaging procedure must be regarded as a
crude approximation because the characteristics of the initial
conditions in a planet and its surroundings may affect $f_\ell$,
$f_{\rm i}$, and $f_{\rm c}$ with high complexity. 

The star formation rate in the Galaxy is estimated to be $3
  \lesssim R_\star/{\rm yr}^{-1} \lesssim
  12$~\cite{Kennicutt:2012ea}. Now, only 10\% of these stars are
  appropriate for harboring habitable planets. This is because the
  mass of the star $M_\star < 1.1 M_\odot$ to be sufficiently
  long-lived and $M_\star > 0.7 M_\odot$ to possess circumstellar
  habitable zones outside the tidally locked region. The frequency
  $\eta_\oplus$ of terrestrial planets in and the habitable zone of
  solar-type stars can be determined using data from the {\it Kepler}
  mission. Current estimates suggest $0.15^{+0.13}_{-0.06} <
  \eta_\oplus < 0.61^{+0.07}_{-0.15}$~\cite{Dressing:2013mid,Kopparapu:2013xpa}. An
  Estimate of the rate of planetary catastrophes that could threaten
  the evolution of life on the surface of these worlds suggests that
  the survival probability of the planet's atmosphere for about 4~Gyr
  is roughly 5\%~\cite{Anchordoqui:2017eog}. The production rate of
  habitable planets with long-lasting ecoshell is then
\begin{equation}
\langle \zeta_{\rm astro} \rangle \sim 3~{\rm yr}^{-1} \times 0.1 \times 0.15 \times
0.05 = 0.002~{\rm yr}^{-1} \, .
\end{equation}
Now, since $\langle \xi_{\rm biotec} \rangle < 1$, it is easily
seen that if the {\it communicative phase} is smaller than 500 years there
would be no paradox. Then, if we assume again that {\it humans}
typify the population of {\it advanced civilizations} in the Galaxy we
would not have yet expected any contact with intelligent beings from
extraterrestrial worlds, but we are on the verge of
discovering evidence of alien life in the cosmos.

\section{Fifty Questions to Ponder}

{\bf 1.}~If a household consumes 300~kWh of electrical energy in a 30 day
month, what is the average actual power usage, assuming utilization
for 16 hour/day?\\

{\bf 2.}~During a normal day of activities, the average person
consumes about $2,200~{\rm Cal}$ (2.5~kWh). {\it (i)}~If 300~Cal are released by
burning 1~oz fat, how much weight (not including initial water loss)
will a person lose by going on a starvation diet for 2 weeks? {\it (ii)}~How long
would it take to lose 15~lb by reducing intake to $1,600~{\rm Cal}$?\\

{\bf 3.}~A typical Detroit automobile averages about 15
miles/gallon in the city. If the energy content of gasoline is about
32,000 Cal/gal (126,944 Btu/gal), compare the energy used in driving a
mile to that used in walking a mile (about 20 Cal), and to that used
as electrical
energy (per day) for an average household (about 10 kWh). \\

{\bf 4.}~For an astronaut sealed inside a space suit, getting rid of
body heat can be difficult. Suppose an astronaut is performing
vigorous physical activity, expending 200 watts of power. An energy of
47.8~Cal is enough to raise the body temperature by $1^\circ$C. If none
of the heat can escape from the space suit, how long will it take
before the body temperature rises by $6^\circ$C ($11^\circ$F), an
amount sufficient to kill.  Express your answer in units of minutes.\\

 {\bf 5.}~The cost of home-heating fuel oil is about
 40\cent/gallon. If the cost of heating a certain home during the
 winer is about \$60/month, calculate the cost of suppling electric
 heat to the same home. Assume the thermal energy content of fuel oil
 to be 140,000 Btu/gal, and the cost of electricity to be 12\cent/kWh.\\

 {\bf 6.}~In 2016, the U.S. imported crude oil at the rate of
 7.8~Mbbl/d (7.8~million barrels/day), at a price of about \$50/bbl. {\it
   (i)}~What was the annual cost to a family of 4 for imported oil?
 (Take the number of such families to be about $1/4$ of the U.S. population, which is about  $322 \times 10^6$). {\it
   (ii)}~The price of gasoline at the pump, \$2.5 per
 gallon, is approximately given by 2.2 times the price of a gallon of
 domestic crude oil, plus taxes. As of 2017, taxes on gasoline amount
 to 18.4\cent{} per gallon. If taxes remain unchanged and the price
 for domestic crude were to rise to present import prices, what would
 be the price of gasoline at the pump? Do you think this price would
 appreciably influence American driving habits?  [{\it Hint:} 1~bbl = 42~gal.] \\

\begin{figure}[t]
    \postscript{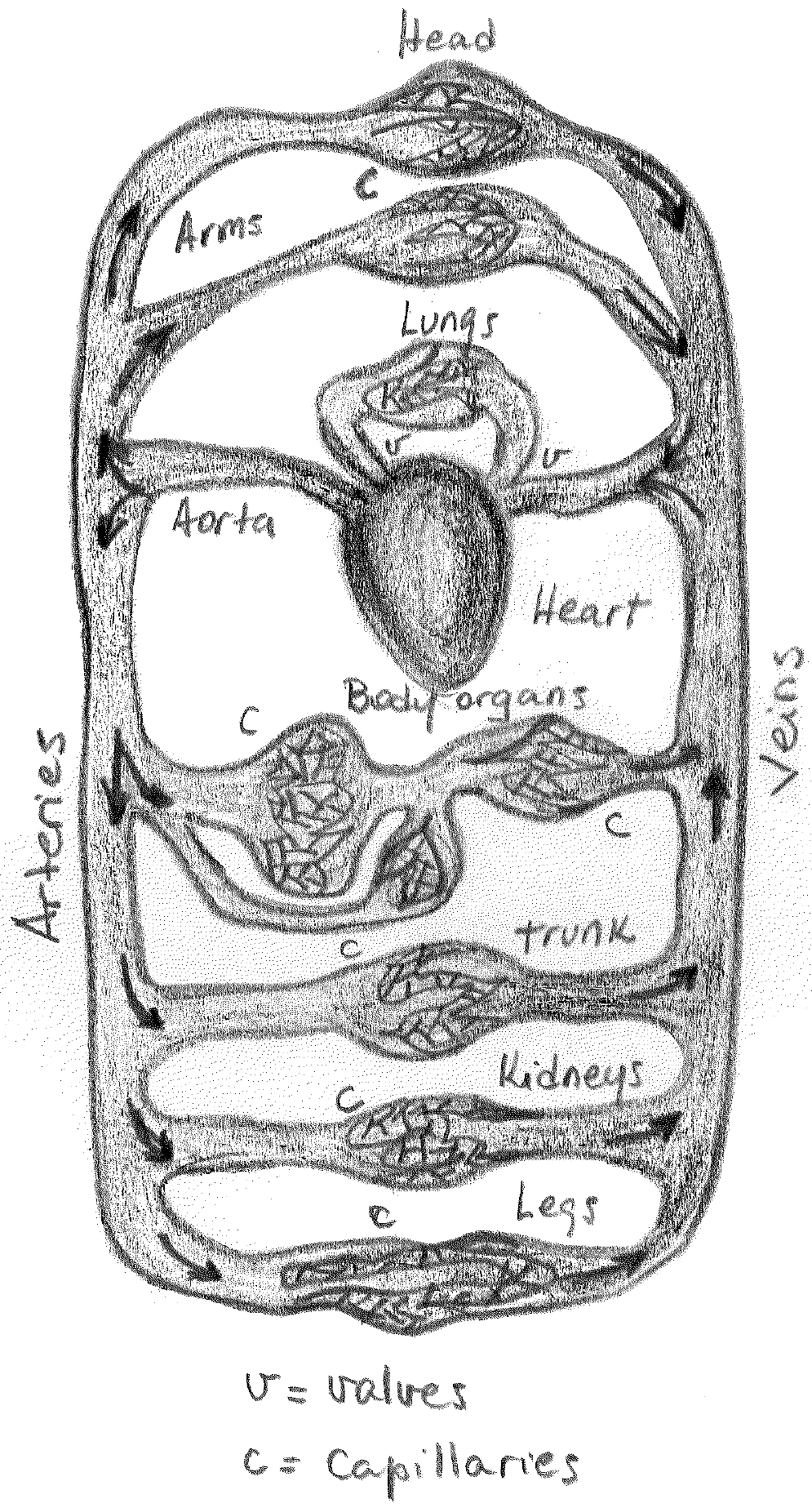}{0.9}
\caption{The situation in question 13.
\label{fig:44}}
\end{figure}

 {\bf 7.}~The rocket shown in Fig.~\ref{fig:44} has an acceleration of
 30~m/s$^2$ until its engines burnout at 100~s. After ignition the
 rocket will coast and slowly lose speed: {\it (i)}~what is its
 instantaneous speed at burnout? {\it (ii)}~what is its 
 average speed during the burnout phase ? {\it (iii)}~what is the altitude at burnout? {\it   (iv)}~what is the altitude at the top of its trajectory. Assume the
 acceleration of gravity is constant, $g = 9.8~{\rm m/s}^2$, up to the
 maximum height.\\

\begin{figure}[t]
    \postscript{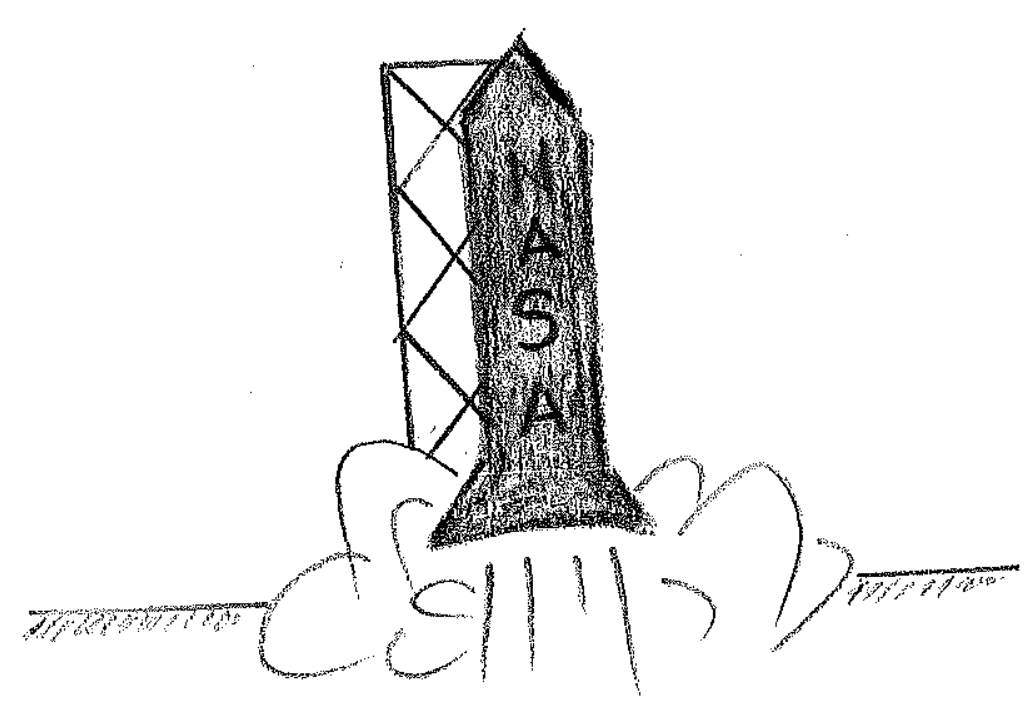}{0.9}
\caption{The situation in question 7.
\label{fig:43}}
\end{figure}

{\bf 8.}~The unnecessary use of electric lights in the home is often
cited as source of energy wastage. Suppose each person in the U.S. is
responsible for the unnecessary use of a 100 watt (0.1 kW) light bulb
for 1 hour each day. {\it (i)}~What is the amount of wasted electrical
energy each day? Give your answer in kWh. At 12\cent/kWh, what is the
added electrical cost per year to a family of 4? Does the saving of
this money provide sufficient incentive to ``turn off lights when not
in use''? {\it (iii)}~How many kWh electric are expended in the
U.S. homes each year on unnecessary lighting? {\it (iv)}~If each kWh
electric demands 3 kWh of thermal heat energy, how many barrels of
oil/day are burn in power plants to provide for this wasted energy?
($1~{\rm bbl} \, {\rm oil} \to 1,700~{\rm kWh}$). Is it worth thinking
about from a societal point of
view?\\

{\bf 9.}~Every few hundred years most of the planets line up on the
same side of the Sun. Calculate the total force on the Earth due to
Venus, Jupiter, and Saturn, assuming all four planets are in a
line. The masses are $M_V = 0.815M_\oplus$, $M_J = 318 M_\oplus$, $M_S
= 95.1M_\oplus$, and their mean distances from the Sun are $r_{\rm SV}
= 108$~million km, $r_{\rm SE} = 150$~million km, $r_{\rm SJ} =
  778$~million km, and $r_{\rm SS} = 1,430$~million km. What fraction
  of the Sun's force on the Earth is this.\\

{\bf 10.}~Given that the acceleration of gravity at the surface of
Mars is 0.38 of what it is on Earth, and that Mars radius is 3400~km,
determine the mass of Mars.\\

{\bf 11.}~At a depth of 10.9~km, the Challenger Deep (in the Marianas
Trench of the Pacific Ocean) is the deepest known site in any
ocean. Yet, in 1960, Walsh and Piccard reached the Challenger Deep in
the bathyscaphe {\it Trieste}. Assuming that seawater has a uniform
density of $1,025~{\rm kg/m}^3$, calculate the force the water would
exert in the ocean floor on a Plexiglas viewing round observation
window of 24~cm diameter.\\

{\bf 12.}~Only a small part of an iceberg protrudes above the water,
while the bulk lies below the surface. The density of ice is $917~{\rm
  kg/m}^3$ and that of seawater is $1,025~{\rm kg/m}^3$. Find the percentage of the iceberg's volume that lies below the surface.\\

{\bf 13.}~In humans, blood flows from the heart into the aorta, from
which it passes into the major arteries. These branch into small
arteries (arterioles), which in turn branch into myriads of tiny
capillaries. The blood returns to the heart via the veins. For an
illustration,  see Fig.~\ref{fig:44}. The radius of
the aorta is about 1.2~cm, and the blood passing through it has a
speed of about 40~cm/s. A typical capilar has a radius of about $4 \times
10^{-4}~{\rm cm}$, and blood flows through it at speed of about 
$5 \times 10^{-4}~{\rm m/s}$. Estimate the number of capillaries that are in the body.\\

{\bf 14.}~Acrobat Sally of mass $m_{\rm S}$ stands on the left end of
a seesaw. Acrobat Harry of mass $m_{\rm H}$ jumps from a height
$h_{\rm H}$ onto the right end of the seesaw, thus propelling Sally
into the air. {\it (i)}~Neglecting inefficiencies (that transform energy
into heat),, how does the
potential energy of Sally at the top of his trajectory compares with
the potential energy of Harry just before he jumps? {\it (ii)}~Shows
that ideally Sally reaches a height $m_{\rm H} h_{\rm H}/m_{\rm S}$. {\it (iii)}~If
Sally's mass is 88~lb, Harry's mass 150~lb, and the height of the
initial jump was 4~m, show
that Sally rises a vertical distance of 7~m.\\

{\bf 15.}~Harry and Sally are at a ski resort with two ski runs, a
beginner's run and an expert's run. Both runs begin at the top of the
ski lift and end at finish line at the bottom of the same lift. Let $h$
be the vertical descent for both runs. The beginner's run is longer
and less steep than the expert's run. Harry and Sally, who is a
much better skier than him, are testing some experimental frictionless
skis. To make things interesting, Harry offers a wager that if she takes
the expert's run and he takes the beginner's run, her speed at the
finish line will not be greater than his speed at the finish
line. Forgetting that Harry study physics, Sally accepts the bet. The
conditions are that both Harry and Sally start from rest at the top of the lift
and they both  coast for the entire trip (i.e., there is no external
work done on the system). Who wins the bet? (Assume
air drag, which may dissipate energy via heat, is negligible).\\

{\bf 16.}~The escape velocity is the minimum speed needed for an object to
``break free'' from the gravitational attraction of a massive body.
More particularly, the escape velocity is the speed at which the sum
of an object's kinetic energy and its gravitational potential energy
is equal to zero.  Show that the escape velocity from Earth is about
25,020 miles per hour.\\

{\bf 17.}~The energy gained by a person climbing to the top of Mt. Washington
is about 300 calories (0.35~kWh). This also happens to be the amount
of energy released by burning about 1 ounce of oil.  {\it (i)}~If for every calorie
of useful work done the body expends 5 calories in heat, what is the
 weight loss resulting from the climb? {\it (ii)}~If you were
to perform this climb, would you expect to regain weight during the
trip down? Why not?\\

{\bf 18.}~A car engine derives its thermal energy from a ``reservoir''
of burning fuel whose average temperature is about $4,000^\circ$F. The
release of the exhaust takes place at a ``cold'' temperature of about
$180^\circ$F. What is the theoretical efficiency of the car engine?
Note that in the absolute scale for Fahrenheit degrees, we use Rankine
($^\circ$R) temperatures. Zero on both the Kelvin and Rankine scales
is absolute zero, but the Rankine degree is defined as equal to one
Fahrenheit degree, rather than the Celsius degree used on the Kelvin
scale. A temperature of $-459.67^\circ$F is exactly equal to $0^\circ$R.\\

{\bf 19.}~Biological energy. {\it (i)} Suppose a person were entirely made of
water (not a bad approximation).  If all the heat generated by the
body during the day ($3,000$~Cal = $12,000$~Btu) were used to warm up
the body, what would be the body temperature at the end of the day for
a person weighing 120~lb? Assume an initial temperature of
$98.6^\circ$F. {\it (ii)}~If  $1,600$~Cal were used to work at 25\%
efficiency, how high could a 120~lb person climb in the day? (1 Btu =
800~ft lb.)\\

{\bf 20.}  It is well known that a heavy rhinoceros is harder to stop
than a small dog at the same speed. We state this fact by saying that
the rhino has more {\it momentum} than the dog. And if two dogs have
the same mass, the faster one is harder to stop than the slower
one. So we also say that the faster moving dog has more momentum than
the slower one. By {\it momentum} we mean the product of the mass of
an object and its velocity
\begin{equation}
{\rm momentum} = {\rm mass} \times {\rm velocity} \, .
\end{equation}
When the direction is not an important factor , we can say
\begin{equation}
{\rm momentum} = {\rm mass} \times {\rm speed} \, .
\end{equation}
Conservation of momentum is a fundamental law of physics, which states
that the momentum of a system is constant if there are no external
forces acting on the system. Momentum conservation is especially
useful for collisions.\footnote{A collision refers to two objects
  hitting one another, interacting with (probably very large) forces
  for some (probably very short) amount of time, and then continuing
  along (probably in radically altered paths, and maybe pretty
  squashed!) In a perfectly {\it elastic} collision the kinetic
  energy {\it is} conserved (no energy is lost to the surroundings or
  participants), whereas in an {\it inelastic} collision the kinetic
  energy {\it is not} conserved (some kinetic energy is converted to
  heat, or sound, or deformation).} The forces involved may be so
complicated that you cannot practically use {\it Newton's second law of
motion} to figure out what happens. And, since some energy may go into
heat, sound, or deformation in a collision, energy conservation may
also not be useful. But as long as there are no {\it outside} forces
involved (only internal ones, no matter how big!) then momentum
conservation often lets you figure out the outcomes! For example, a
crater in Arizona is thought to have been formed by the impact of a
meteorite with the Earth over 20,000~years ago. The mass of the
meteorite is estimated at $5 \times 10^{10}~{\rm kg}$ and its speed
$7,200~{\rm m/s}$.  {\it (i)}~Judging from a frame of reference in
which the Earth is initially at rest, what speed would such a meteor
impart to the Earth in a head-on collision? Assume the pieces of the
shattered meteor stayed with the earth as it moved. {\it (ii)}~What
fraction of meteor's kinetic energy was transformed to kinetic energy
of Earth?  {\it (iii)}~By how much did Earth kinetic energy change as
a result of this collision?
The mass of the Earth is $M_\oplus = 6 \times 10^{24}~{\rm kg}$.\\

{\bf 21.}~{\it (i)}~If your heart is beating at 76.0 beats per minute,
what is the frequency of your heart's oscillations in hertz? {\it
  (i)}~What is the oscillating period of your heart when the frequency
increases by a factor of 1.3?  {\it (iii)}~A sewing machine needle
moves in simple harmonic motion with a frequency of 2.5~Hz and an
amplitude of 1.27~cm;  how long does it take the tip of the needle
to move from the highest point to the lowest point in its travel?
{\it (iv)}~How long does it take the needle tip to travel a total distance of
11.43 cm?\\

{\bf 22.}~Astronaut Harry has landed on Pluto and conducts an
experiment to determine the acceleration due to gravity on the dwarf
planet. He uses a simple pendulum that is 0.640~m long and measures 10
complete oscillations in 63.8~s. What is the acceleration of gravity
on Pluto?\\

{\bf 23.}~If you drop a stone into a mineshaft that is $\ell = 122.5~{\rm m}$ deep, how soon after you drop the stone do you hear it hit the bottom of the shaft? The temperature in the mineshaft is $10^\circ$C.\\

\begin{figure}[t]
    \postscript{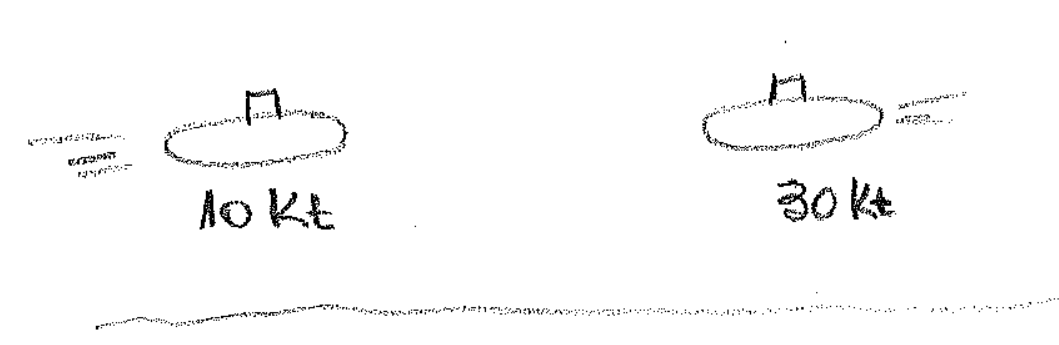}{0.9}
\caption{The situation in question 24.
\label{fig:45}}
\end{figure}

{\bf 24.} Two submarines are underwater and approaching each other
head-on as shown in Fig.~\ref{fig:45}. Sub $A$ has a speed of 10~knots
and sub $B$ has a speed of 30~knots. Commander Harry on Sub $A$ is
pinging on $B$ with an active sonar frequency of 10~kHz.  The speed of
sound underwater is $2,912~{\rm knots}$, that is roughly 4.3 times as
fast as in air.  {\it (i)}~What frequency will receive Sally on
submarine $B$ from Harry's
sonar?
{\it (ii)}~What is the frequency of the echo Harry receives from
Sally's submarine?\\

{\bf 25.} The security alarm on a parked car goes off and produces a
frequency of $960$~Hz. The speed of sound is 343~m/s. As you drive
toward this parked car, pass it, and drive away, you observe the
frequency to change by 95~Hz. At what speed are you driving?\\

{\bf 26.} {\it (i)}~Compare the electric force holding the electron in
orbit ($r=0.53 \times 10^{-10}~{\rm m}$) around the proton nucleus of
the hydrogen atom, with the gravitational force between the same
electron and proton. What is the ratio of these two forces? {\it
  (ii)}~Would life be different if the electron were positively
charged and the proton were negatively charged? Does the choice of
signs have any
bearing on physical and chemical interactions? \\

{\bf 27.}~A typical 1.5 volt flashlight battery can deliver a current
of 1 ampere for 1 hour. {\it (i)}~What is the power (in watts) when
the current is 1 ampere? {\it (ii)}~What is the total energy (in kWh)
delivered by the battery in 1~hour. {\it (ii)}~If the cell costs
50\cent, what is the cost for 1~kWh of this form of chemical energy?\\

{\bf 28.}~Kerosene has a fuel value of $1,400$~Btu/oz. At what rate
(i.e., how many oz/hr) must it be burned in order to give off as much
heat as a $1,000$~watt electric heater?\\

{\bf 29.}~An air conditioner operating on a 110~volt line is rated at
750~watts. {\it (i)}~What is the current (in amperes) drawn by this appliance?
{\it (ii)}~At 12\cent/kWh, what is the cost of running the air
conditioner for 8 hours?\\

\begin{figure}[t]
    \postscript{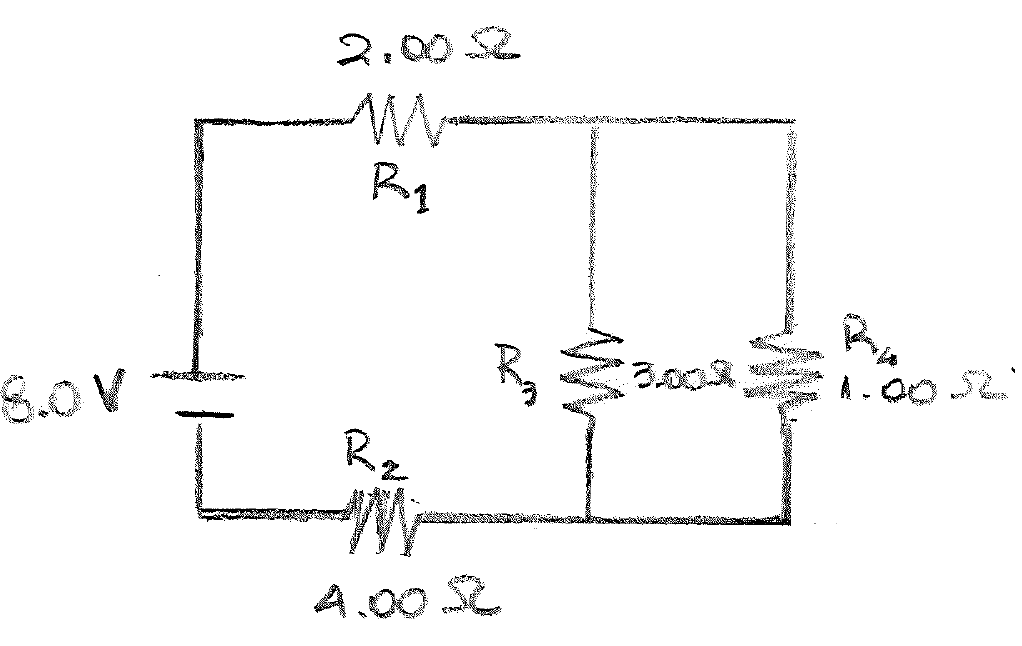}{0.9}
\caption{The situation in question 30.
\label{fig:46}}
\end{figure}

{\bf 30.}~Calculate the power delivered to each resistor in the
circuit shown in Fig.~\ref{fig:46}.\\

{\bf 31.}~{\it (i)}~The distance to the North Star, Polaris, is
approximately $6.44 \times 10^{18}~{\rm m}$. If Polaris were to burn
out today, in what year would we see it disappear? {\it (ii)}~How long
does it take for sunlight to reach the Earth? {\it (iii)}~How long
does it take for a microwave radar signal to travel from the Earth to
the Moon and back? {\it (iv)}~How long does it take for a radio wave
to travel once around the Earth in a great circle, close to the
planet's surface?  {\it (v)} How long does it take for light to reach
you from a lightning stroke 10~km away?\\

\begin{figure}[t]
    \postscript{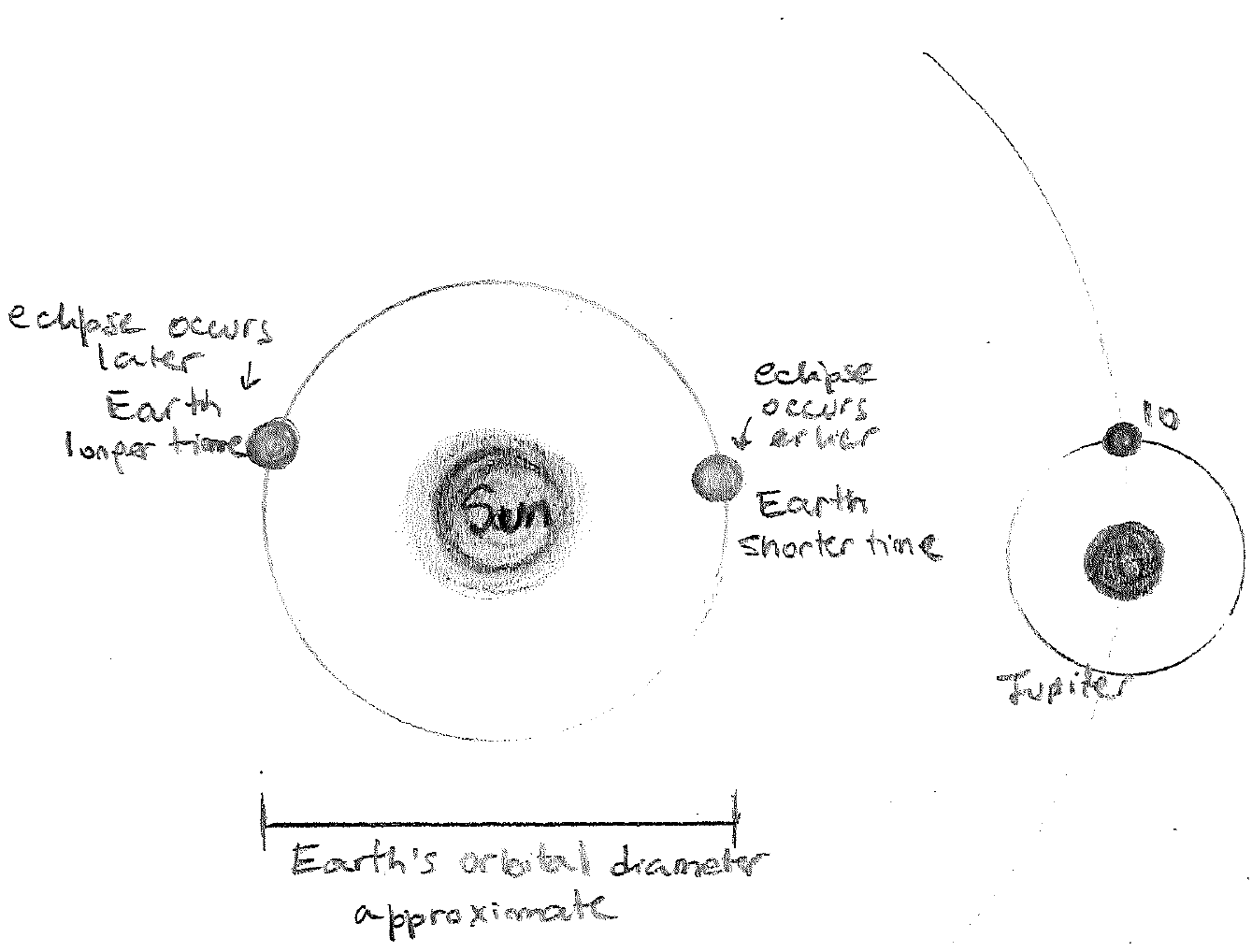}{0.9}
\caption{Earth's orbital diameter causes the eclipse of Io to occur at
  different times because of the extra distance the light must travel
  when Earth is farthest from Jupiter.
\label{fig:47}}
\end{figure}

  {\bf 32.} {\it (i)}~As a result of his observations, Roemer concluded that
  eclipses of Io by Jupiter were delayed by 22 min during a 6 month
  period as the Earth moved from the point in its orbit where it is
  closest to Jupiter to the diametrically opposite point where it is
  farthest from Jupiter~\cite{Roemer}. Using $1.5 \times 10^8~{\rm
    km}$ as the average radius of the Earth’s orbit around the Sun,
  calculate the speed of light from these data; see
  Fig.~\ref{fig:47}. {\it (ii)}~The Apollo 11 astronauts set up a
  panel of efficient corner-cube retroreflectors on the Moon's
  surface. The speed of light can be found by measuring the time
  interval required for a laser beam to travel from Earth, reflect
  from the panel, and return to Earth. If this interval is measured to
  be 2.51~s, what is the measured speed of light? Take the
  center-to-center distance from Earth to Moon to be $3.84 \times
  10^{8}$~m, and do not ignore the sizes of the Earth and Moon. The
  Earth radius is $R_\oplus = 3,959$~miles and $R_{\rm moon} = 1,079$~miles.\\

\begin{figure}[t]
    \postscript{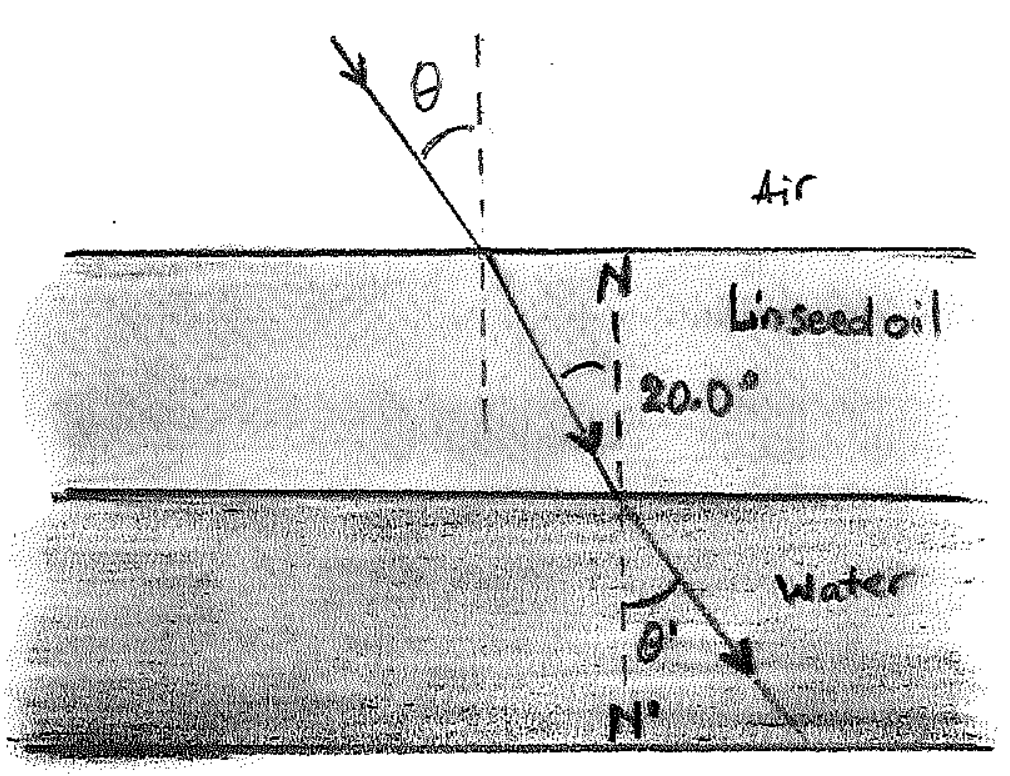}{0.9}
\caption{The situation in question 33.
\label{fig:48}}
\end{figure}

{\bf 33.}~The light beam shown in Fig.~\ref{fig:48} makes an angle of
$20.0^\circ$ with the normal line $NN'$ in the linseed oil. Determine
the angles $\theta$ and $\theta'$. (The index of refraction of air is
$1.00029$, the one of water is $1.33$, and that of linseed oil is
$1.48$.)\\

\begin{figure}[t]
    \postscript{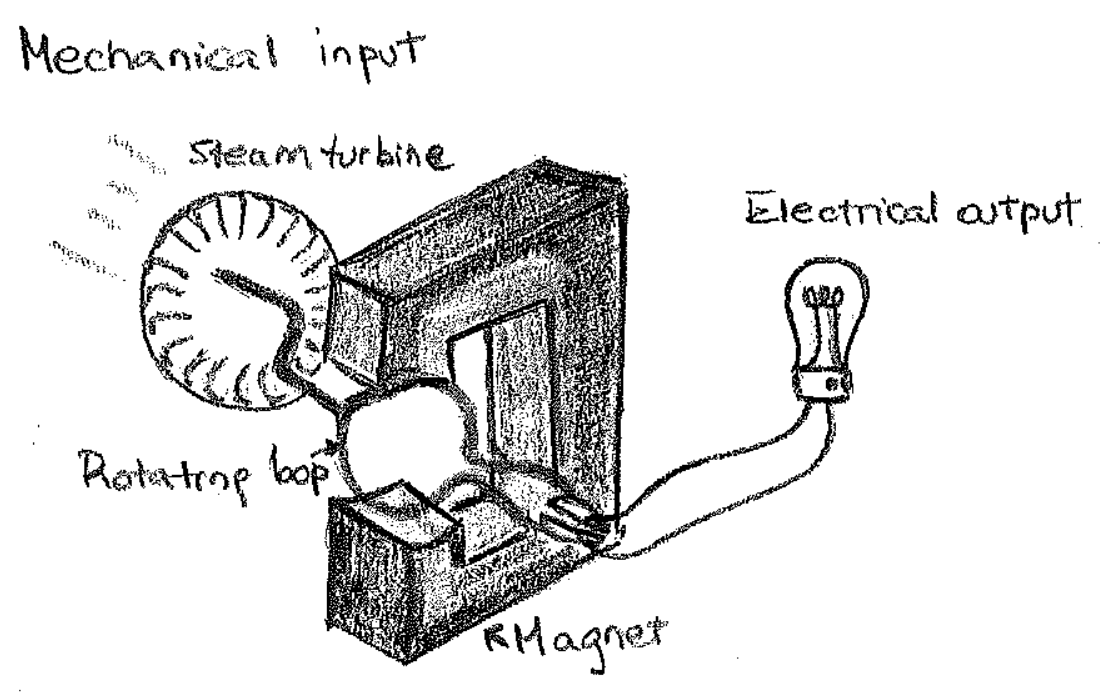}{0.9}
\caption{A simple generator. Voltage is induced in the loop when it is
  rotated in the constant magnetic field. Steam drives the turbine,
  which provides the mechanical input.
\label{fig:49}}
\end{figure}

{\bf 34.}~A (thin) converging lens of focal length 10~cm forms images
of objects placed at {\it (i)}~30~cm, {\it (ii)}~10~cm, {\it
  (iii)}~5~cm from the lens. In each case, find the image distance and
describe the image characteristics. A (thin) diverging lens of focal length 10~cm forms images
of objects placed at {\it (iv)}~30~cm, {\it (v)}~10~cm, {\it
  (vi)}~5~cm from the lens. Repeat the calculations to find the image distance and
describe the image.\\

{\bf 35.}~Faraday's law states that the induced voltage in a coil is
proportional to the product of its number of loops, the
cross-sectional area of each loop, and the rate at which the magnetic
field lines change within those loops~\cite{Faraday}. For example,
when one end of a magnet is repeatedly plunged into and back out of a
coil of wire, the direction of the induced voltage alternates. As the
magnetic field strength inside the coil is increased (as the magnet
enters the coil), the induced voltage in the coil is directed one
way. When the magnetic field strength diminishes (as the magnet leaves
the coil), the voltage is induced in the opposite direction. The
frequency $\nu$ of the alternating voltage that is induced equals the
frequency of the changing magnetic field within the loops. It is more
practical to induce voltage by moving a coil than by moving a
magnet. This can be done by rotating the coil in a stationary magnetic
field, as shown in Fig.~\ref{fig:49}. This arrangement is called a
generator. The voltage induced in the loop of area $A$ when it is
rotated in the magnetic field $B$ varies with time $t$ according to
\begin{equation}
V = B \ A \ \omega \ \, \sin (\omega t) \,,
\end{equation}
where $\omega = 2 \pi \nu$ is known as the angular
frequency.\footnote{Actually, the modern view of electromagnetism
  states that the induction is at the more basic level of electric and
  magnetic fields, which are at the root of both voltages and
  currents. Hence, induction occurs whether or not a conducting wire
  or any material medium is present. In this more general sense,
  Faraday's law states that {\it an electric field is induced in any
    region of space in which a magnetic field is changing with
    time}. As we noted at the beginning of Sec.~\ref{sec6}, there is a
  second effect in which the roles of electric and magnetic fields are
  interchanged. Maxwell's counterpart to Faraday's law states that {\it
    a magnetic field is induced in any region of space in which an
    electric field is changing with time~\cite{Maxwell:1865zz}.}}
Assume the circular loop of Fig.~\ref{fig:49} (with radius of $r =
0.25~{\rm meters}$) is rotated about the axis at a constant rate of
120 revolution per minute in a uniform magnetic field that has a
magnitude of 1.3~Tesla. {\it (i)}~Calculate the maximum value of the
emf induced in the loop. {\it (ii)}~Determine the times for which the
voltage will be at the maximum. {\it (iii)}~This is another example of
periodic motion. What is the period?\\

{\bf 36.}~{\it (i)}~How many moles of copper atoms are in a copper
penny, which has a mass $\simeq 3.1~{\rm g}$?  {\it (ii)}~How many
copper atoms are in the penny? {\it (iii)}~What is the mass in grams
of $10^{12}$ (a trillion) gold atoms?  {\it (iv)}~How many moles of
carbon atoms and oxygen atoms are in 0.25~mol of CO$_2$ (or carbon
dioxide)? {\it (v)}~If $5 \times 10^9$~bbl oil are burn in the
U.S./yr, and each barrel has a mass of about 300~lb, how many pounds
of matter
disappear/yr due to oil consumption in the U.S.?\\

{\bf 37.}~Feynman pointed out that if two persons stood at arm's
length from each other and each person had 1\% more electrons than
protons, the force of repulsion between the two people would be enough
to lift a ``weight'' equal to that of the entire Earth~\cite{Feynman:1963uxa}. Carry out an
order-of-magnitude calculation to substantiate this assertion.\\

{\bf 38.} If 1 gram of carbon extracted from the soot on a cave wall
is 40\% as radioactive as 1~gram of carbon extracted from a living
tree, estimate the age of the soot. Before proceeding you must  convince yourself  that
$0.5^{1.3} = 0.4$.\\

{\bf 39.} The average solar energy falling on each square meter of a
roof top is about 700~watts (= 0.7~kW). If a house has 10~m$^2$ of
collector, how long would it take to bring 60 gallons of water (1
bathtub-full) to $150^\circ$F from an initial temperature of
$50^\circ$F. Remember: 1 kW = 1kWh/hour = 3,412 Btu/hour, 1 gallon of
H$_2$O = 8 lb, and to heat 1~lb of H$_2$O $1^\circ$F takes 1~Btu.\\

\begin{figure}[t]
    \postscript{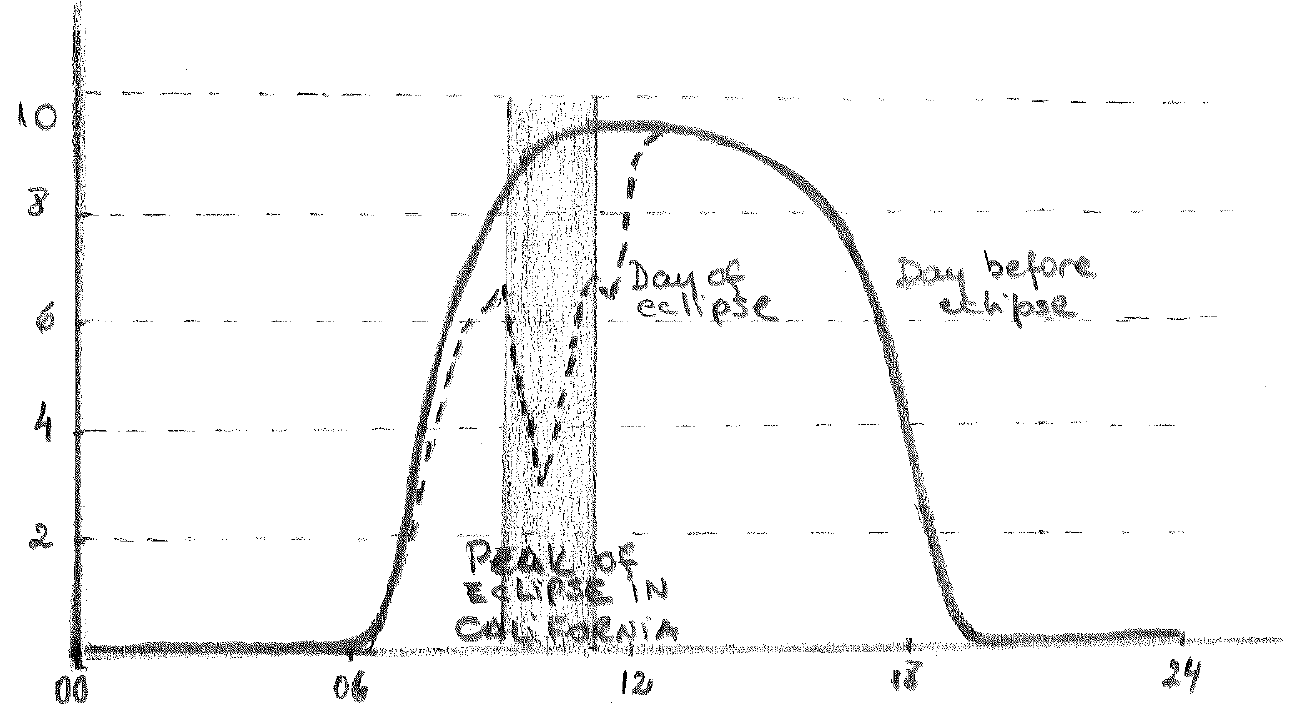}{0.9}
\caption{The California solar power output dropped during 21st August
  2017  eclipse. The vertical axis shows the power in units of $1,000$~MW and the
  horizontal axis indicates the Pacific daylight time.
  \label{fig:50}}
\end{figure}

{\bf 40.}~On the 21st August 2017, a giant shadow moving west to east
temporarily removed a large amount of the photovoltaic resources from the
U.S. This was the first total solar eclipse to darken the skies of the
country in a generation, and forced utilities to draw up contingency
plans for an electric grid increasingly powered by the Sun. The
previous total solar eclipse crossed the U.S. in 1979, when president
Jimmy Carter bemoaned an energy crisis and renewable technology was in
its infancy. The state of California lost on average about $3,400~{\rm MW}$ of
output during the event, a big chunk of the 10,000~MW of solar power that currently provides one-tenth of the state’s electricity;  see
Fig.~\ref{fig:50}. At 12\cent/kWh, what was the total electrical cost during the
eclipse?\\

\begin{figure}[t]
    \postscript{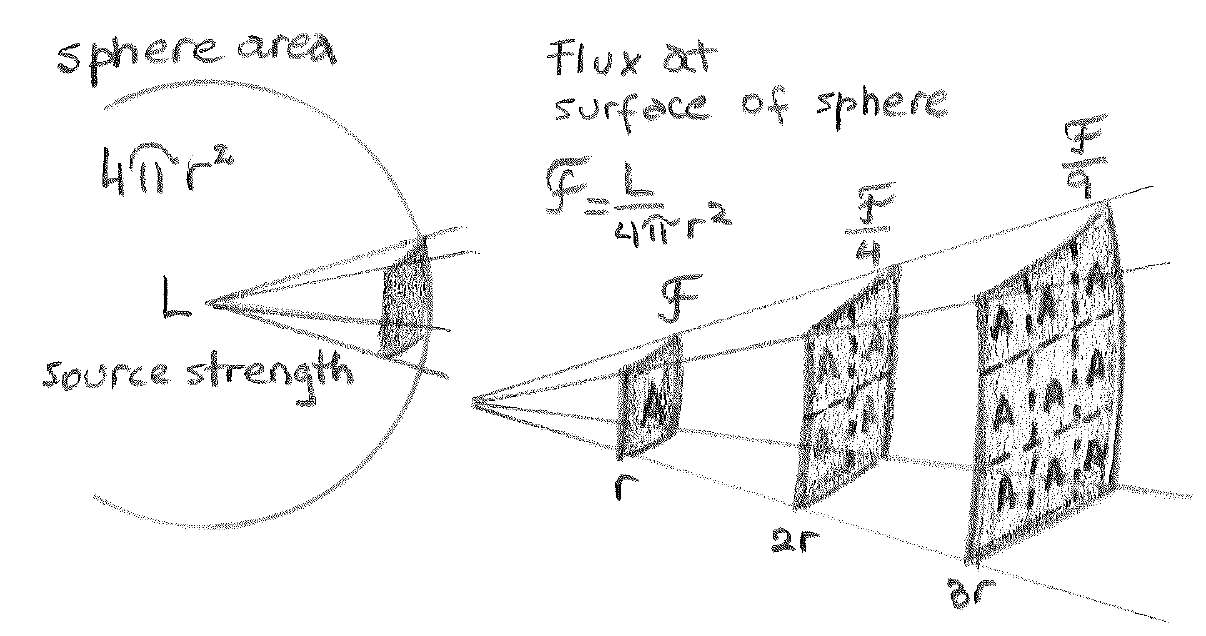}{0.8}
\caption{Inverse square law. The energy twice as far from the source is spread over four
  times the area, hence one-fourth the flux.
  \label{fig:51}}
\end{figure}

{\bf 41.}~Consider a source which emits energy at a rate of $L$ units
per second (the type of source, and the units of $L$ are actually
irrelevant for this discussion). This situation is shown in the
diagram of Fig.~\ref{fig:51}.  Consider a sphere centered on the source,
and surrounding it at a radius $r$. If we assume the energy flows out
isotropically (this means the flux is the same in all directions) from
the source, then the energy received at any point on the sphere should
be the same. It is easy to calculate the flux on the sphere, which is
the energy as it passes through the sphere (energy/ unit area). It is
just the total energy divided by the surface of the sphere, see (\ref{Lodot}).
Now, extend this idea to spheres at different radii: the surface area
of each sphere increases as $r^2$, so the flux of the energy (energy per unit
area) must reduce as $1/r^2$. This is known as the inverse square
law. Approximately $1.6 \times 10^{38}$ neutrinos are produced by the $pp$ chain
in the Sun every second~\cite{Serenelli:2011py}. Using the inverse square law calculate the
number of neutrinos from the Sun that are passing through your brain
every second. Imagine your brain as a circle with a diameter of 15~cm. \\

{\bf 42.}~Consider a $2 M_\odot$ neutron star. The mass of a neutron
is $1.67 \times 10^{-24}~{\rm g}$, and $M_\odot = 2 \times
10^{33}~{\rm g}$. {\it (i)}~How many neutrons are in this neutron
star? {\it (ii)}~Assuming the energy released during core bounce and
supernova phase of this neutron star was $3.037 \times 10^{-16}~{\rm
  Btu}$ per neutron, calculate the total energy output of the
supernova in joules. {\it (iii)}~Show that the total energy output of
the supernova delivered over the course of a few seconds may be as
much as the total output of our Sun
during its 10 billion year lifetime. Note that $1~{\rm Btu} = 1,055~{\rm J}$.\\

{\bf 43.}~The Schwarzschild radius is the radius of a sphere such
that, if all the mass of an object were to be compressed within that
sphere, the escape velocity from the surface of the sphere would equal
the speed of light~\cite{Schwarzschild:1916uq}. If a stellar remnant
were to collapse to or below this radius, light could not escape and
the object is no longer directly visible outside, thereby forming a
black hole. Estimate the size of the supermassive black hole candidate
at the center of our Galaxy. The black hole mass has been estimated to
be $4 \times 10^6 M_\odot$~\cite{Ghez:2008ms}.\\

{\bf 44.}~Compare how much $^{235}$U is required to fission
and how much gasoline is required to burn in order
to boil a bathtub (which is approximately 280~liters of water).  Consider
the initial temperature of the water at $15^\circ$C. Recall that 
in 1 gram of $^{235}$U there are $N_A/235 = 2.6 \times 10^{21}$ atoms.\\

{\bf 45.}~The escape velocity from Earth is $4 \times
10^4$~km/h. What would be the percent decrease in length of a 95.2~m
long spacecraft traveling at that speed? \\

{\bf 46.}~At what speed do the relativistic formulas for length and
time intervals differ from the classical values by 1\%?\\

{\bf 47.}~Space explorer Harry sets off at a steady $0.95c$ to a
distant star. After exploring the star for a short time, he returns at
the same speed and gets home after a total absence of $80~{\rm yr}$
(as measured by earth-bound observers). How long do Harry's clocks say
he was gone, and by how much has he aged as compared to his twin Sally 
who stayed behind on Earth. [Note: This is the famous ``twin
paradox.'' It is fairly easy to get the right answer by judicious
insertion of a factor of $\gamma$ in the right place, but to
understand it, you need to recognize that it involves three uniformly
moving reference frames: the earth-bound frame $S$, the frame $S'$ of
the outbound rocket, and the frame $S''$ of the returning
rocket. Write down the time dilation formula for the two halves of the
journey and then add. Noticed that the experiment is not symmetrical
between the two twins: Sally stays at rest in the single uniformly moving frame $S$, but Harry occupies at least two different frames. This is what allows the result to be unsymmetrical.]\\

{\bf 48.}~A distant galaxy is observed to have a redshift $V/c =
0.1$, where $V$ is the recession velocity of the galaxy, and $c$ is
the speed of light. {\it (i)}~What is the recession velocity of the
galaxy in units of km/s?  {\it (ii)}~Using the Hubble expansion
formula calculate the distance to the galaxy in units of ly? {\it
  (iii)}~How long ago was the light we are now seeing from the galaxy
emitted?\\
 
{\bf 49.}~The electromagnetic spectrum includes a wide range of light
waves, some that we cannot see. Some of the non-visible types of waves
are radio waves, microwaves, infrared rays, and X-rays. In the visible
spectrum of light, the color of the light depends on the frequency. We
have seen in Sec.~\ref{sec6} that the visible spectrum is always the same for
a rainbow, or the separated light from a prism. The order of colors is
red, orange, yellow, green, blue, indigo, and violet. Wien's Law tells
us that objects (such as stars) of different temperature emit spectra
that peak at different wavelengths~\cite{Wien:1894}. Hotter objects
emit most of their radiation at shorter wavelengths; hence they will
appear to be {\it bluer}. Cooler objects emit most of their radiation
at longer wavelengths; hence they will appear to be {\it redder}. The
wavelength of the peak of the spectrum emitted by a star gives a
measure of the temperature,
\begin{equation}
{\rm maximum \ intensity \ wavelength } =
\frac{0.29}{T}~{\rm cm} \, ^\circ{\rm K} \, .
\label{wien}
\end{equation}
{\it (i)}~Use (\ref{Lodot}) and (\ref{SB3}) to determine the temperature of the
sun, which has a radius $R_\odot = 432,288$~miles. {\it (ii)}~Using
(\ref{wien}) convince yourself that while the sun does emit
ultraviolet radiation, the majority of solar energy comes in the form
of {\it light} in the visible regions of the electromagnetic spectrum,
$390 \lesssim \lambda/{\rm nm} \lesssim 700$, where one nanometer (nm)
equals $10^{-9}~{\rm m}$.\\

{\bf 50.}~We have seen that the average translational kinetic energy
of molecules in a gas is directly proportional to the temperature of
the gas. We can invert (\ref{k_b}) to find the average speed of molecules in a
gas as a function of the temperature,
\begin{equation}
{\rm average \ speed} = \sqrt{\frac{3 \times k_B \times {\rm
      temperature}}{{\rm average \ mass}}} \, .
\label{prob45}
\end{equation}
If the average speed of a gas is greater than about 15\% to 20\% of
the escape speed of a planet, virtually all the molecules of that gas
will escape the atmosphere of the planet. {\it (i)}~At what
temperature is average speed for O$_2$ equal to 15\% of the escape
speed of the Earth? {\it (ii)}~At what temperature is the average
velocity for H$_2$ equal to 15\% of the escape speed for Earth? {\it
  (iii)}~Temperatures in the upper atmosphere reach
$1,000^\circ$K. How does this help account for the low abundance of
hydrogen in Earth's atmosphere? {\it (iv)}~Compute the temperatures
for which the average speeds of O$_2$ and H$_2$ are equal to 15\% of
the escape speed at the surface of the moon, where the mass of the
moon is $1.6 \times 10^{23}~{\rm lb}$ and its radius $1,080~{\rm
  miles}$. How this account for the absence of an atmosphere on the
moon? The average mass of a hydrogen atom is $1.674 \times
10^{-24}$~g and that of a oxygen atom is $2.657 \times 10^{-23}~{\rm
  g}$.

\acknowledgments{This work has been partially supported by the U.S. National
  Science Foundation (NSF) Grant No. PHY-1620661 and by the National
  Aeronautics and Space Administration (NASA) Grant No. NNX13AH52G.}

\onecolumngrid

\section*{Answers and Comments on the Questions}

\begin{enumerate}

\item The total time-of-use is 480~hours, so the average power is 625~watts.

\item {\it (i)}~In 14 days a person will consume $30,800~{\rm Cal}$,
  which corresponds to about 103~oz. Now, since $1~{\rm oz} =
  0.0625~{\rm lb}$, the person will lose about 6.4~pounds. {\it
    (ii)}~Now if the intake is $1,600~{\rm Cal}$, and it burns
  $2,200~{\rm Cal}$ per day, it will lose 2~oz per day, or
  equivalently 600~Cal. So if the person loses 0.125~lb per day, to
  lose 15~lb it will take 120 days.

\item The energy used in driving a mile is $2,133~{\rm Cal}$ and that used
  as electrical energy (per day) for an average household is $8,600~{\rm
    Cal}$.

\item To raise the temperature $6^\circ$C the astronaut will
  need to burn 286.8~Cal, or equivalently 0.333~kWh. If the astronaut
  is expending 200~watts of power,  in 333/200~hr which is about
  100 minutes, would reach critical damage temperatures. 

\item If the cost of heating in a house is \$60 per month, this means
  that in a month the family uses 150 gallons of oil, or equivalently
  $2.1 \times 10^7~{\rm Btu}$. Now, since $1~{\rm kWh} = 3,412~{\rm
    Btu}$, it is easily seen that the family uses $6,155~{\rm kWh}$. Therefore
  to heat the same house with electricity will cost  \$739.

\item {\it (i)}~The annual cost per family is $ \$50  \times 7.8 \times
  10^6 \times 365 \times 4/ (3.22 \times 10^8) = \$1,768$.  {\it
    (ii)}~The price would be $\$50 \times 2.2/42 + 0.184 = \$2.8$ per
  gallon. This would not change substantially the American driving
  habits.

\item {\it (i)}~The burnout instantaneous speed $= 30~{\rm m/s}^2 
  \times 100~{\rm s} =
  3,000~{\rm m/s}$. {\it (ii)}~The average speed = $1,500~{\rm m/s}$. {\it
    (iii)}~The  height at burnout = $1,500~{\rm m/s} 
  \times  100~{\rm s} =
  150~{\rm km}$. {\it (iv)}~The coasting time = $(3,000~{\rm
  m/s})/(9.8~{\rm m/s^2})= 306.12~{\rm s}$, and the average coasting
speed = $1,500~{\rm m/s}$, yielding a coasting height = $1,500~{\rm m/s} \times
306.12~{\rm s} = 459,180~{\rm m}$. The maximum height is then 609.18~km.

\item  {\it (i)}~The energy wastage per day is $32$~MkWh. {\it
    (ii)}~The added electrical cost per year is \$17. {\it (iii)}~The
  energy wastage per year is $10^{10}~{\rm kWh}$. {\it (iv)}~This
  corresponds to $48,348$~bbl per day.

\item To calculate the force on Earth we need the distance of each
  planet from Earth: $r_{\rm EV} = (150 - 108) \times 10^6~{\rm km} =
  4.2 \times 10^{10}~{\rm m}$, $r_{\rm EJ} = (778 - 150) \times
  10^6~{\rm km} = 6.28 \times 10^{11}~{\rm m}$, and $r_{\rm ES} =
  (1430 - 150) \times 10^6~{\rm km} = 1.28 \times 10^{12}~{\rm
    m}$. Jupiter and Saturn will exert a rightward force, while Venus
  will exert a leftward force. Take the right direction as positive, 
  and so force due to the planets on the Earth is
\begin{eqnarray}
F_{\rm Ep} & = &  G \frac{M_\oplus M_J}{r^2_{\rm EJ}} + G \frac{M_\oplus M_{\rm S}}{ r^2_{\rm
  ES}} - G \frac{M_\oplus M_{\rm V}}{ r_{\rm EV}^2} \nonumber \\
& = & GM_\oplus^2
\left[\frac{318}{6.28 \times 10^{11}~{\rm m})^2} + \frac{95.1}{(1.28 \times
    10^{12}~{\rm m})^2} - \frac{0.815}{(4.2 \times 10^{10}~{\rm m})^2}
  \right] \nonumber \\ & = & (6.67 \times 10^{-11}~{\rm N \,
    m^2/kg^2}) \times (5.97 \times
  10^{24}~{\rm kg})^2 \times (4.02 \times 10^{-22}~{\rm m^{-2}}) = 9.56
  \times 10^{17}~{\rm N} \, . \nonumber
\end{eqnarray}
The force of the Sun on Earth is as follows
\begin{equation}
F_{\rm ES} = G \frac{M_\oplus M_\odot}{r_{\rm SE}^2} = (6.67 \times
10^{-11}~{\rm N \, m^2/kg^2}) \times \frac{(5.97 \times
  10^{24}~{\rm kg}) \times (1.99 \times 10^{30}~{\rm kg})}{(1.50 \times
  10^{11}~{\rm m})^2} = 3.52 \times 10^{22}~{\rm N} \nonumber \, .
\end{equation}
Finally, the ratio is $F_{\rm Ep}/F_{\rm ES} = 9.56 \times
10^{17}~{\rm N}/ 3.52 \times 10^{22}~{\rm N} = 2.71 \times
10^{-5}$, which is 27 millionths.

\item The expression for the acceleration due to gravity at the
  surface of the Earth is given by (\ref{Earth-g}). For Mars, $g_{\rm M} = 0.38
  g$ and so  $G M_{\rm M}/R^2_M = 0.38 G M_\oplus/R_\oplus^2$. It
  follows that $M_{\rm M} = 0.38 M_\oplus (R_{\rm M}/R_\oplus)^2 = 6.4
  \times 10^{23}~{\rm kg}$.

\item The pressure at the bottom of the ocean is $P = \rho g h =
  10^8~{\rm N/m^2}$, and so the force on the window is $F = P \pi r^2
  \approx 5 \times 10^6~{\rm N}.$ The descent into the Challenger Deep took
  nearly five hours. Once the Bathyscaphe Trieste reached the sea
  floor, Walsh and Piccard observed their surroundings. The ship's
  light allowed them to see what they described as a dark brown
  ``diatomaceous ooze'' covering the sea floor, along with shrimp and
  some fish that appeared to resemble flounder and sole. Since the
  Plexiglas viewing window had cracked during the descent, the men
  were only able to spend about twenty minutes on the sea floor. Then,
  they unloaded the ballasts (nine tons of iron pellets, and tanks
  filled with water) and began to float back to the ocean's
  surface. The ascent was much quicker than the dive, taking only
  three hours and fifteen minutes.

\item Because the iceberg is in equilibrium, the buoyant force equals
  its weight, $B = F_g$, and so $\rho_{\rm ice} V g = \rho_{\rm sea-water} V_{\rm sub} g$. It follows that the fraction which is
  submerged is $f= V_{\rm
    sub}/V = \rho{\rm ice}{\rho_{\rm sea-water} }=0.89$, i.e. 89\%.

\item From (\ref{continuity2}) we have $v_{\rm cap} N \pi r^2_{\rm
    cap} = v_{\rm aorta} \pi r^2_{\rm aorta}$, and so $N = v_{\rm
    aorta} r_{\rm aorta}^2/(v_{\rm cap} r_{\rm cap}^2) = 7 \times 10^9$.

\item  {\it (i)} Neglecting inefficiencies (that  transform energy into
  heat) we can consider the system to be isolated. Then, the entire potential
  energy of Harry before he drops goes into the potential energy of
  Sally rising to her peak, that is at Sally's  moment of zero kinetic
  energy. {\it (ii)} ~The potential energy of Harry equals the
  potential energy of Sally, i.e., $U_{\rm H} = U_{\rm
    S}$. This means that $m_{\rm H}gh_{\rm H} = m_{\rm S} gh_{\rm S}$, and
  so $h_{\rm S} = m_{\rm H} h_{\rm H}/m_{\rm S}$. {\it (iii)}~Since 1~lb =
  0.45~kg, we have $h_{\rm
    S} = (70~{\rm kg}/40~{\rm kg}) \times 4~{\rm m} = 7~{\rm m}$.

\item The final speed is related to the total kinetic energy at the
  bottom of the mountain, which in turn is related to the total
  gravitational potential energy at the top of the hill, as the system
  is isolated. The potential
  energy Harry has at the top of the mountain is $U_{\rm H} = m_{\rm
    H} gh$, where $m_{\rm H}$ is Harry's mass. The initial potential
  energy of Sally is $U_{\rm S} = m_{\rm S} gh$, where $m_{\rm S}$ is
  her mass. The final kinetic energies of Harry and Sally are $K_{\rm
    H} = m_{\rm H} v_{\rm H}^2/2$ and $K_{\rm S} = m_{\rm S} v_{\rm
    S}^2/2$, respectively. By equating the initial potential energy to
  the final kinetic energy it is easily seen that both Harry and Sally
  will have the same speed at the bottom of the hill. Harry wins,
  because the bet was that she would not be going faster than him.

\item The escape velocity on Earth is the minimum velocity with which
  a rocket (of mass $m$) has to be projected vertically upwards from
  the earth's surface so that it just crosses the earth's
  gravitational field and never returns. Just after the rocket is
  launched, the potential energy of the system is $U_i = - G m
  M_\oplus/R_\oplus$ and its kinetic energy is $K_i = mv^2/2$. When it
  escapes the earth's gravitational field (at an infinite height above
  the earth's surface) the potential energy is zero. At the critical
  escape velocity $v_{\rm esc}$, the velocity of the spacecraft at
  this point drops to zero. The total energy at escape is therefore
  zero.  By energy conservation we know that the initial total energy
  of the system must equal the final total energy, that is $U_i + K_i
  = U_f + K_f = 0$. Therefore, $mv_{\rm esc}^2/2 = G m
  M_\oplus/R_\oplus$.  One thing we notice is there is a single $m$ on
  both sides of the equation, so they cancel. What does this tell us?
  The escape velocity does not depend upon the mass of the rocket:
  $v_{\rm esc}^{\rm Earth} = \sqrt{2 G M_\oplus/R_\oplus}$. The escape velocity at
  Earth is 11.2~km/s.

\item {\it (i)}~Since 1~oz = 0.062~lb, the weight loss is
  0.31~lb. {\it (ii)}~No, because the process is irreversible.

\item The theoretical efficiency is $\eta_{\rm max} = 0.85$.

\item {\it (i)}~A Btu raises the temperature of a 1~lb of water
  $1^\circ$F. Then $12,000~{\rm Btu}$ will raise 120~lb of water
  $100^\circ$F. Since the initial temperature is $98.6^\circ$F the
  final temperature would be $198.6^\circ$F. {\it (ii)}~Since 1~Cal =
  4~Btu, it follows that $1,600$~Cal with 25\% efficiency amounts to
  $1,600$~Btu. Now, 1~Btu = 779~ft lb, and so the total high $= 1,600
  \times 779/120 \approx 10,387$ feet $\approx 3,166$~m.

\item {\it (i)}~The initial momentum of the system is $p_i = m_m v_m$,
  where $m_m$ and $v_m$ are the mass and velocity of the meteor,
  respectively. The final momentum of the system is $p_f = (m_m +
  M_\oplus) v_f$. Since the momentum during the collision is conserved
  $p_i = p_f$ and so the recoil speed is $v_f = 6 \times
  10^{-11}~{\rm m/s} \approx 2~{\rm mm}$ per year. {\it (ii)}~The fraction of the meteor's kinetic
  energy that was transformed to kinetic energy of the Earth is equal
  to $K_f^{\rm Earth}/K_i^{\rm meteor} \approx 8 \times 10^{-15}$. {\it
    (iii)}~The change in the Earth's kinetic energy is $\Delta K_{\rm
    Earth} = K_f^{{\rm Earth}} - K_i^{{\rm Earth}} = M_\oplus v_f^2/2 =
  10,800~{\rm J}$.

\item {\it (i)}~The frequency is $\nu = 1.27~{\rm Hz}$. {\it (ii)}~The
  period is ${\cal T} = 0.60~{\rm s}$. {\it (iii)} It takes
  0.20~s. {\it (iv)}~It takes 0.90~s.

\item The oscillating period is ${\cal T} = 6.38~s$, so from
  (\ref{pendulum-period}) we have $g_{\rm Pluto} = \ell \times (2 \pi/{\cal
    T})^2 = 0.62~{\rm m/s^2}$.

\item The total time is the sum of the time it takes the rock to reach
  the bottom  and the time it takes the sound wave to reach you. The
  average velocity of the rock is $\langle v_{\rm rock} \rangle = g
  t_{\rm rock}/2$, so the time it takes is $t_{\rm rock} = \sqrt{2 \ell/g} = 5~{\rm
    s}$. The time for the sound wave to go up is $t_{\rm sound} =
  \ell/v_{\rm sound}
  = 0.36~{\rm s}$, where using (\ref{v-sound}) we have taken $v_{\rm sound} = (331.5 + 0.6 10)~{\rm
    m/s}$. The total time is 5.36~seconds.

\item Because both the source and observer are in motion, there will
  be two Doppler shifts: first for the emitted sound with the sub A as
  the source and sub B as the observer, and then the reflected sound
  with sub B as the source and sub A as the observer. {\it (i)}~Using
  (\ref{ptown1}) we have $$\nu_{\rm received}^{\rm target} = \nu_{\rm
    emitted} \ \frac{v_{\rm sound} + V_B}{v_{\rm sound} - V_A} =
  10.14~{\rm kHz} \, .$$ 
{\it (ii)}~Using (\ref{ptown2}) we have $$\nu_{\rm echo} =
  \nu_{\rm received}^{\rm target}  \ \frac{(v_{\rm sound} +
    V_A)}{(v_{\rm sound} - V_B)} = \nu_{\rm emitted}  \ \frac{(v_{\rm
      sound} +V_B)}{(v_{\rm sound}
    - V_A)} \
  \frac{(v_{\rm sound} + V_A)}{(v_{\rm sound} - V_B)} = 10.28~{\rm kHz}.$$

\item The frequency heard when you move towards a sound source at rest is 
$\nu_{\rm received}^{\rm towards} = \nu_{\rm emitted}  (1+ V_{\rm
  receiver}/v_{\rm sound})$, where $\nu_{\rm emitted}$ is the emitted frequency,
$V_{\rm receiver}$ is the  velocity of receiver, and $v_{\rm sound}$ is the speed of sound. 
The frequency heard when moving away from the sound source is 
$\nu_{\rm received}^{\rm away} = \nu_{\rm emitted} (1 - V_{\rm
  receiver}/v_{\rm sound})$. The difference between these frequencies
 $\nu_{\rm received}^{\rm towards} - \nu_{\rm received}^{\rm away} =
 95~{\rm Hz}$ and so $V_{\rm receiver} = 16.97~{\rm m/s}$.

\item {\it (i)}~Take the ratio of the electric force
divided by the gravitational force, that is $$\frac{F_e}{F_g} =
\frac{k_e \ e^2}{G \, m_p m_e}  =
\frac{8.99 \times 10^9~{\rm N} \cdot {\rm m}^2/{\rm C}^2 \times (1.602
  \times 10^{-19}~{\rm C})^2}{6.67 \times 10^{-11}~{\rm N} \cdot {\rm
    m}^2/{\rm kg}^2  \times 9.11 \times 10^{-31}~{\rm kg} \times 1.67 \times
  10^{-27}~{\rm kg}} \simeq 2.3 \times 10^{39} \, .
$$ The electric force is
about $2.3 \times 10^{39}$ times stronger than the gravitational force
for the given scenario. {\it (ii)}~No. Life would be no different if electrons were
positively charged and protons were negatively charged. Opposite
charges would still attract, and like charges would still repel. The
designation of charges as positive and negative is merely a
definition.

\item {\it (i)}~Power = 1.5 volts $\times$ 1 ampere = 1.5~watts = $1.5
  \times 10^{-3}~{\rm kW}$. {\it (ii)}~Energy = Power $\times$ time =
  $1.5 \times 10^{-3}~{\rm kW} \times 1~{\rm hr} = 1.5 \times
  10^{-3}~{\rm kWh}$. {\it (iii)}~Cost/kWh = $\$0.5/1.5 \times 10^{-3} =
  \$333$. 

\item $1,000$~W = 1 kW = 1~kWh/hr = 3410~Btu/Hr. At $1,400$~Btu/oz we
  have $3,410/1,400 = 2.4$~oz/hr.

\item {\it (i)}~Current = power/volts = 750/150~amperes =
  6.8~amperes. {\it (ii)}~Energy used = power $\times$ time = 0.75~kWh
  $\times 8$~hr = 6 kWh. Now, at $12\cent$/kWh we have $72\cent$.

\item $R_3$ and $R_4$ are connected in parallel with equivalent
  resistance given by $R^{-1}_{\rm eq} = R_3^{-1} + R_4^{-1}$,
  yielding $R_{\rm eq} = 0.75~\Omega$. The total resistance is 
  $R_{\rm total} = R_1 + R_2 +  R_{\rm eq}  = 6.75~\Omega$. The current
  circulating through $R_1$ and $R_2$ is  $i= V/R_{\rm total} =
  1.18~{\rm A}$.  This current splits into $i_3$ and $i_4$ satisfying
  $i_3 R_3 = i_4 R_4$ and $i = i_3 + i_4$. Therefore, $i_3 = R_4
  i/(R_3 + R_4) = 0.295~{\rm A}$ and $i_4 = R_3 i /(R_3 +R_4) =
  0.885~{\rm A}$. The power delivered to $R_1$ is  $i^2R_1 =
  2.78$~watts, to $R_2$ is $i^2 R_2= 5.57$~watts, to $R_3$ is $i_3^2
  R_3 = 0.26$~watts, and to $R_4$ is $i_4^2 R_4 =0.78$~watts.

\item The speed of light is $c = 3 \times 10^8~{\rm m/s}$ and since $t
  = d/c$: {\it (i)}~using the distance to Polaris $d = 6.44 \times
  10^{18}$~meters we obtain $t \approx 2.15 \times 10^{10}~{\rm
    seconds} \approx 5.972 \times 10^6~{\rm hours} = 248,840~{\rm days}
  \approx 682~{\rm years}$; {\it (ii)}~the distance from the Sun to
  the Earth is $d = 150 \times 10^9$~meters and therefore $t =
  500~{\rm seconds} = 8$ minutes and 20 seconds. The distance from the
  Earth to the Moon is $d= 3.84 \times 10^{8}~{\rm meters}$, thus $t =
  2.56$~seconds.  {\it (iii)}~The Earth radius is $R_\oplus =
  6,371~{\rm km}$, and the distance of maximum circle around the Earth
  is $d = 4 \times 10^7$~meters, so $t = 0.13~{\rm seconds}$. {\it (iv)}~
  For $d = 10~{\rm kilometers}$, we have $t = 3 \times 10^{-5}~{\rm seconds}$.

\item {\it (i)}~According to Roemer measurements the speed of light is $c = 2 \times {\rm distance}_{\rm SE}/(22~{\rm minutes}) =
  2.27 \times 10^8~{\rm m/s}$. {\it (ii)}~A more precise determination of the speed of light gives $c = 2 \times {\rm distance}_{\rm ME}/ (2.51~{\rm seconds}) =
  2.99516 \times 10^8~{\rm m/s}$.

\item Using (\ref{Snell-ley2}) we can write $\sin \theta = 1.48 \sin
  20^\circ$ and so $\theta = 30^\circ$. Likewise, $1.48 \sin 20^\circ = 1.33
  \sin \theta'$, and so $\theta' = 22^\circ$.

\item {\it (i)}~The thin lens equation (\ref{thinlens}) can be used to
  find the image distance,
$$\frac{1}{30~{\rm cm} } + \frac{1}{\rm image \ distance} = \frac{1}{10~{\rm cm}}
\Rightarrow {\rm image \ distance } = 15~{\rm cm}.$$ The positive sign
for the image distance tells us that the image is indeed real and on
the back side of the lens. From (\ref{Magnification}) we obtain the
magnification of the image $= 
-0.5$. Hence, the image is reduced in height by one half, and the
negative sign for the magnification tells us that the image is
inverted. {\it (ii)}~No calculation is necessary for this case because
we know that, when the object is placed at the focal point, the image
is formed at infinity. This is readily verified by substituting an
object distance of 􏱉10~cm into the thin lens equation.  {\it (iii)}~We
now move inside the focal point. In this case the lens acts as a
magnifying glass; that is, the image is magnified, upright, on the
same side of the lens as the object, and virtual. Because the object
distance is 5~cm, the thin lens equation gives
$$\frac{1}{5~{\rm cm} } + \frac{1}{\rm image \ distance} = \frac{1}{10~{\rm cm}}
\Rightarrow {\rm image \ distance } = -10~{\rm cm}$$ and the
magnification of the image $= 2$. The
negative image distance tells us that the image is virtual and formed
on the side of the lens from which the light is incident, the front
side. The image is enlarged, and the positive sign for the magnification tells us
that the image is upright. {\it (iv)}~We use the thin lens equation to
  find the image distance,
$$\frac{1}{30~{\rm cm} } + \frac{1}{\rm image \ distance} = - \frac{1}{10~{\rm cm}}
\Rightarrow {\rm image \ distance } = -7.5~{\rm cm}.$$ The
magnification of the image $= 0.25$. We conclude that the image is
virtual, smaller than the object, and upright. {\it (v)}~When the
object is at the focal point we have
$$\frac{1}{10~{\rm cm} } + \frac{1}{\rm image \ distance} = - \frac{1}{10~{\rm cm}}
\Rightarrow {\rm image \ distance } = -5~{\rm cm}.$$ The magnification
of the image =  0.5~cm. Notice the difference between this
situation and that for a converging lens. For a diverging lens, an
object at the focal point does not produce an image infinitely far
away. {\it (vi)}~When the object is inside the focal point we have
$$\frac{1}{5~{\rm cm} } + \frac{1}{\rm image \ distance} = -\frac{1}{10~{\rm cm}}
\Rightarrow {\rm image \ distance } = -3.33~{\rm cm}$$ and the
magnification of the image = 0.667. In this case the virtual image is
upright and shrunken.

\item {\it (i)}~$V_{\rm max} = B A \omega = B \pi r^2 \omega =
  1.3~{\rm T} \times \pi \times (0.25~{\rm m})^2 \times 4 \times
  \pi~{\rm s}^{-1} = 3.2~{\rm V}$. {\it (ii)}~The maxima occur when
  the argument of the sine equals $(4\varkappa +1)/2$, with $\varkappa
  = 0,1,2,3,\cdots$. In other words, $\omega t = (4 \varkappa+1)\pi/2$
  and so $t = (4 \varkappa+ 1)/8~{\rm s}$. {\it (ii)}~The periodic
  motion has a frequency $\nu = 2~{\rm Hz}$ and so the period is
  ${\cal T} = 0.5~{\rm s}$.

\item {\it (i)}~Since we have $63.546~{\rm g/mol}$ of $^{63}$Cu there
  are 0.0488~mol. {\it (ii)}~We know there are $6.022 \times
  10^{23}~{\rm atoms/mol}$, then in 0.0488~mol we have $2.94 \times
  10^{22}$~{\rm atoms}. {\it (iii)}~Since 1 mol has $6.022 \times
  10^{23}$ atoms, in $10^{12}$ atoms there are $1.66 \times
  10^{-12}~{\rm mol}$. There are $196.967~{\rm g/mol}$ of $^{197}$Au,
  so the mass of a trillion gold atoms is $3.27 \times 10^{-10}~{\rm
    g}$. {\it (iv)}~Since one molecule of CO$_2$ contains one carbon
  atom, one mole of CO$_2$ molecules will contain one mole of carbon
  atoms. If we have 0.25~mol of CO$_2$, there will be 0.25 mol of
  carbon atoms present. Since there are two oxygen atoms in one CO$_2$
  molecule, there are $2 \times 0.25~{\rm mol} = 0.5~{\rm mol}$ of
  oxygen atoms present in this amount. {\it (v)}~$1.5 \times 10^{12}$
  pounds of oil are burned in the U.S. per year, from which $15,000$
  pounds of mass disappears.

\item If each person has a mass of about 70~kg and is (almost)
  composed of water, then each person contains
\begin{equation}
N \simeq \left(\frac{70,000~{\rm grams}}{18~{\rm grams/mol}} \right)
\times \left( 6.023 \times 10^{23}~\frac{{\rm molecules}}{\rm mol}
\right) \times
\left(10~\frac{\rm protons}{\rm molecule}\right) \simeq 2.3 \times
10^{28}~{\rm protons} \, . \nonumber
\end{equation}
With an excess of 1\% electrons over protons, each person has a charge
$q = 0.01 \times 1.602 \times 10^{-19}~{\rm C} \times 2.3 \times
10^{38} \simeq 3.7 \times 10^7~{\rm C}$, and so the electric force is
\begin{equation}
F_e = k_e \frac{q_1 q_2}{r^2} = 9 \times 10^9 \times \frac{(3.7 \times
  10^7)^2}{0.6^2}~{\rm N} = 4 \times 10^{25}~{\rm N} \sim 10^{26}~{\rm
  N} \, . \nonumber
\end{equation} 
This force is almost enough to lift a weight equal to that of the Earth
$M_\oplus g = 6 \times 10^{24}~{\rm kg} \times 9.8~{\rm m/s}^2 = 6
\times 10^{25}~{\rm N} \sim 10^{26}~{\rm N}$.

\item Since $(1/2)^{1.3} = 0.4$ it must have taken about 1.3
half-lives to get them to the present ratio. So we conclude the soot is
$7,449$ years old.

\item The total heat falling on a $10~{\rm m}^2$ roof is equal to $ 7~{\rm kW} =
  23,884~{\rm Btu/hr}$. We want to raise the temperature of 480~lb of
  water by $100^\circ$F. To heat 1~lb of water $1^\circ$F takes
  1~Btu. A straightforward calculation pinpoints that we need
  $48,000~{\rm Btu}$, and so the time required is about 2~hours.

\item The cost of electricity is \$120/MWh. From Fig.~\ref{fig:49} we
  see that the event lasted about 5 hours or $18,000~{\rm seconds}$, and so about $3,400~{\rm
    MW} \times 18,000~{\rm s} = 6.12 \times 10^{13}~{\rm J} =
  17,000~{\rm MWh}$
  came off the system during the eclipse. This gives a cost of
  $17,000~{\rm MWh} \times \$120/{\rm MWh} \approx  2$ million dollars.

\item The flux density of neutrinos at Earth is
\begin{equation}
\mathscr{F}_\nu = \frac{1.6 \times 10^{38}~{\rm neutrinos/s}}{4 \pi d^2} = 6 \times
10^{10}~\frac{\rm neutrinos}{\rm cm^2 \, s} \, , \nonumber
\end{equation} 
where $d$ is the Sun-Earth distance. Thus, the flux of neutrinos
passing through the brain per second is
\begin{equation}
\frac{\Delta N_\nu}{\Delta t} =  \mathscr F_\nu \, A_{\rm brain} = 6 \times
10^{10}~\frac{\rm neutrinos}{\rm cm^2 \, s} \frac{\pi D_{\rm brain}^2}{4}
\approx  10^{13}~\frac{\rm neutrinos}{\rm s} \,, \nonumber
 \end{equation}
where we have assumed that the  diameter of the brain is
$D_{\rm brain} \simeq 15~{\rm cm}$.

\item {\it (i)}~The ${\rm number \ of \ neutrons} = {\rm mass \ of \
    neutron \
    star}/{\rm neutron \ mass} = 2 \times M_\odot/m_n = 2 \times 2 \times 10^{33}~g/(1.67 \times
  10^{24}~{\rm g}) = 2.4 \times 10^{57}$. {\it (ii)}~The total energy output of the supernova $=
  ({\rm number \ of \ neutrons}) \times ({\rm energy \ per \ neutron}) = 2.4 \times
  10^{57} \times 3.037 \times 10^{-16}~{\rm Btu} = 7.3 \times 10^{41}~{\rm
    Btu} = 7.7 \times 10^{44}~{\rm J}$. {\it (iii)}~From (\ref{Lodot}) we know
  that the rate at which the sun emits energy (its luminosity) is
  around $4 \times 10^{23}~{\rm Btu/s}$. Assuming a constant rate
  during the estimated lifetime of $10^{10}~{\rm yr}$ we find that the
  total energy output of the sun = $1.3 \times 10^{44}~{\rm J}$.

\item The escape velocity from the surface ({\it i.e.}, the event
  horizon) of a black hole is exactly $c$, the speed of
  light. Particularizing this to the black hole candidate at
  the center of our Galaxy we have $v_{\rm esc}^{\rm BH} = c = \sqrt{2 G M_{\rm
      BH}/R_{\rm S}}$, where $M_{\rm BH} = 4 \times 10^6 M_\odot$ is the black
  hole mass and $R_{\rm S}$ is the Schwarzschild radius.  The size of
  the black hole is characterized by $R_S = 2 GM/c^2 = 7.4$ million miles.

\item We have seen that a Cal raises the temperature of 1 liter of
  water $1^\circ$C. Then, the heat required to raise the temperature
  from $15^\circ$C to $100^\circ$C is $Q = ({\rm variation \ of \
    temperature}) \times ({\rm number \ of \ liters}) = 21,000~{\rm
    Cal} \approx 9 \times 10^7~{\rm J}$. Now, since the energy
  released by the fission of one uranium-235 nucleus is approximately $3 \times
  10^{-11}~{\rm  J}$  and there are roughly $2.6 \times 10^{21}$ atoms in 1 gram
  of $^{235}$U, the amount of energy released when
  1 gram of uranium-235 undergoes fission is about $8\times 10^{10}~{\rm
    J}$. Thus, the heat required to boil the bathtub can be obtained
  through fission of $10^{-3}$~grams, or 1~mg (a speck) of
  $^{235}$U. Recalling that the energy content of gasoline is about 
$32,000~{\rm Cal/gal} = 1.32 \times 10^8~{\rm J/gal}$ and that 1 gallon =
3.79~liters, we have that the energy content of gasoline = $3.4\times
10^7~{\rm J/l}$. This means we need about 3 liters of gasoline. Since
the mass of 1 liter of water is 1~kg, by comparison we see that
burning gasoline would require $10^6$ times more mass. 

\item~The fractional decrease in length of the spacecraft is $\delta d'
  = (d' - d)/d'$, where $d'$
  is the length measured by observers relative to whom the vessel is
  at rest and $d$ is the contracted length of the vessel along the direction of its motion.
For a spacecraft moving at speed of $1.1
  \times 10^4~{\rm m/s}$, we have $\delta d' =
  1 - d/d' = 1 - \sqrt{1-v^2/c^2} = 7 \times 10^{-10}$, where we have
  used (\ref{l-contraction}).  This corresponds to
 $7 \times
  10^{-8}\%$. 

\item~For a 1\% change, $\sqrt{1-v^2/c^2} =  0.99$, which gives $v =
  0.14 c$. 

\item~For $v/c = 0.95$,  the Lorentz factor for both the outward and return trip is $\gamma = (1 - v^2/c^2)^{-1/2} = 3.20$. The times for the two halves of the journey satisfy $\Delta t_{\rm S}^{\rm out} = \gamma \Delta t_{\rm H}^{\rm out}$ and $\Delta t_{\rm S}^{\rm back} = \gamma \Delta t_{\rm H}^{\rm back}$, so by addition, the times for the whole journey satisfy the same relation. Thus $\Delta t_{\rm H} = \Delta t_{\rm S}/\gamma = 25~{\rm yr}$, which is the amount by which Harry has aged.\\

\item {\it (i)}~The recession velocity is $V = 0.1 c = 0.1 \times 3
  \times 10^5~{\rm km/s} = 30,000~{\rm km/s}$. {\it (ii)}~$d = V/H_0 =
  1,339~{\rm million \ light \ years}$. {\it (iii)}~The light we are
  seeing today was emitted $1,339$ million years ago.\footnote{Relativistic effects  must be taken into account for the
optical Doppler effect. Besides the ordinary Doppler effect, of order
$V/c$, a relativistic correction of order $(V/c)^2$ contribute to the
Doppler signal. For source and observer moving away from each other,
the relativistic correction leads to 
\begin{eqnarray}
\lambda' & = & \lambda
  \sqrt{(1+V/c)/(1-V/c)} \nonumber \\ 
& \approx & \lambda \left[ 1 + \frac{V}{2c} + {\cal O}
    \left(\frac{V^2}{c^2}\right) \right] \left[ 1 - \left(-\frac{V}{2c}\right) + {\cal O}
    \left(\frac{V^2}{c^2}\right) \right] \nonumber \\
& \approx & \lambda  \left[1 + \frac{V}{c} + {\cal O}
  \left(\frac{V^2}{c^2} \right) \right], \nonumber
\label{rel-doppler}
\end{eqnarray}
where $\lambda$ is the emitted wavelength
as seen in a reference frame at rest with respect to the source and
$\lambda'$ is the wavelength measured in a frame moving with velocity
$V$ away from the source along the line of sight. Note that in the
second rendition we used the binomial expansion.
For relative motion
toward each other, $V < 0$ in (\ref{rel-doppler}). For the problem at
hand, relativistic corrections are of order $(V/c)^2 = 0.01$, and can
be safely neglected.}

\item {\it (i)}~The temperature at the surface of the Sun is found to
  be $T = [L_\odot/(4 \pi \sigma_{\rm SB} R_\odot^2)]^{1/4} =
  5,914~{\rm K}$. {\it (ii)}~The solar radiation has a peak wavelength
  $\lambda = 0.29~{\rm cm}/5,914 = 5 \times 10^{-7}~{\rm m} = 500~{\rm
    nm}$,
  which is in the visible part of the electromagnetic spectrum.

\item {\it (i)}~The escape velocity from Earth is $v_{\rm esc}^{\rm  Earth} =
  \sqrt{2GM_\oplus/R_\oplus} \approx 11~{\rm km/s}$ and the average speed of oxygen molecules
  is given by (\ref{prob45}). By equating (\ref{prob45}) to 15\% of
  $v_{\rm esc}^{\rm Earth}$
  we obtain the critical temperature,
\begin{equation}
{\rm temperature}_{{\rm crit},\, {\rm O}_2}^{\rm Earth} = \frac{2}{3} \times (0.15)^2 \times G \times
M_\oplus \times \frac{1}{R_\oplus} \times  {\rm average \
  mass \ O}_2 \ {\rm molecule} \times \frac{1}{k_B} \sim
3,628^\circ{\rm K} \, , \nonumber
\end{equation}
where ${\rm average \
  mass \ O}_2 \ {\rm molecule} = 5.314 \times 10^{-26}~{\rm kg}$.
{\it (ii)}~Duplicating the procedure for H$_2$ we have
\begin{equation}
{\rm temperature}_{{\rm crit},\, {\rm H}_2}^{\rm Earth}  = \frac{2}{3} \times (0.15)^2 \times G \times
M_\oplus \times \frac{1}{
R_\oplus} \times  {\rm average \
  mass \ H}_2 \ {\rm molecule} \times \frac{1}{k_B} \sim 228^\circ{\rm
  K} \, , \nonumber
\end{equation}
where ${\rm average \
  mass \ H}_2 \ {\rm molecule} = 3.348 \times 10^{-27}~{\rm kg}$
{\it (iii)}~At a temperature of $1,000^\circ$K the average speed of
H$_2$ molecules is much larger than the escape velocity from Earth.
{\it (iv)} The escape velocity from the Moon is $v_{\rm esc}^{\rm
  Moon} =
\sqrt{2GM_{\rm Moon}/R_{\rm Moon}} \approx 2.4~{\rm km/s}$, where we
have taken $M_{\rm Moon} = 7.3 \times 10^{22}~{\rm kg}$ and $R_{\rm Moon}
= 1,737~{\rm km}$. The
critical temperature for oxygen is
\begin{equation}
{\rm temperature}_{{\rm crit},\, {\rm O}_2}^{\rm Moon}  = \frac{2}{3} \times (0.15)^2 \times G \times
M_{\rm Moon} \times \frac{1}{R_{\rm Moon}} \times  {\rm average \
  mass \ O}_2 \ {\rm molecule} \times \frac{1}{k_B} \sim 162^\circ{\rm
  K} \, , \nonumber
\end{equation}
and that of hydrogen is
\begin{equation}
{\rm temperature}_{{\rm crit},\, {\rm H}_2}^{\rm Moon}  = \frac{2}{3} \times (0.15)^2 \times G \times
M_{\rm Moon} \times \frac{1}{R_{\rm Moon}} \times  {\rm average \
  mass \ H}_2 \ {\rm molecule} \times \frac{1}{k_B} \sim 10^\circ{\rm
  K} \, . \nonumber
\end{equation}
The mean surface temperature of the Moon is $107^\circ$C
during the day and $-153^\circ$C during night, so molecules of
O$_2$ and H$_2$ would escape the attraction of the Moon.
\end{enumerate}

\end{document}